\documentclass[a4paper,12pt,final,colorlinks=true,citecolor=blue,pdfencoding=auto,psdextra]{article}
\pdfoutput=1

\usepackage{silence}
\newcommand{\BR}{\mathop{\text{BR}}\nolimits}

\newcommand{\event}{\text{\rm events}}
\newcommand{\pr}{\text{\rm prod}}
\newcommand{\decay}{{\text{\rm decay}}}
\newcommand{\ship}{\textsc{ship}}
\newcommand{\mat}{{\textsc{mat}}} %
\newcommand{\CL}{\mathcal{L}}

\usepackage{booktabs}
\usepackage[normalem]{ulem}
\usepackage{units}
\usepackage[english]{babel}
\usepackage{xspace}
\usepackage{paralist}

\usepackage{amsmath,amssymb,bm} 
\usepackage{graphicx}
\usepackage[dvipsnames,usenames]{color,xcolor}
\usepackage{jheppub} 

\usepackage[numbers,sort&compress]{natbib}

\graphicspath{{./plots/}}
\usepackage{lineno}
\usepackage{longtable}
\usepackage[]{hyperref}
\usepackage[abs]{overpic}

\title{Sensitivity of the intensity frontier experiments for neutrino and scalar portals: analytic estimates}

\author[1]{Kyrylo~Bondarenko,} \author[1]{Alexey~Boyarsky,}
\author[1]{Maksym~Ovchynnikov,} \author[2]{Oleg~Ruchayskiy}

\affiliation[1]{Intituut-Lorentz, Leiden University, Niels Bohrweg 2, 2333 CA
  Leiden, The Netherlands} 
  \affiliation[2]{Discovery Center, Niels Bohr
  Institute, Copenhagen University, Blegdamsvej 17, DK-2100 Copenhagen,
  Denmark}
  
\emailAdd{bondarenko@lorentz.leidenuniv.nl} \emailAdd{boyarsky@lorentz.leidenuniv.nl} \emailAdd{ovchynnikov@lorentz.leidenuniv.nl} \emailAdd{oleg.ruchayskiy@nbi.ku.dk}

\begin{document}

\abstract{
In recent years, a number of intensity frontier experiments have been proposed to search for feebly interacting particles with masses in the GeV range. We discuss how the characteristic shape of the experimental sensitivity regions  -- upper and lower boundaries of the probed region, the maximal mass reach -- depends on the parameters of the experiments.
We use the SHiP and the MATHUSLA experiments as examples. 
We find a good agreement of our estimates with the results of the Monte Carlo simulations. This simple approach allows to cross-check and debug Monte Carlo results, to scan quickly over the parameter space of feebly interacting particle models, and to explore how sensitivity depends on the geometry of experiments. 
}

\maketitle

\section{Introduction: searching for feebly coupled particles}

The construction of the Standard Model has culminated with the confirmation of one of its most important predictions -- the discovery of the Higgs boson. The quest for new particles has not ended, however. The observed but unexplained phenomena in particle physics and cosmology (such as neutrino masses and oscillations, dark matter, baryon asymmetry of the Universe) indicate that other particles exist in the Universe. It is possible that these particles evaded detection so far because they are too heavy to be created at accelerators. Alternatively, some of the hypothetical particles can be sufficiently light (lighter than the Higgs or $W$ boson), but interact very weakly with the Standard Model sector (we will use the term \emph{feeble interaction} to distinguish this from {the weak interaction of the Standard Model}). In order to explore this latter possibility, the particle physics community is turning its attention to the so-called \emph{Intensity Frontier experiments}, see \textit{e.g.}~\cite{Beacham:2019nyx} for an overview. 
Such experiments aim to create high-intensity particle beams and use large detectors to search for rare interactions of feebly interacting hypothetical particles.

New particles with masses much lighter than the electroweak scale may be directly responsible for some of the BSM phenomena, or can serve as mediators (or ``\emph{portals}''), coupling to states in the ``hidden sectors'' and at the same time interacting with the Standard Model particles.
Such portals can be renormalizable (mass dimension $\le 4$) or be realized as higher-dimensional operators suppressed by the dimensional couplings $\Lambda^{-n}$, with $\Lambda$ being the new energy scale of the hidden sector.
In the Standard Model there can only be three {\em renormalizable} portals:
\begin{compactitem}[--]
\item
 a scalar portal that couples gauge singlet scalar to the $H^\dagger H$ term constructed from a Higgs doublet field $H_a$, $a=1,2$;
\item a neutrino portal that couples new gauge singlet fermion to the $\epsilon_{ab} \bar L_a H_b$ where $L_a$ is the SU(2) lepton doublet and $\epsilon_{ab}$ is completely antisymmetric tensor in two dimensions;
\item a vector portal that couples the field strength of a new U(1) field to the U(1) hypercharge field strength.
\end{compactitem}

Let us denote a new particle by $X$. The interaction of $X$ with the SM is controlled by the \textit{mixing angle} $\theta_{X}$ --- a dimensionless parameter that specifies the mixing between $X$ and the corresponding SM particle: the SM neutrinos for the neutrino portal, the Higgs boson for the scalar portal and the hyperfield for the vector portal. The searches for such particles are included in the scientific programs of many existing experiments~\cite{Hyun:2010an,Lees:2012ra,Adams:2013qkq,Lees:2014xha,Aaij:2014aba,Aad:2015xaa,Khachatryan:2015gha,TheBelle:2015mwa,Banerjee:2016tad,Lees:2017lec,Mermod:2017ceo,Izmaylov:2017lkv,Aaij:2017rft,Dobrich:2017yoq,Sirunyan:2018mtv}. 
Although the LHC is a flagship of the \emph{Energy Frontier} exploration, its high luminosity (especially in the Run 3 and beyond) means that huge numbers of heavy flavored mesons and vector bosons are created. This opens the possibility of supplementing the High Luminosity phase of the LHC with \textit{Intensity Frontier} experiments associated with the existing interaction points. 
Several such experiments have been proposed: CODEX-b~\cite{Gligorov:2017nwh}, MATHUSLA~\cite{Chou:2016lxi,Curtin:2018mvb}, FASER~\cite{Feng:2017uoz,Ariga:2019ufm}, and AL3X~\cite{Gligorov:2018vkc}.
\emph{Given that all these experiments can probe similar parameter spaces, it is important to be able to assess their scientific reach in a consistent way, under clearly specified identical assumptions.}

Detailed Monte Carlo (MC) simulations of both production and decays, complemented with background studies and detector simulations,  offer ultimate sensitivity curves for each of the experiments.

Such simulations are however difficult to reproduce and modify.
The modifications are nevertheless routinely  needed because
\begin{compactenum}[\bf (a)]
\item Geometrical configurations of most experiments are not fully fixed yet and it is important to explore changes of the science reach with the modification of experimental designs; 
\item Production or decays of GeV-mass feeble interacting particles involving quarks and mesons often requires the description outside of the validity range of both perturbative QCD and low-energy meson physics and is, therefore, subject to large uncertainties. This is the case for example for both scalar and neutrino portals (see \textit{e.g.}~\cite{Bezrukov:2009yw,Bondarenko:2018ptm,Monin:2018lee,Bezrukov:2018yvd,Winkler:2018qyg,Boiarska:2019jym} as well as the discussion in Section~\ref{sec:comparison-simulations}). In particular,
\begin{compactitem}[--]
\item Different groups use different prescription for  scalar production \cite{Krnjaic:2015mbs,Curtin:2018mvb,Winkler:2018qyg,Beacham:2019nyx,Boiarska:2019jym}
\item the  decay width and hadronic branching fractions for scalars with masses from $\sim 0.5$ GeV to few GeV are subject to large uncertainties, see~\cite{Monin:2018lee,Bezrukov:2018yvd};
\item  multi-hadronic HNL decays are not accounted for by any of the existing simulation tools. Yet they account for the largest part of the HNLs with masses around few~GeV~\cite{Bondarenko:2018ptm,Dib:2019ztn}.
\end{compactitem}

\item Monte Carlo simulations are done for a limited set of model parameters  and it is difficult to explore the overall parameter space and/or modify the sensitivity estimates for extended models (see \textit{e.g.} the discussion and approach in~\cite{SHiP:2018xqw})
\end{compactenum}
With this in mind we gathered in one place a sufficiently simple and fully controlled (semi)analytic estimates. 
Such estimates emphasize the main factors that influence the sensitivity:
\begin{inparaenum}[\it (i)]
\item dependence on the model (parameters, physical assumptions);
\item dependence on the geometry of the experiment;
\item factors, related to the beam energy, etc. 
\end{inparaenum}
We present the final number as a convolution of these factors, which allows to modify any of them at will.
As a result one can efficiently compare between several experimental designs; to identify the main factors that influence the sensitivity reach of a particular experiment/model; to  \textit{reuse} existing Monte Carlo sensitivities by separating them into the experimental efficiencies and physical input (model, production/decay phenomenology) with the subsequent modification of one of these factors; to scan over the  parameter space of different models as compared to those used in the MC simulations.

It turns out that the ratio between the sensitivities of the experiments to a great extent does not depend on the specific model of new physics, and is determined mainly by the geometry and collision energies of the experiments, which allow a comparison of the sensitivities in a largely model-independent way. To illustrate this point, we compare the potentials of two proposed experiments: the LHC-based MATHUSLA experiment~\cite{Curtin:2018mvb,Chou:2016lxi,Curtin:2017izq,Evans:2017lvd,Helo:2018qej} and a proton fixed target experiment using the proton beam of the Super Proton Synchrotron (SPS) at CERN -- SHiP~\cite{Anelli:2015pba,Alekhin:2015byh,SHIP:2018yqc}.
We analyze their sensitivity to the neutrino~\cite{Minkowski:1977sc,Yanagida:1979as,Glashow:1979nm,GellMann:1980vs,Mohapatra:1979ia,Mohapatra:1980yp} and scalar~\cite{Silveira:1985rk,McDonald:1993ex,Binoth:1996au,Burgess:2000yq,Schabinger:2005ei,Strassler:2006im,Patt:2006fw,OConnell:2006rsp} portals.
For particle masses $M_X \lesssim m_{B_c}$\footnote{By $m_{\dots}$ we denote the masses of lightest flavour mesons, for example, kaons ($m_K$), $D^+$ ($m_D$), $B^+$ ($m_B$), etc.} the main production channels are decays of heavy flavored mesons and $W$ bosons~\cite{Alekhin:2015byh,Evans:2017lvd} (see also Appendix~\ref{sec:production} for a brief overview). 
We concentrate on the mass range $M_X\gtrsim m_{K}$, since the domain of lower masses for the HNL and Higgs-like scalar is expected to be probed by the currently running NA62 experiment~\cite{Dobrich:2017yoq,Drewes:2018gkc}.

The sensitivity of the experiments is determined by the number of events that one expects to detect for a set of given parameters. In realistic experiments such events should be disentangled from the ``background'' signals.

For SHiP, detailed simulations have shown that the number of background events is expected to be very low, so that the experiment is ``background free''~\cite{Anelli:2015pba,SHiP:2015gkj,Baranov:2017chy,Akmete:2017bpl}. For MATHUSLA, the background is also expected to be low~\cite{Chou:2016lxi,Curtin:2018mvb}, although no simulation studies of background have been performed.
Even in the most favorable case of $N_{\rm bg} \ll 1$ one needs \emph{on average} $\bar N_{\event} = 2.3$ expected signal events to observe at least one event with the probability higher than $90\%$.\footnote{To obtain $95\%$ confidence limit one should assume $\bar N_\event = 3$, as the Poisson probability to see \emph{at least one} event, while expecting 3 ``on average'' is $0.9502$.}
However, due to the lack of spectrometer, mass reconstruction and particle identification at MATHUSLA, the meaning of the discovery of 2.3 events in the two experiments is very different as there is no way to associate the signal with a model in MATHUSLA and further consolidate the discovery.

\begin{figure}[!t]
    \centering
    \includegraphics[width=0.5\textwidth]{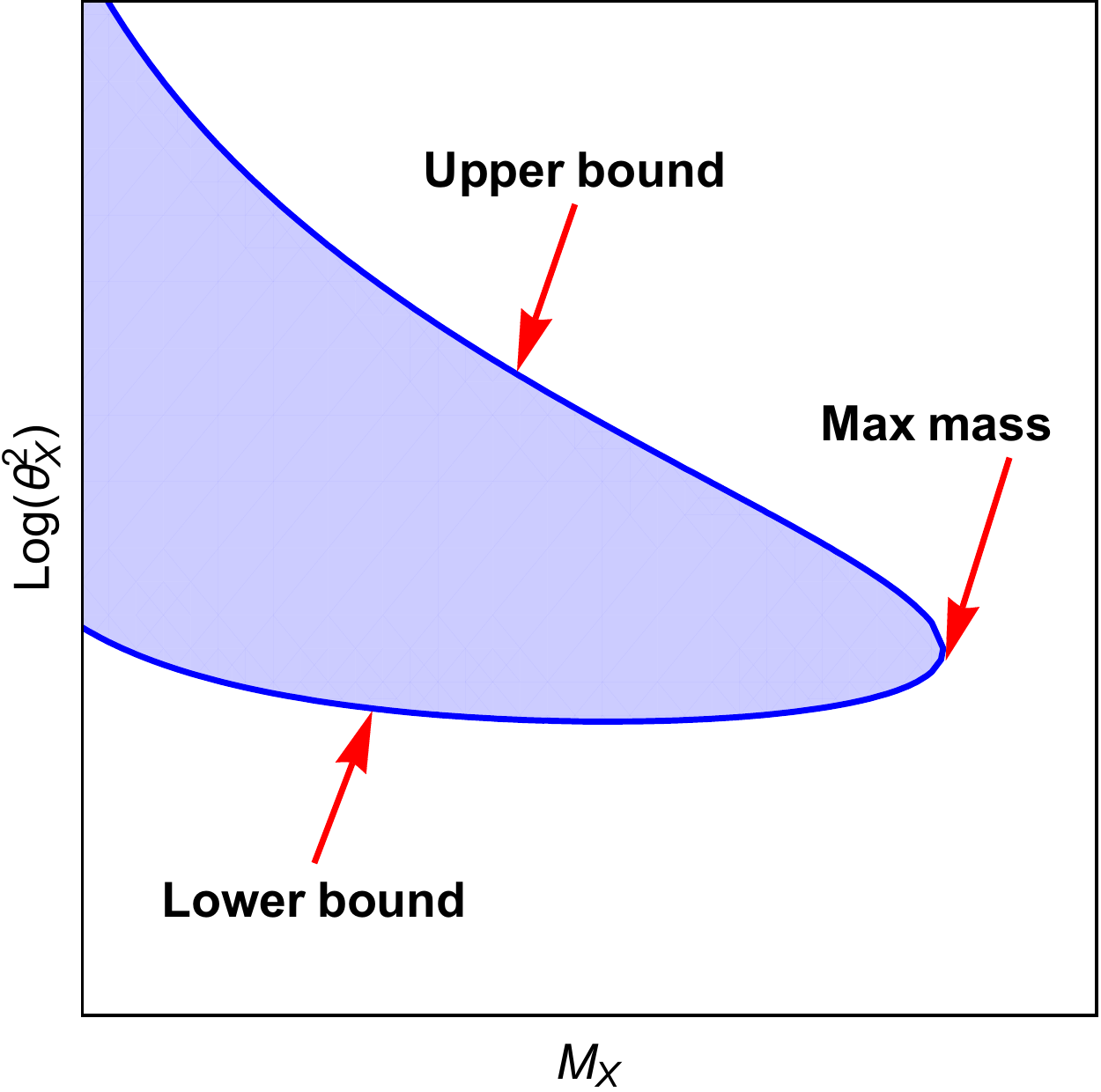}
    \label{fig:cigar}
    \caption{A typical cigar-like shape of the sensitiviy region of Intensity Frontier experiments. The \emph{upper boundary} is determined by the condition  $l_\decay \sim l_{\text{target-det}}$, \textit{i.e.} particles do not reach the detector. The \emph{lower boundary} of the sensitivity region is determined by the parameters at which decays become too rare.}
\end{figure}

For both experiments considered here the production point (``target'') is separated from the detector decay volume (of length $l_{\det}$) by some macroscopic distance $l_{\text{target-det}}$  (see Appendix~\ref{sec:geometry-experiments}).
For such experiments the sensitivity curve has a typical ``cigar-like shape'' in the plane ``mass vs.\ interaction strength'', see Fig.~\ref{fig:cigar}.

The number of decay events in the decay volume factorizes into
\begin{equation}
  N_\event = \sum_{M} N_{\pr, M} \times P_{\decay,M},
  \label{eq:matrix1}
\end{equation}
where $N_{\pr, M}$ is the number of particles $X$ that are produced from a mother particle $M$ and $P_{\decay, M}$ is the decay probability. For $N_{\pr, M}$ we have
\begin{equation}
  N_{\pr, M} \approx N_{\text{M}}\times \BR_{M\to X}\times \epsilon_{\decay,M}
  \label{eq:nprod}
\end{equation}
Here, $N_{M}$ is the number of parent particles produced at the experiment; in the case of mesons $N_M = N_{\text{meson}} = 2 N_{\bar{q}q}\times f_{\text{meson}}$, where $f_{\text{meson}}$ is the fragmentation fraction of a quark $q$ into a given hadron, and $N_M = N_W$ in the case of the $W$ bosons. $\BR_{M\to X}$ is the total branching ratio of decay of the parent particle into $X$ (see Appendix~\ref{sec:production}). Finally, $\epsilon_\decay$ is the \emph{decay acceptance} -- the fraction of particles $X$ whose trajectory intersects the decay volume, so that they \textit{could} decay inside it.

The probability of decay into a state that can be detected is given by\footnote{Here we ignored that particles travel slightly different distances depending on their off-axis angle. Eq.~\eqref{eq:matrix:2}  also neglects the energy distribution of the produced particles, assuming that all of them travel with the same average energy. This is essential for proper determining of the upper boundary and we will return to this in Section~\ref{sec:upper-bound}.}
\begin{equation}
  \label{eq:matrix:2}
  P_{\decay,M} = \left[\exp\left(-\frac{l_{\text{target-det}}}{l_{\decay}}\right) - \exp\left(-\frac{l_{\text{target-det}} + l_{\text{det}}}{l_{\decay}}\right)\right]\times \epsilon_{\det}\times \text{BR}_{\rm vis},
\end{equation}
where the branching ratio $\BR_{\rm vis}$ is the fraction of all decays producing final states that can be registered. Finally, $\epsilon_{\det}\le 1$ is the \emph{detection efficiency} -- a fraction of all decays inside the decay volume for which the decay products could be detected. 
In the absence of detector simulations  we  optimistically assume a detector efficiency of MATHUSLA of $\epsilon_{\det}=1$.
The decay length $l_\decay$ in Eq.~\eqref{eq:matrix:2} is defined as
\begin{equation}
  \label{eq:matrix:3}
  l_\decay = c \tau_X \beta_{X} \gamma_{X},
\end{equation}
where $\tau_X$ is the lifetime of the particle $X$ (see Appendix~\ref{sec:decay}), $\beta_{X}$ is its velocity and $\gamma_{X}$ is the $\gamma$ factor (which depends on the mother particle that produces $X$).

The production branching ratio and the lifetime behave with the mixing angle as
\begin{equation}
\BR_{\text{meson}\to X} \propto \theta_{X}^{2}, \quad \tau_{X} \propto \theta_{X}^{-2}
\label{eq:production-decay-mixing-angle}
\end{equation} 
At the lower bound of the sensitivity the decay probability behaves as $P_{\decay} \propto l_{\det}/l_{\decay}$, and as a consequence of~\eqref{eq:production-decay-mixing-angle} the number of events scales as
\begin{equation}
N_{\event,\text{lower}} \propto \theta_{X}^{4}/\gamma_{X}
\label{eq:nevents-lower}
\end{equation}
At the upper bound $P_{\decay}\approx e^{-l_{\text{target-det}}/l_{\decay}}$, and 
\begin{equation}
N_{\event,\text{upper}} \propto \theta_{X}^{2}e^{-C \theta_{X}^{2}/\gamma_{X}},
\label{eq:nevents-upper}
\end{equation}
where $C$ is some numerical factor (that depends on properties of $X$).

Larger $\gamma$ factor suppresses the exponents in the expression for the decay probability~\eqref{eq:matrix:2}. From~\eqref{eq:nevents-lower},~\eqref{eq:nevents-upper} we see that this affects the upper and lower bounds of the cigar-like sensitivity plots in the opposite ways. For the lower bound, an experiment with the smaller average $\gamma$ factor is sensitive to small coupling constants. For sufficiently large couplings, larger $\gamma$ factor ensures that particles do not decay before reaching the detector, thus increasing the sensitivity to the upper range of the sensitivity curve.

\bigskip

The paper is organized as follows. In Sections~\ref{sec:lower-bound}--\ref{sec:maximal-mass} we  discuss the lower and upper boundaries of the sensitivity region, the maximal mass that can be probed and experimental parameters that affect them. 
In Sec.~\ref{sec:energy-distributions} we discuss the total amount and energy distribution of charm- and beauty mesons at both  SHiP and MATHUSLA experiments, as well as the contribution from the $W$ bosons. In Sec.~\ref{sec:results} we summarize and discuss our results, while in Sec.~\ref{sec:comparison-simulations} we compare our approach with results of official simulations. Finally, in Sec.~\ref{sec:conclusion} we make conclusions.
Appendices~\ref{sec:coupling}--\ref{app:HNLsatMATHUSLA} provide details of computations and relevant supplementary information.

\section{Lower boundary of the sensitivity region: main factors}
\label{sec:lower-bound}

As we will see later (see Section~\ref{sec:results}), the production from the $W$ bosons does not give a contribution to the lower bound of the sensitivity curve for neither of the two experiments, and for neither of the two models discussed. So, in this Section we will consider only the production from the mesons.

Let us first estimate the \textit{lower boundary} of the sensitivity region, where $l_\decay \gg l_{\det},l_{\text{target-det}}$.  
For the number of events~\eqref{eq:matrix1} we have
\begin{equation}
N_{\event,\text{lower}} \approx N_{\text{meson}} \times \BR_{M\to X}\times \frac{\langle l_{\det}\rangle}{c\tau_{X}\langle \gamma_{X}\rangle}\times \epsilon_{X},    
\label{eq:nevents-lower-bound-explicit}
\end{equation}
where $\epsilon_{X} \equiv \epsilon_{\text{prod}}\times \epsilon_{\text{decay}}\times \BR_{\text{vis}}$ is the overall efficiency and $\tau_X$ is the lifetime of the particle $X$ (see the discussion below Eq.~\eqref{eq:matrix:3}). The particles are assumed to be relativistic (we will see below when this assumption is justified), so that $\beta_{X}\approx 1$. 
We estimate the $\gamma$ factor $\gamma_X$ from that of the parent meson:
\begin{equation}
    \gamma_{X} \approx \gamma_{\text{meson}} \frac{\langle E_{X}^{\text{rest}}\rangle }{M_{X}},
    \label{eq:5}
\end{equation}
The average formula~\eqref{eq:5} does not take into account the distribution of HNLs (scalars) in the meson rest frame -- some of the new particles fly in the direction of the parent meson and have $\gamma_X$ larger than~\eqref{eq:5}, while the other fly in the opposite direction.
We show below that this \emph{does not play a role} for the lower boundary of the sensitivity curve while  the \emph{upper} boundary is exponentially sensitive to the high $\gamma$-factor tail of the distribution and therefore cannot be determined from Eq.~\eqref{eq:5}.
For the experiments like FASER this difference plays an essential role, see~\cite{Boiarska:FASER}.

Since at the lower bound $N_{\event}\propto \theta_{X}^{4}$ (see Eq.~\eqref{eq:nevents-lower}), for the ratio of the mixing angles at the lower bound, we have
\begin{equation}
  \label{eq:events-ratio}
  \boxed{\frac{(\theta^{\text{SHiP}}_{X,\text{lower}})^{2}}{(\theta^{\text{MAT}}_{X,\text{lower}})^{2}}= \sqrt{\frac{N_{\text{events}}^{\mat}}{N_{\text{events}}^{\ship}}} \simeq \sqrt{\frac{N_{\text{meson}}^{\mat}}{N_{\text{meson}}^{\ship}} \times \frac{l_{\det}^{\mat}}{l_{\det}^{\ship}}\times \frac{\langle\gamma_{\text{meson}}^{\ship}\rangle}{\langle\gamma_{\text{meson}}^{\mat}\rangle}\times \frac{\epsilon_{\mat}}{\epsilon_{\ship}}}},
\end{equation}
where we assumed that the same meson is the main production channel at both the SHiP and MATHUSLA experiments for the given mass $M_X$ of the new particle, so the branching ratio $\BR_{\text{meson}\to X}$ from Eq.~\eqref{eq:nevents-lower-bound-explicit} disappears. Therefore, to make a comparison between the experiments we only need to know the total number of mesons, their average $\gamma$ factor, the decay volume length and the overall efficiency.

\section{Upper boundary of the sensitivity curve}
\label{sec:upper-bound}

\begin{figure}[t!]
    \centering
    \includegraphics[width=0.5\textwidth,draft=false]{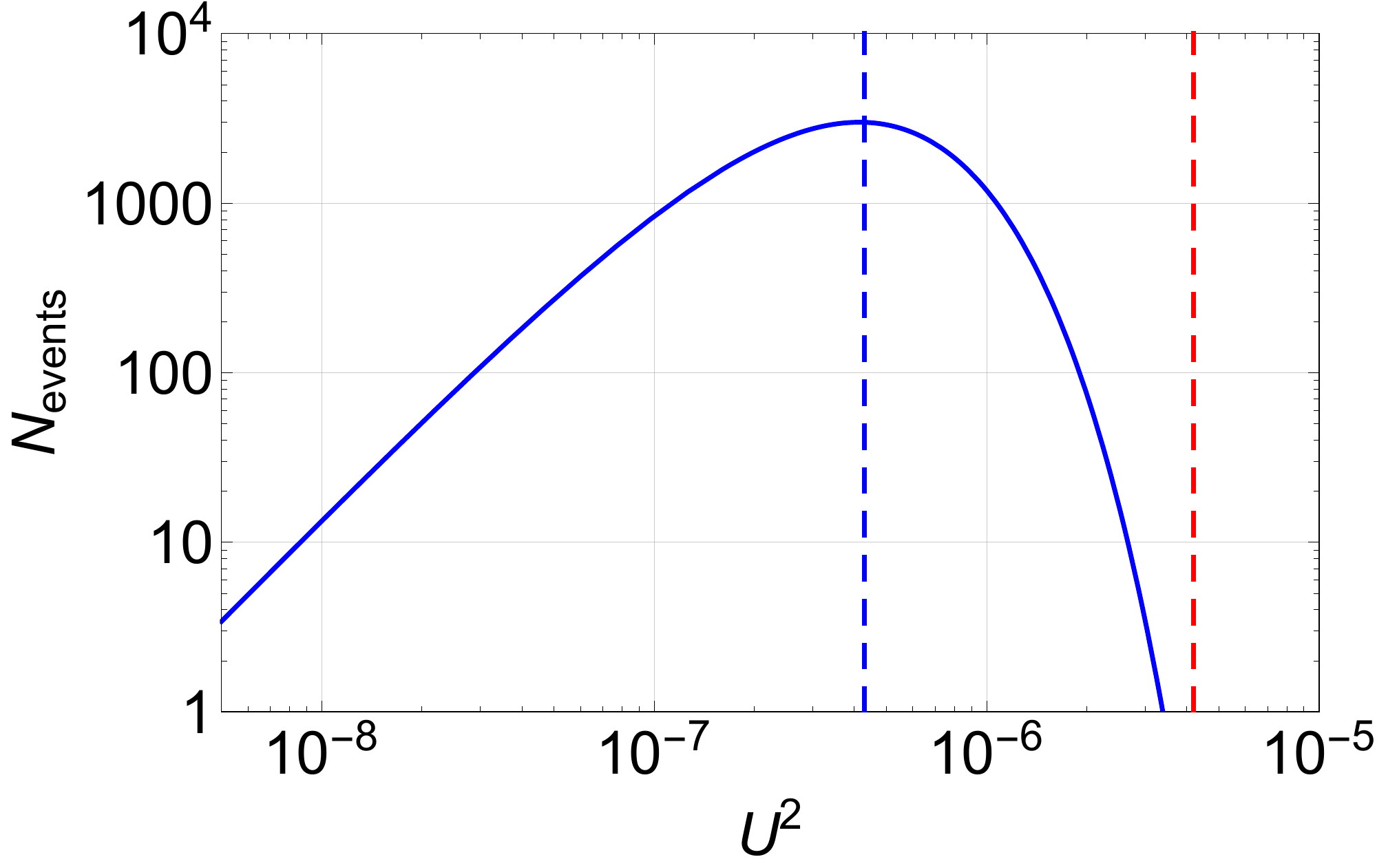}
  \caption{The number of decay events for  HNL with mass $M_{N} = 3\text{ GeV}$ as a function of $U_e^{2}$. 
  The number of mesons is taken $N_{\text{meson}} = 10^{14}$, the $\gamma$ factor is  $\langle \gamma_{N}\rangle = 15$, the efficiency $\epsilon = 1$, and the distances $l_{\text{target-det}} =l_{\det} =50\text{ m}$. The decay width can be found from Eq.~\protect\eqref{eq:HNLdecaywidth}. The dashed blue line corresponds to $U^{2}_{\text{max}}$ (Equation~\protect\eqref{eq:max-events}), while the dashed red line corresponds to the estimate of the upper bound based on Eq.~\protect\eqref{eq:upper-bound-estimation}. Small discrepancy between the position of the upper bound and the estimate is caused by logarithmic errors in~\protect\eqref{eq:upper-bound-estimation}.}
  \label{fig:umax}
\end{figure}

If particles have sufficiently large interaction strength (i.e., the mixing angles), they decay before reaching the decay volume. This determines the upper bound of the sensitivity curve, that we call $\theta_{X, \text{upper}}^2$.

A useful quantity to consider is a mixing angle for which the amount of decays inside the decay volume is maximal, $\theta_{X, \text{max}}$. It can be found using the asymptotic behavior for the number of events $N_{\event}$ from the estimations~\eqref{eq:nevents-lower},~\eqref{eq:nevents-upper}. In the domain $l_{\decay} \gg l_{\text{target-det}}$ for a fixed mass $M_{X}$ it follows that $N_{\event}$ monotonically grows as $\theta_{X}^{4}$ with the increase of $\theta_{X}$, while in the domain $l_{\decay} \ll l_{\text{target-det}}$ it falls exponentially. The position of the maximum $\theta_{\text{max}}$ can be found from
\begin{equation}
  l_{\decay}\bigl(M_{X},\theta^{2}_{\text{max}}\bigr) \simeq \left\{\begin{aligned}
  1.5l_\text{target-det}, & \quad\text{if}\quad l_\text{target-det} \simeq l_{\text{det}}\\
  0.5l_\text{target-det},& \quad\text{if}\quad l_\text{target-det} \gg l_{\text{det}}\\
  \end{aligned}
  \right.
  \label{eq:max-events}
\end{equation}
Using $\theta_{\text{max}}$, we can estimate the value of $\theta_{\text{upper}}$ assuming that all the particles $X$ have the same (average) energy  $\langle E_{X}\rangle$. If we neglect the second exponent in the expression for the decay probability~\eqref{eq:matrix:2}, then the formula for the number of events~\eqref{eq:matrix1} becomes
\begin{equation}
 N_{\event}\simeq N_{\text{prod}}\times \epsilon_{\det}\times \BR_{\text{vis}}\times e^{-l_{\text{target-det}}/l_{\decay}} 
 \label{eq:nevents-upper-explicit}
\end{equation}
We can estimate the exponent in~\eqref{eq:nevents-upper-explicit} as $l_{\text{target-det}}/l_{\decay} \approx \theta_{X}^{2}/\theta^{2}_{\text{max}}$, see Eq.~\eqref{eq:max-events}. So imposing the condition $N_{\event} \simeq 1$ in Eq.~\eqref{eq:nevents-upper-explicit} with the logarithmic precision we get
\begin{equation}
\boxed{\theta^{2}_{\text{upper}}\simeq \theta^{2}_{\text{max}}\times \log\Bigl[N_{\text{prod}}(\theta^{2}_{\text{max}})\, \epsilon_{\det}\BR_{\text{vis}}\Bigr]}.
\label{eq:upper-bound-estimation}
\end{equation}
An example of the dependence of the number of events on $\theta_{X}^{2}$ for the fixed mass $M_{X}$, together with the estimation of the $\theta_X$ for the maximal number of events given by~\eqref{eq:max-events} and the upper bound predicted by~\eqref{eq:upper-bound-estimation}, is shown in Fig.~\ref{fig:umax}.

Of course, it is not sufficient to use only  the average energy $\langle E_{X}\rangle$ to estimate the position of the upper boundary. 
Indeed, the decrease of $c\tau_{X}$ with the growth of $\theta_{X}^2$  can be compensated by the increase of the energy $E_{X}$ and, therefore, of the $\gamma$-factor. 
As a result the particles with $E_{X} > \langle E_{X}\rangle$ can reach the detector even if the mixing angle $\theta_{X}$ is larger than the estimate~\eqref{eq:upper-bound-estimation}.

The expression~\eqref{eq:upper-bound-estimation} helps to estimate how the sensitivity curve depends on the parameters of the experiment and on various assumptions.
In particular, we can now estimate how large is a mistake from using $\langle E_X^{\rm rest}\rangle$ in Eq.~\eqref{eq:5} rather than the actual $E_X$ distribution. 
In order to do that we replaced $\langle E_X^{\rm rest}\rangle \to m_{\rm meson}$ -- the maximal energy of the particle $X$  in the meson's rest frame. 
This substitution increases the $\gamma_X$ by a factor of $2$. The estimates~\eqref{eq:max-events}--\eqref{eq:upper-bound-estimation} show that $\theta_{\text{max}}^2$ and as a result $\theta_{\text{upper}}^2$  will shift by the same factor of $2$. 
This number indicates an upper bound on the possible error, introduced by the approximate treatment.

Next, we turn to the exact treatment. 
To this end we consider the energy distribution of the $X$ particles,
\begin{equation}
    f_{X}(E_X)= \frac{1}{N_{X}} \frac{dN_{X}}{dE_X}.
\end{equation}
Taking into account this distribution, the formula for the decay probability~\eqref{eq:matrix:2} at the upper bound should be modified as
{
\begin{equation}
  \label{eq:decay-probability-upper-bound}
  P_{\decay} =  \epsilon_{\det}\times \BR_{\text{vis}} \times \int\limits_0^{\infty} dE_X\, f_{X}(E_{X})
 \times \pi\left(\frac{\tau_X p_X}{l_{\text{target-det}} M_X} \right),
\end{equation}
where an argument of $\pi$ function is $l_{\decay}/l_{\text{target-det}}$ and we used the expression for the decay length~\eqref{eq:matrix:3}. The function $\pi(y)$, defined via
\begin{equation} 
 \pi(y) \equiv \exp\left(-\frac{1}{y}\right) - \exp\left(-\frac{l_{\text{target-det}}+l_{\text{det}}}{l_{\text{target-det}}} \frac{1}{y}\right),
 \label{eq:pi}
\end{equation}
determines a ``window'' of energies in which the shape of $f_X(E_X)$ distribution (rather than the averange number of particles) contributes to the overall probability. 
$\pi(y)$ is shown in Fig.~\ref{fig:pi}.
For small energies (small $y$) $\pi(y)$ is exponentially small, while for large energies (large $y$) $\pi(y)$ is inversely proportional to energy and decreases slowly. 
Therefore, a sufficiently long ``tail'' of high-energy mesons can contribute to the integral in~\eqref{eq:decay-probability-upper-bound}, but this range cannot be estimated without knowledge of the distribution function $f_{X}$. We will discuss $f_{X}$ for mesons and $W$ bosons in Sec.~\ref{sec:energy-distributions}.

\begin{figure}[t!]
    \centering
    \includegraphics[width=0.5\textwidth,draft=false]{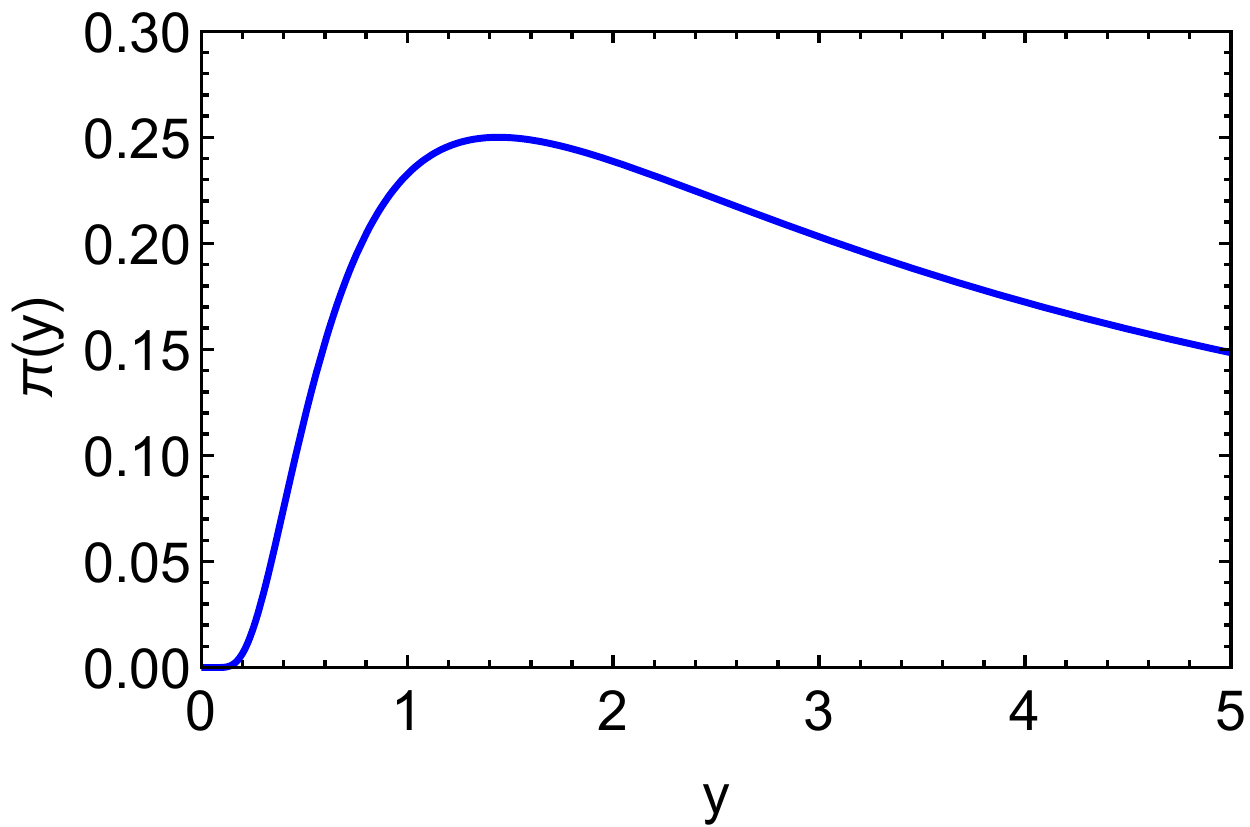}
  \caption{The function $\pi(y)$ that determines the position of the upper boundary (see Eq.~\protect\eqref{eq:pi}). 
  We assumed $l_{\text{target-det}}=l_{\text{det}}$.}
  \label{fig:pi}
\end{figure}
}

\section{Maximal mass probed}
\label{sec:maximal-mass}
The maximal mass probed by the experiment is defined as the mass at which the lower sensitivity bound meets the upper sensitivity bound. It can be estimated from the condition that the decay length, calculated at the lower bound $\theta_{\text{lower}}$ (see Sec.~\ref{sec:lower-bound}), is equal to the distance from the target to the decay volume of the given experiment:
\begin{equation}
  \label{eq:max_mass}
 l_{\decay}(M_{X,\text{max}},\theta^{2}_{\text{lower}}(M_{X,\max})) \simeq  l_{\mathrm{target-det}}.
\end{equation}
The decay length~\eqref{eq:matrix:3} depends on the mass as $l_{\decay}\propto M_{X}^{-\alpha-1}$, where the term $\alpha$ in the exponent approximates the behaviour of the lifetime with the mass, and the term $1$ comes from the $\gamma$ factor. 

Using the condition~\eqref{eq:max_mass}, the maximal mass probed can be estimated as
\begin{equation}
\label{eq:max_massX}
  M_{X,\max} \propto \left(\frac{\langle E_{X}\rangle}{|\theta_{\text{lower}}|^2 l_{\mathrm{target-det}}}\right)^{\frac{1}{\alpha+1}},
\end{equation}
which results in the following ratio of the maximal mass probed at the SHiP and MATHUSLA experiments:
\begin{equation}
  \label{eq:maximally-probed-mass-ratio}
  \boxed{\frac{M_{X,\max}^{\ship}}{M_{X,\max}^{\mat}} \simeq
  \left(
    \frac{\langle E_{X}\rangle ^{\ship}}{\langle E_{X}\rangle^{\mat}} \times
    \frac{|\theta^{\mat}_{X,\text{lower}}|^{2}}{|\theta^{\ship}_{X,\text{lower}}|^{2}} \times
    \frac{l_{\text{target-det}}^{\mat}}{l_{\text{target-det}}^{\ship}}
  \right)^{\frac1{\alpha+1}}}.
\end{equation}
For Higgs-like scalars we have $\alpha \approx 2$, while for HNLs it is $\alpha \approx 5$, see Appendix~\ref{sec:decay}.

The estimate of the maximal mass probed~\eqref{eq:max_massX} is applicable only if the result does not exceed the kinematic threshold; for the production from $B$ mesons for the HNLs it is $m_{B_{c}} - m_{l}$ or $m_{B} - m_{l}$ depending on whether amount of produced $B_{c}$ mesons is large enough to be relevant for the production (see the discussion in Sec.~\ref{sec:meson-distributions}), and for the scalars it is $m_{B} - m_{\pi}$.

\section{Number and momentum distribution of mesons and $W$'s at SHiP and MATHUSLA}
\label{sec:energy-distributions}

In this Section, we discuss the number and distribution of charm and beauty mesons and of $W$ bosons at SHiP and MATHUSLA experiments.
As we have seen, to estimate the lower boundary we need only the number of parent particles and their \textit{average} $\gamma$ factors (see Eqs.~\eqref{eq:nevents-lower-bound-explicit},~\eqref{eq:maximally-probed-mass-ratio}). On the other hand, for the estimation of the upper boundary we also need the energy distribution of the mesons and $W$ (see Sec.~\ref{sec:upper-bound}). 

\subsection{$B$ and $D$ mesons}
\label{sec:meson-distributions}
The main production channel of HNLs in the mass range $M_N \lesssim m_{D_{s}}$ is the two-body leptonic decay of $D_s$ mesons. For masses $m_{D_{s}}\lesssim M_{N}\lesssim m_{B_{c}}$ the main contribution comes from decays of $B$ mesons, see, e.g.,~\cite{Bondarenko:2018ptm}.\footnote{This statement is true for HNLs with dominant mixing with $\nu_{e/\mu}$. For dominant mixing with $\nu_{\tau}$ the main production channel is from $\tau$ leptons for $M_{N}\lesssim m_{\tau}$ and from $B$ mesons for larger masses~\cite{Bondarenko:2018ptm}.} For masses  $M_{N} \gtrsim 3\text{ GeV}$ the main HNL production channel is determined by the value of the fragmentation fraction of $B_{c}$ mesons, $f_{B_{c}}$: in the case $f_{B_{c}}\gtrsim 10^{-4}$ it is the two-body decay of the $B_{c}$ meson, while for smaller values it is the two-body decay of the $B^{+}$ meson~\cite{SHiP:2018xqw}. For the scalars the production from $D$ mesons is negligible as compared to the $B^{+/0}$ mesons decays even for masses $m_K \lesssim M_S \lesssim m_D$. The $B_{c}$ mesons are not relevant for their production (see, e.g.,~\cite{Bezrukov:2009yw,Boiarska:2019jym}). The branching ratios of the production of the HNLs and the scalars used for our estimations are given in Appendix~\ref{sec:production}. 

For the LHC energies the fragmentation fraction $f_{B_{c}}$ was measured at the LHCb~\cite{Aaij:2017kea} and found to be $f_{B_c} \approx (2.6\pm 1.3) \times 10^{-3}$. Earlier measurements at the Tevatron give a similar value $f_{B_c} \approx (2\pm 1) \times 10^{-3}$~\cite{Abe:1998wi,Abe:1998fb,Cheung:1999ir}, which is in good agreement with~\cite{Aaij:2017kea}. 
Therefore, at the LHC the $B_{c}$ decay is the main production channel for heavy HNLs. However, at the energies of the SHiP experiment, $\sqrt{s}\simeq 30\text{ GeV}$, currently there is no experimental data on $f_{B_{c}}$. 
Additionally, the theoretical  predictions of $f_{B_{c}}$ (see, e.g.,~\cite{Berezhnoy:2004gc,Kolodziej:1997um,Kolodziej:1995nv}) disagree with the LHC and Tevatron measurements at least by an order of magnitude, which also makes them untrustable at SHiP's energies. 
As a result, the value of $f_{B_{c}}$ at SHiP experiment is unknown. 
In order to estimate the effect of this uncertainty, we perform our analysis of the sensitivity of the SHiP experiment for two extreme cases: \emph{(i)} SHiP's $f_{B_c}$ at the same level as at the LHC, and \emph{(ii)} $f_{B_c}= 0$, ``no $B_c$ mesons''.

Let us now discuss the available data. For the SHiP experiment, the amounts of produced charmed and beauty mesons (except the $B_{c}$ mesons) were obtained in detailed PYTHIA simulations; the corresponding numbers can be found in~\cite{CERN-SHiP-NOTE-2015-009} and are reproduced in Table~\ref{tab:b-d-parameters}. We estimate the spectrum of the $B_{c}$ mesons from the spectrum of the $B^{+}$ mesons by rescaling the energy $E_{B_{c}} = (m_{B_{c}}/m_{B})E_{B}$ for the events with $B^{+}$ mesons. For MATHUSLA experiment, the situation is different: there is no available data with detailed simulations that give us the relevant properties of the mesons, so we discuss them below.

\subsection{Mesons at MATHUSLA}
\label{sec:mesons_mathusla}
\begin{figure}[!t]
  \begin{minipage}{0.5\textwidth}
    \centering \includegraphics[width=\textwidth]{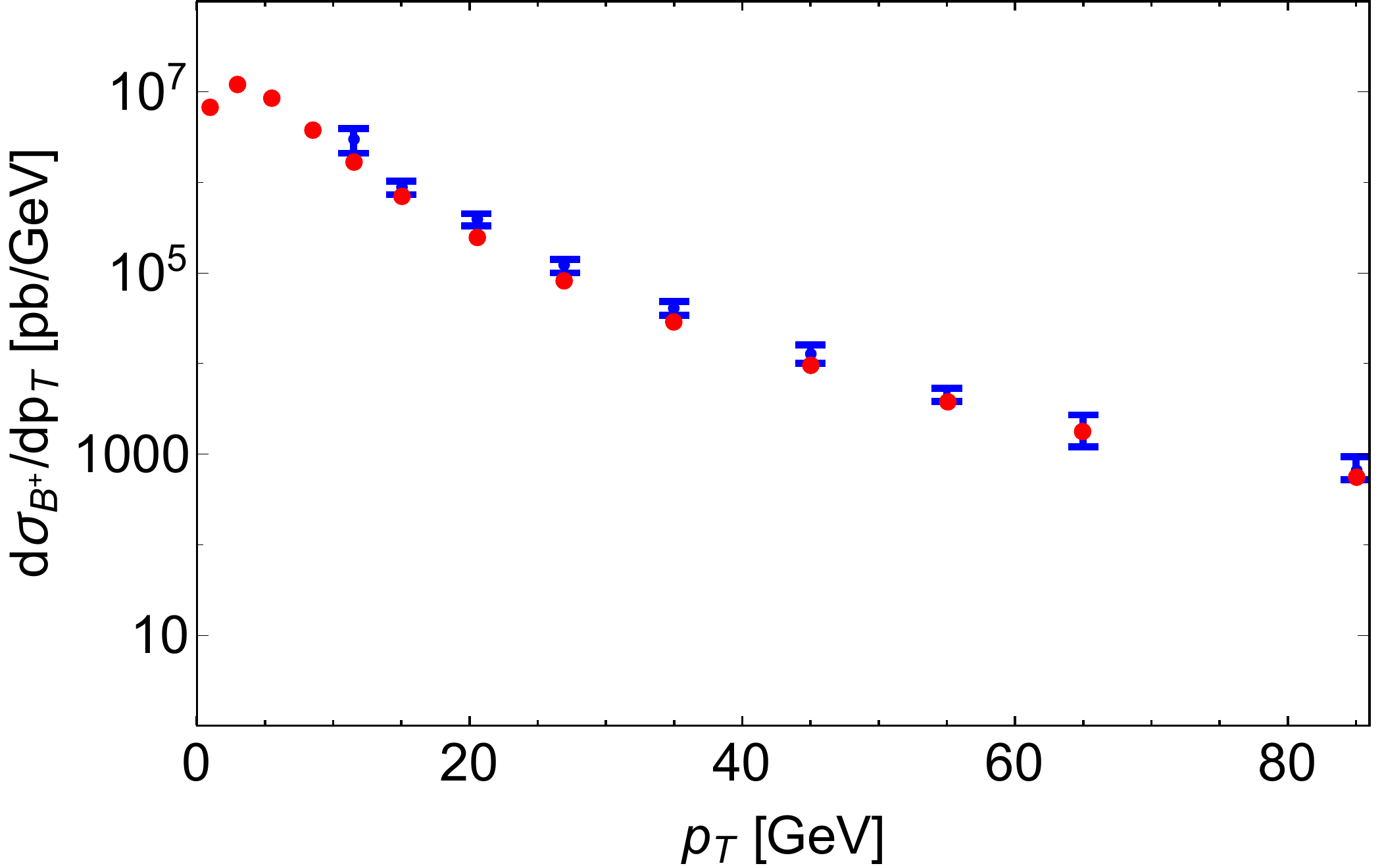}
  \end{minipage}\hfill
  \begin{minipage}{0.5\textwidth}
    \centering \includegraphics[width=\textwidth]{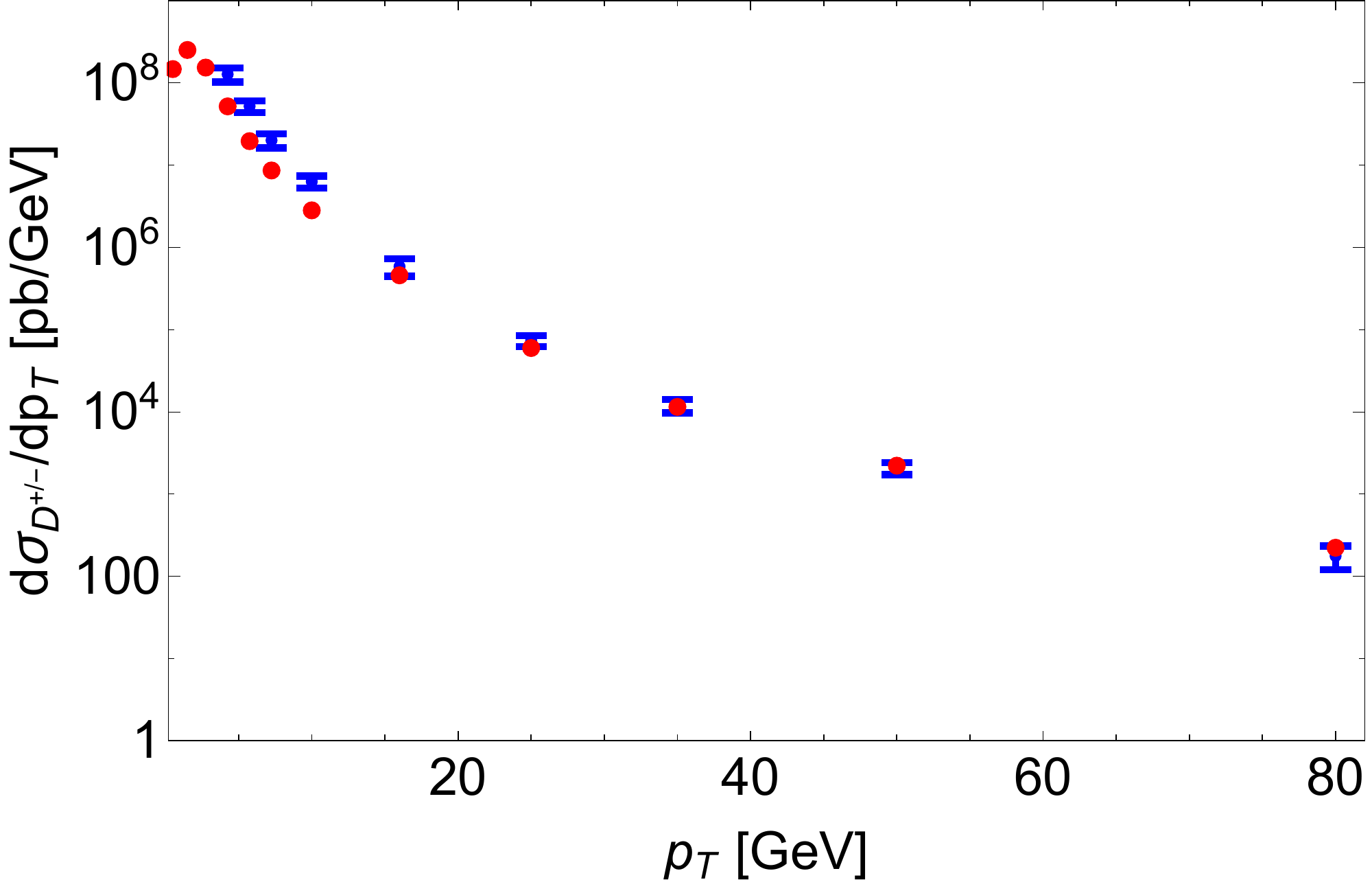}
  \end{minipage}
  \caption{Comparison of the $p_{T}$ spectra of $B^{+},D^{+}$ mesons predicted by the FONLL simulations (red points) with the measurements of the ATLAS and CMS collaborations~\cite{Khachatryan:2016csy,Aad:2015zix} (blue points with uncertainties bars). Only the central values of the  FONLL predictions are shown. See text for details.}
  \label{fig:b-d-mesons-simulations}
\end{figure}

In order to estimate the number of mesons and their $\gamma$ factors for the MATHUSLA experiment, one needs to know their $p_T$ distribution at ATLAS/CMS in the MATHUSLA pseudorapidity range $0.9<\eta < 1.6$ (see Appendix~\ref{sec:geometry-experiments}). The relevant distributions were measured for $B^+$ mesons by the CMS collaboration~\cite{Khachatryan:2016csy} (13~TeV) with the $p_T$ cut $p_{T}^{B}\geqslant 10\text{ GeV}$, and for $D^+/D^0$ mesons by the ATLAS collaboration~\cite{Aad:2015zix} (7 TeV) for $p_T^{D}\geqslant 3.5\text{ GeV}$. We show the spectra obtained in these papers in Fig.~\ref{fig:b-d-mesons-simulations}.

The low $p_T$ mesons, unaccounted for these studies, are the most relevant for the MATHUSLA sensitivity estimate because of two reasons. Firstly, the $p_T$ spectrum of the hadrons produced in $pp$ collisions has a maximum at $p_T\sim{}$few GeV (see, e.g., experimental papers~\cite{Aaij:2013noa,Khachatryan:2010us}, theoretical paper~\cite{Cacciari:2012ny} and references therein), and therefore we expect that most of the $D$ or $B$ mesons have $p_T$s below the LHC cuts. Secondly, low $p_T$ mesons produce decay products with the smallest $\gamma$ factor, and therefore with the shortest decay length~\eqref{eq:matrix:3} and the largest probability to decay inside the decay volume (here we consider the case $l_\decay\gg l_{\text{target-det}}$). Therefore, by shifting the position of the peak to smaller $p_{T}$s, we increase the number of mesons and decrease their average $\gamma$ factor, and both of these effects enhance the number of events at the lower bound~\eqref{eq:nevents-lower-bound-explicit}. Therefore an accurate prediction of the distribution $d\sigma/dp_T$ in the domain of low $p_{T}$s is very important.

In order to evaluate the distribution of heavy flavored mesons at low $p_T$ and also to estimate  $D$ meson production cross-section at $\sqrt{s} = 13$~TeV we use FONLL (Fixed Order + Next-to-Leading Logarithms) -- a model for calculating the single inclusive heavy quark production cross section which convolutes perturbative cross section with  non-perturbative fragmentation function, see~\cite{Cacciari:1998it,Cacciari:2001td,Cacciari:2012ny,Cacciari:2015fta} for details.

Predictions of FONLL have been calibrated against the accelerator data and were found to be in very good agreement,
see e.g.~\cite{Aad:2015zix,Khachatryan:2016csy,Aad:2015gna,Aaij:2015bpa,Aaij:2013lla}. In particular, comparison of the FONLL simulations of the production of the $B^{+}$s with the measurements at the Tevatron and at the LHC showed that FONLL predicts the low $p_T$ distribution accurately.
We show the central values of the FONLL predictions down to $p_T =0$, confronted with the measurements of the CMS~\cite{Khachatryan:2016csy} and ATLAS~\cite{Aad:2015zix} collaborations in Fig.~\ref{fig:b-d-mesons-simulations}. As expected, the distributions have maxima, after which they fall. We see, however, that the central predictions of FONLL for the differential cross-sections typically lie below the uncertainty range of the experimental cross-section, which results in a somewhat lower total cross-sections. Namely, integrating the central predictions over the experimentally measured $p_{T}$s, we have $\sigma_{D,\text{FONLL}}/\sigma_{D,\text{exp}} \approx 0.4$ and $\sigma_{B,\text{FONLL}}/\sigma_{B,\text{exp}} \approx 0.7$. However, as is demonstrated in the same papers~\cite{Khachatryan:2016csy,Aad:2015zix}, the agreement between the FONLL predictions and the experimental data is much better if one uses the upper bound of the FONLL predictions defined by the theoretical uncertainties.

Using the results of the FONLL simulations, we find the amounts of low $p_T$ mesons traveling in the MATHUSLA direction:
\begin{equation}
  \frac{N_{D}|_{p_{T}<3.5\text{ GeV}}}{N_{D}|_{p_{T}>3.5\text{ GeV}}} = 3.8, \quad \frac{N_{B}|_{p_{T}<10\text{ GeV}}}{N_{B}|_{p_{T}>10\text{ GeV}}} = 5.7
\end{equation}
This justify our statement that most of the $B$ and $D$ mesons have the $p_{T}$ below the cuts in the currently available experimental papers~\cite{Aad:2015zix,Khachatryan:2016csy}.

FONLL does not provide the distributions of the $D_{s}$ and the $B_{c}$ mesons. 
We approximate their distributions by those the $D^{+}$ and $B^{+}$ distributions. In the case of the $B_{c}$ mesons we justify this approximation by comparing the distributions provided by BCVEGPY 2.0 package~\cite{Chang:2003cq} (that simulates the distribution of the $B_c$ mesons and was tested at the LHC energies) for the $B_{c}$ meson with that of FONLL for the $B^+$ meson. We conclude that the $p_T$ and $\eta$ distributions of $B_c$ and $B^+$ have similar shapes.

The relevant parameters --- the total number of mesons, the average $\gamma$ factor of the mesons that are produced in the direction of the decay volume of the experiments and the geometric acceptances $\epsilon_{\text{geom},\text{meson}}$ for the mesons --- are given in Table~\ref{tab:b-d-parameters}.
\begin{table}[!t]
    \centering
 \begin{tabular}{|c|c|c|c|c|c|c|c|}
    \hline
    \textbf{Experiment} & $N_{D}$ & $N_{B}$ & $\langle\gamma_{D}\rangle$ & $\langle\gamma_{B}\rangle$& $\epsilon_{\text{geom},D}$ & $\epsilon_{\text{geom}, B}$\\
    \hline
    MATHUSLA & $4.4 \times 10^{16}$& $3\times 10^{15}$ & 2.6  & 2.3 & $1.3\times 10^{-2}$ & $1.8\times 10^{-2}$ \\
    \hline
    SHiP & $1.6\times 10^{18}$ & $1.1\times 10^{14}$ & 19.2 & 16.6 & $-$ & $-$ \\
    \hline
    \end{tabular}
    \caption{Parameters of the  SHiP and MATHUSLA experiments: the total number of all charmed/beauty hadrons; the average $\gamma$ factor of mesons flying in the direction of the decay volumes of the experiment;
    the geometric acceptances for these hadrons. 
    We take $B_c$ meson distribution to be proportional to that of $B^+$ mesons, scaled by $f_{B_c}$. 
    As a result, $B_c$ gamma factor is the same as for $B^+$ mesons for SHiP and scaled by $m_B/m_{B_c}$ for MATHUSLA, see discussion in Section~\ref{sec:meson-distributions} and~\ref{sec:mesons_mathusla}
    For SHiP we assumed 5~years of operation ($2\times 10^{20}$ protons on target) and for MATHUSLA we took the luminosity of the HL phase, $\mathcal{L}_{h} = 3000\text{ fb}^{-1}$.
    Predictions are based on the FairSHiP simulations (SHiP) and on the FONLL simulations (MATHUSLA). See text for details.}
    \label{tab:b-d-parameters}
\end{table}

\subsection{$W$ bosons}
\label{sec:w-boson-distributions}
The production channel from the decays of $W$ bosons is only relevant for the MATHUSLA experiment since the center of mass energy at SHiP experiment is not enough to produce on-shell $W$ bosons.

The total $W$ boson production cross-section at the LHC energies $\sqrt{s} = 13\text{ TeV}$ was measured in~\cite{Aad:2016naf} as $\sigma_{W\to N+l} \approx 20.5\text{ nb}$. The corresponding number of $W$ bosons produced during the high luminosity phase of the LHC is 
\begin{equation}
N_{W,\text{total}} \approx 6\cdot 10^{11}
\label{eq:nw-lhc}
\end{equation}

The $p_{T}$ distribution of the $W$ bosons at the LHC in the pseudorapidity range $|\eta|< 2.5$ and for energies $\sqrt{s} = 7-8\text{ TeV}$ was measured by the ATLAS and CMS collaborations~\cite{Aad:2011fp,Khachatryan:2016nbe}. 
Their results show that most of the vector bosons are produced with low $p_T$ (below 10~GeV or so). 
However, these results do not give us the magnitude of the $W$'s average momentum $\langle p_{W}\rangle $, needed to  estimate  the decay acceptance and the average momentum of HNLs.

\begin{figure}[!t]
 \begin{minipage}{0.5\textwidth}
    \centering
    \includegraphics[width=\textwidth,draft=false]{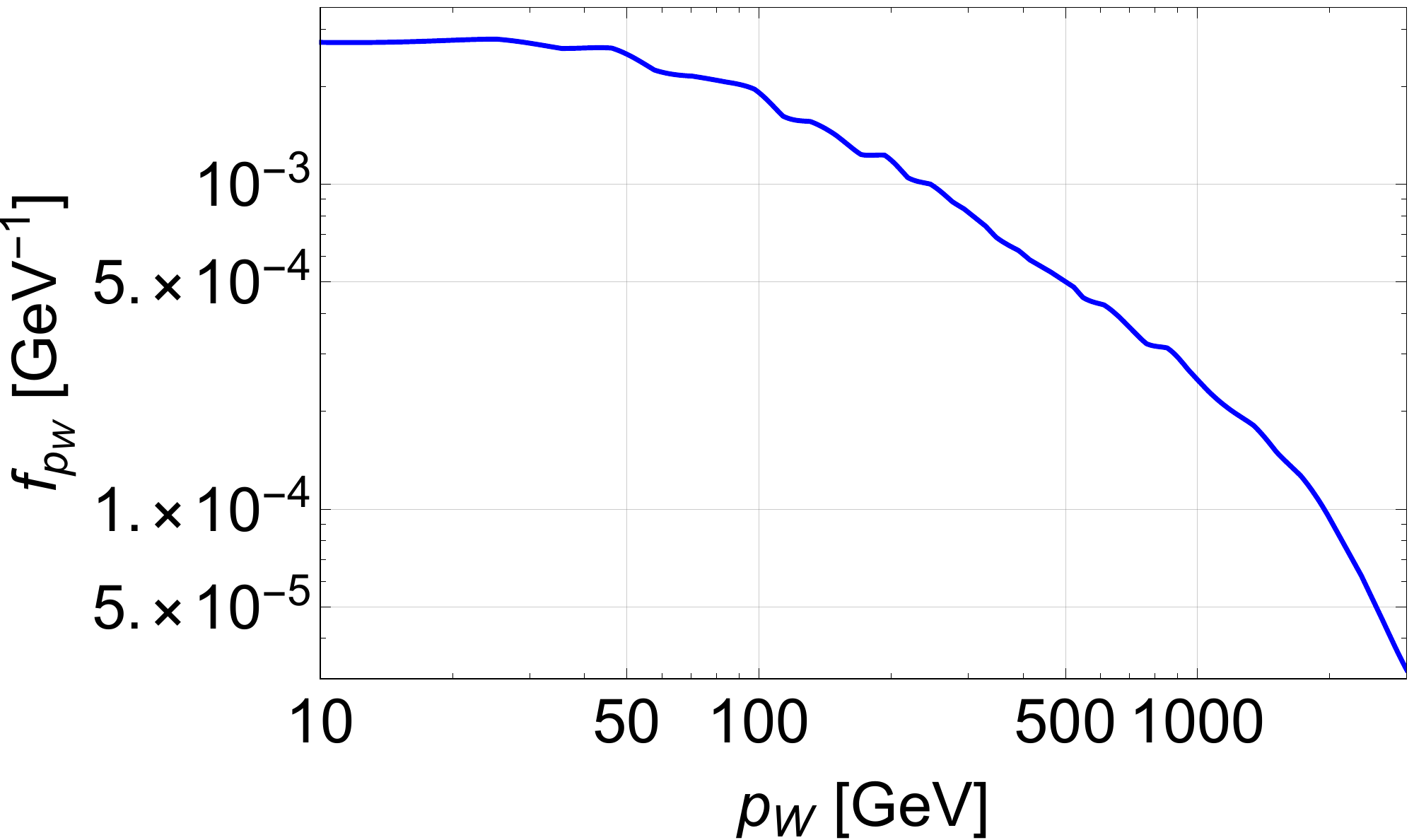}
 \end{minipage}\hfill
  \begin{minipage}{0.5\textwidth}
    \centering
    \includegraphics[width=\textwidth,draft=false]{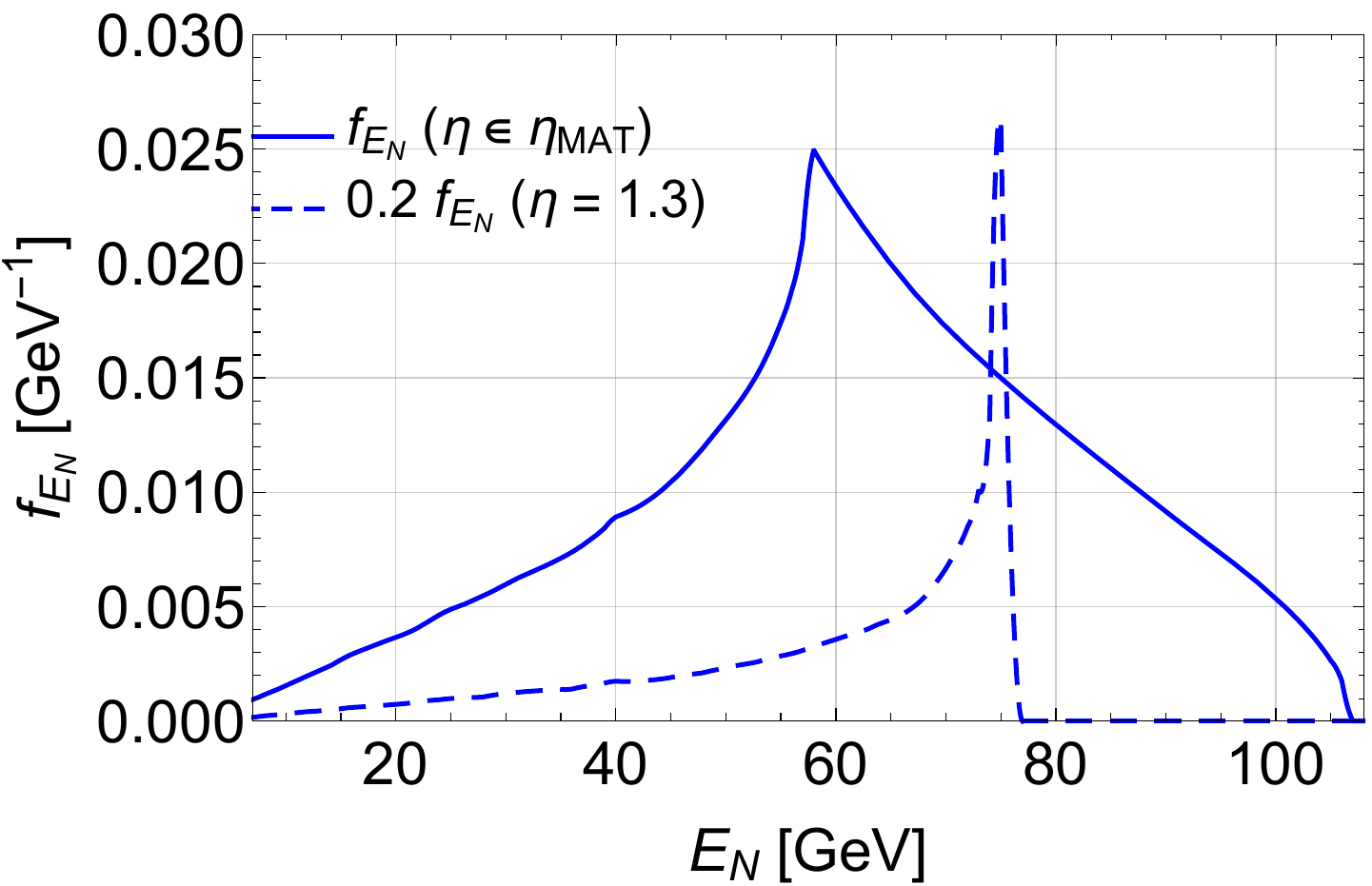}
 \end{minipage}
  \caption{Left: momentum spectrum of $W$ bosons produced in the $pp$ collisions at $\sqrt{s} = 13\text{ TeV}$ that is predicted by MadGraph5. Right: the energy spectrum of the HNLs produced in the decay of the $W$ bosons and flying in the direction of the decay volume of the MATHUSLA experiment. The solid line corresponds to the spectrum obtained for the pseudorapidity range of the MATHUSLA experiment $\eta \in (0.9,1.6)$, while the dashed line --- to the spectrum for the HNLs flying in the direction $\eta \approx 1.3$.}
  \label{fig:hnl-energy-spectrum-w}
\end{figure}

In order to obtain $\langle p_{W} \rangle$ we have simulated the process $p + p \to W^{\pm}$ in MadGraph5~\cite{Alwall:2014hca}. In the leading order we have obtained $\sigma_{W\to \nu+l}\approx 15.7\text{ nb}$, which is in reasonable agreement with the prediction~\cite{Aad:2016naf}. The resulting momentum distribution of $W$ bosons is shown in Fig.~\ref{fig:hnl-energy-spectrum-w} (left). A remark is in order here: at the leading order MadGraph5 does not predict the $p_{T}$ distribution of $W$s, since the production process is $2 \to 1$ process and the colliding partons have $p_{T} = 0$; therefore, all of the $W$ bosons in the simulations fly along the beam line, and the magnitude of their momentum is given by the longitudinal momentum $p_{L}$. The realistic $p_{T}$ spectrum can only be obtained after implementation of the parton showering. However, based on the above-mentioned measurements~\cite{Aad:2011fp,Khachatryan:2016nbe}, the typical $p_{T}$'s of $W$ bosons are significantly smaller than their typical $p_{L}$ and therefore we chose to neglect the $p_{T}$ momentum of the $W$ bosons in what follows.

Having the $W$ boson distribution $dN_{W}/dp_{W}$, we can obtain the $\epsilon_{\decay,W}$ and the average HNL momentum $\langle p_{X}\rangle$ by calculating the distribution of the particles in the energy $E_{X}$ and the angle $\theta_{X}$ between the direction of motion of the $X$ and the beam:
\begin{equation}
\label{eq:hnls-from-w}
    \frac{d^{2}N_{X}^{W}}{dE_{X}d\cos(\theta_{X})} = \int dp_{W}\frac{dN_{W}}{dp_{W}}\times \frac{d^{2}\BR_{W\to X}}{d\theta_{X}dE_{X}}\times P(\theta_{X})
\end{equation}
Here $d^{2}\BR_{W\to X}/d\theta_{X}dE_{X}$ is the differential production branching ratio, and $P(\theta_{X})$ is a projector which takes the unit value if $\theta_{X}$ lies inside MATHUSLA's polar angle range and zero otherwise.

Let us compare the amounts of the $X$ particles produced from the $W$ bosons and from $B$ mesons and flying in the direction of the decay volume. We have
\begin{equation}
     N_{\text{prod},W}/N_{\text{prod},B} \approx \frac{N_{W}}{N_{B}} \times \frac{\BR_{W\to X}}{\BR_{B\to X}}\times \frac{\epsilon_{\decay,W}}{\epsilon_{\decay,B}} \approx \begin{cases} 10^{-3}\epsilon_{\decay,W}, \quad \text{scalars}\\ 10\ \epsilon_{\decay,W}, \quad \text{HNLs},  \end{cases}
     \label{eq:w-b-comparison}
\end{equation}
where we used the amount of $B$ mesons at the LHC and the decay acceptance from the Table~\ref{tab:ship-mathusla}, the number of $W$s at the LHC~\eqref{eq:nw-lhc} and the branching ratios of the scalar and HNL production from Appendix~\ref{sec:production}. Therefore we conclude that for scalars the production from the $W$s is not relevant, while for HNLs careful estimation is needed.

In the case of HNL, the differential branching ratio in the Eq.~\eqref{eq:hnls-from-w} is
\begin{equation}
\frac{d^{2}\BR_{W\to N}}{d\theta_{N}dE_{N}} = \frac{1}{\Gamma_{W}}\frac{|\mathcal{M}_{W \to e + N}|^{2}}{8 \pi}\frac{p_{N}}{E_{W}}\delta(M_{N}^{2}+m_{W}^{2} -2E_{N}E_{W}+2|\mathbf p_{N}||\mathbf p_{W}|\cos(\theta_{N}))
\end{equation}
The energy and angular distributions of the HNLs from the $W$ bosons at MATHUSLA are almost independent of the HNL mass in the mass range of interest, $M_N \ll m_{W}$. It is an expected result because the kinematic in this limit should not depend on small HNL masses. The energy distribution for $M_{N} = 1\text{ GeV}$ is shown in Fig.~\ref{fig:hnl-energy-spectrum-w}. The decay acceptance was found to be $\epsilon_{\decay,W}\simeq 2\%$, while the average momentum of the produced HNLs is $\langle p_{N}\rangle \approx 62\text{ GeV}$.

The shape of the energy spectrum of the HNLs can be qualitatively understood in the following way. For a given value of the angle $\theta_{N}$ of the HNL, the energy distribution has a maximum at $E_{N,\text{max}}(\theta_{N}) = m_{W}/2\sin(\theta_{N})$,\footnote{This formula is valid for 2-body decay into massless particles.} which corresponds to HNLs produced from the $W$ bosons with some momentum $p_{W,\max}(\theta)$. As a consequence, the largest amount of HNLs flying in the direction $\theta_{N}$ has an energy close to their maximum, see the dashed line at the right panel of Fig.~\ref{fig:hnl-energy-spectrum-w}. The total energy spectrum is a superposition of different angles and has a peak at $E_{N,\text{peak}} \approx 58\text{ GeV}$ corresponding to the maximal angle possible at MATHUSLA, $\theta^{\max}_{\text{MAT}} \approx 44^{\circ}$. From the other side, the maximal energy possible for HNLs at MATHUSLA is defined by the minimal angle $\theta^{\text{min}}_{\text{MAT}} \approx 22^{\circ}$, which explains why the spectrum tends to zero near the energy $E_{N,\text{max}} \approx 106\text{ GeV}$.

\section{Calculation of sensitivities}
\label{sec:results}

\subsection{Efficiencies}
\label{sec:efficiency}
Using the results of Sec.~\ref{sec:energy-distributions}, we have almost all ingredients needed to estimate the lower bound, the upper bound, the maximal mass probed and the total sensitivity curve. The only questions remaining are the following. The first one is the relation between the mesons spectra and the $X$ particles spectra. The second one is the value of the overall efficiency 
\begin{equation}
\epsilon = \epsilon_{\decay}\times\epsilon_{\det}\times \BR_{\text{vis}},
\label{eq:overall-efficiency}
\end{equation}
where the quantities $\epsilon_{\decay},\epsilon_{\det},\BR_{\text{vis}}$ are the decay acceptance, detection efficiency and the visible branching correspondingly; they are defined by Eqs.~\eqref{eq:nprod},~\eqref{eq:matrix:2}.

We approximate the spectra of the $X$ particles originating from the mesons and flying to the decay volume by the distributions of the mesons flying in the direction of the decay volume. To take into account the kinematics of the meson decays, we use the relation~\eqref{eq:5} between $\gamma$ factors of the $X$ particle and the meson in the expressions~\eqref{eq:matrix:2},~\eqref{eq:decay-probability-upper-bound} for the decay probability.

Let us discuss the efficiencies. For the HNLs at SHiP experiment, we used the values of $\epsilon_{\decay}$ and $\epsilon_{\det}$ provided by detailed FairSHiP simulations~\cite{zenodo}. The results of the SHiP collaboration on the sensitivity to the scalars are not currently available, and for the product of $\epsilon_{\decay}\cdot \epsilon_{\det}$ we used the value for the HNL averaged over its mass, $\overline{\epsilon_{\decay}\cdot \epsilon_{\det}} \approx 0.2$.

For the MATHUSLA experiment there currently is no such detailed analysis of the efficiencies and background. In~\cite{Curtin:2017izq,Curtin:2018mvb} it is claimed that all the SM background can be rejected with high efficiency, but detailed simulations are needed for the justification of this statement. Here we optimistically use $\epsilon_{\det} = 1$. For the decay acceptance of the particles produced from the mesons we use the geometric acceptance of the mesons at MATHUSLA, which we obtained using FONLL.\footnote{For the geometric acceptance as MATHUSLA we use the definition $\epsilon_{\text{geom}} = N_{\text{meson}}^{\eta \in \eta_{\mat}}/N_{\text{meson}}\times \Delta \varphi/(2\pi)$, where $\eta_{\mat} \in (0.9;1.6)$ and $\Delta \varphi = \pi/2$ are correspondingly pseudorapidity range of the MATHUSLA experiment and azimuthal size, see Sec.~\ref{sec:geometry-experiments}.} For the decay acceptance of the HNLs produced in the decays of the $W$ bosons we used the value $\epsilon_{\decay,W} \approx 0.02$ obtained in Sec.~\ref{sec:w-boson-distributions}. All the parameters above, together with geometrical properties of the experiments are summarized in Table~\ref{tab:ship-mathusla}. We estimate $\langle l_{\det}\rangle$ and $\langle l_{\text{target-det}}\rangle$ using an assumption that the angular distribution of the $X$ particles in the angular range of the decay volume is isotropic, see Appendix~\ref{sec:geometry-experiments} for details.

\begin{figure}
    \begin{minipage}{0.5\textwidth}
        \centering
        \includegraphics[width=\textwidth]{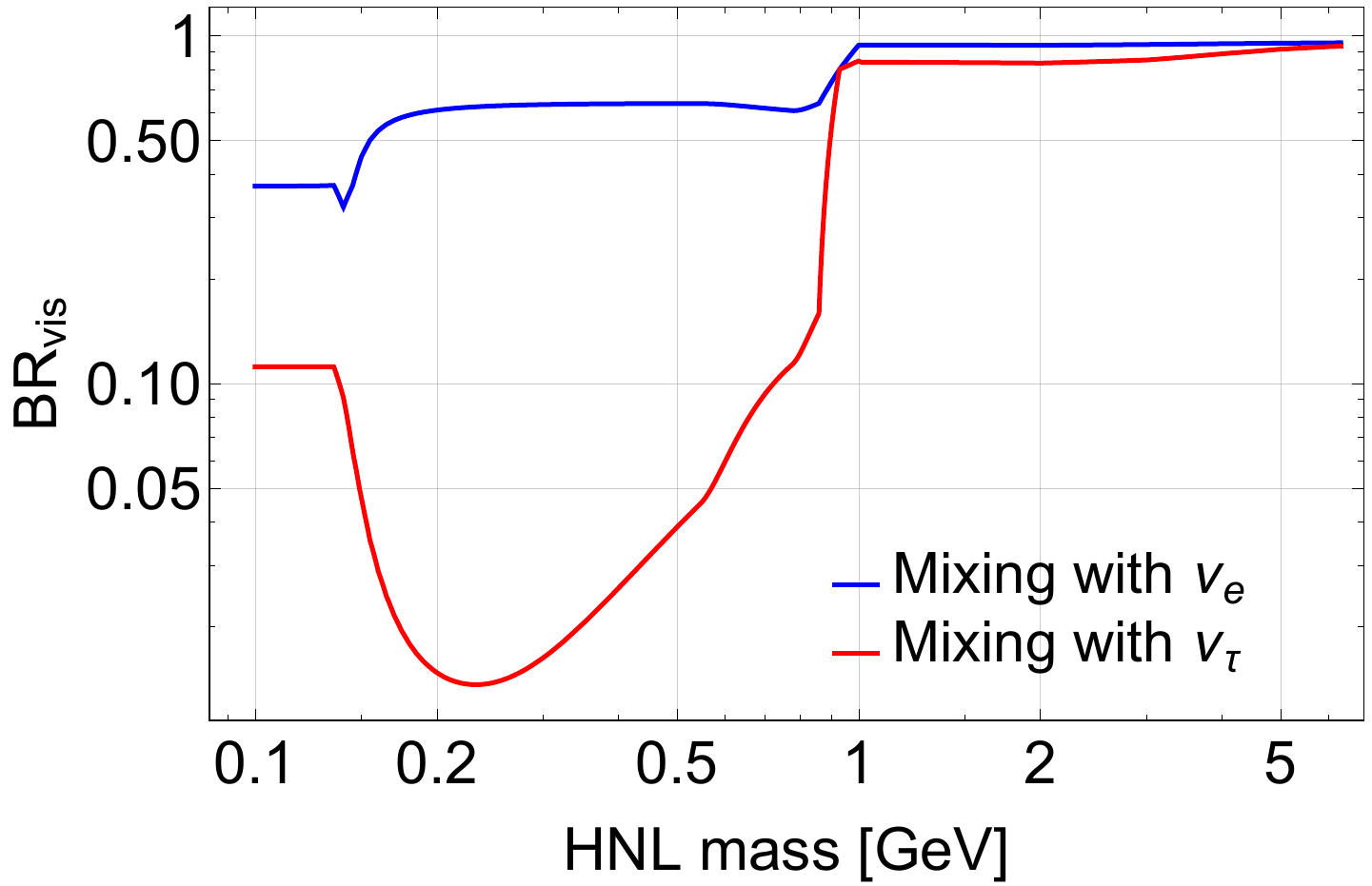}
        \end{minipage}
        \begin{minipage}{0.5\textwidth}
        \centering
        \includegraphics[width=\textwidth]{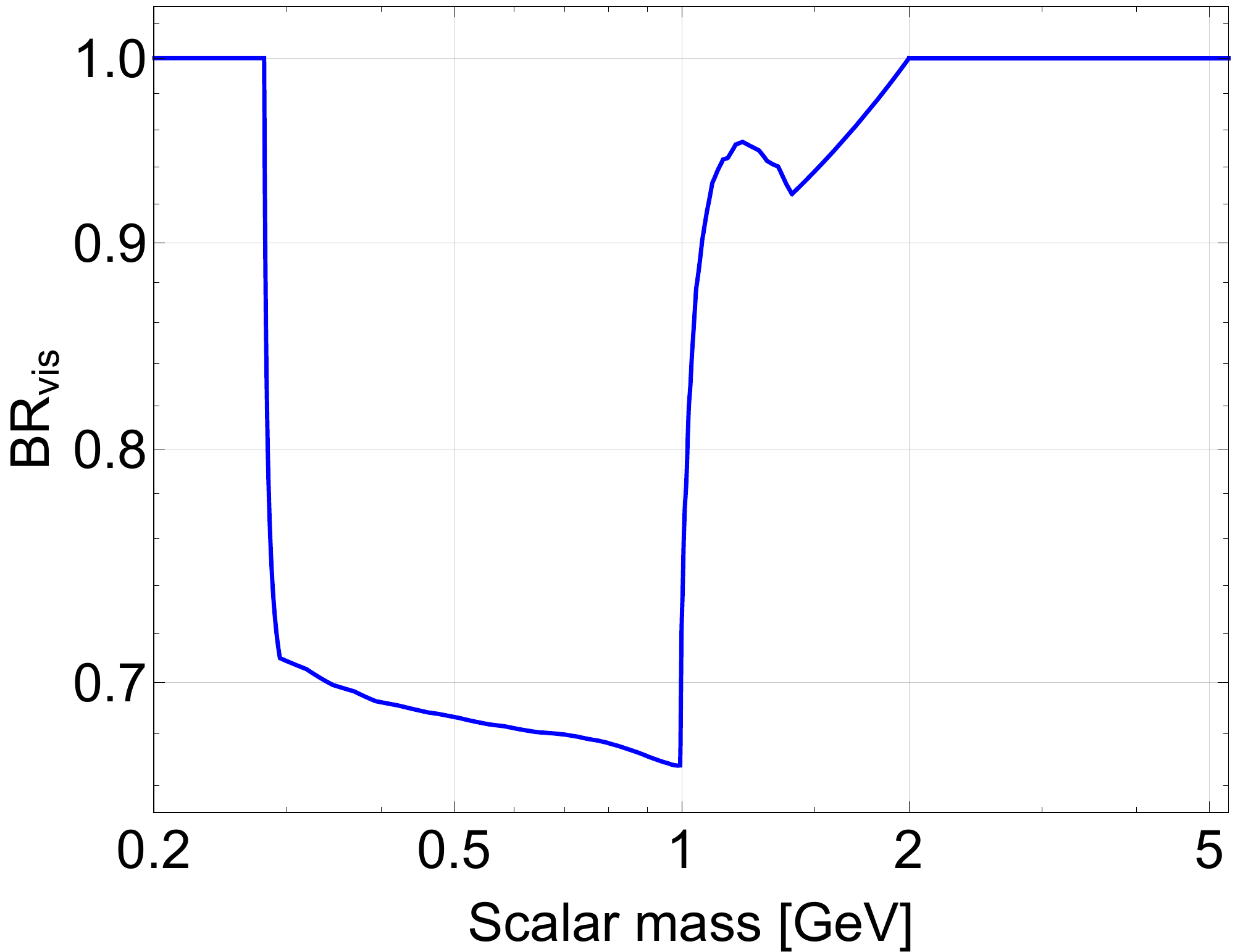}
    \end{minipage}
    \caption{The branching ratio of decays of the HNLs (left) and the scalars (right) in visible states. The drop of the branching ratio for the HNLs mixing with $\nu_{\tau}$ in the domain of HNL masses $\lesssim 1\text{ GeV}$ is caused by the dominant invisible decay $N\to \pi^{0}\nu_{\tau}$, while for the scalars of the same masses --- by the decay $S\to \pi^{0}\pi^{0}$.}
    \label{fig:visible-br-ratios}
\end{figure}

The last needed parameter is the visible decay branching fraction. Following~\cite{SHiP:2018xqw,Curtin:2018mvb}, for the visible decay branching fractions for both MATHUSLA and SHiP experiments we include only the decay channels of the $X$ particle that contain at least two charged tracks. Our estimation of $\BR_{\text{vis}}$ is described in Appendix~\ref{sec:visible-br}. The plots of the visible branching ratios for the HNLs and for the scalars are shown in Fig.~\ref{fig:visible-br-ratios}.

\begin{table}[t]
\centering
  \begin{tabular}{|c|c|c|c|c|c|c|c|c|c|c|}
    \hline
    \textbf{Exp.} & $\langle l_{\text{target-det}}\rangle$ & $\langle l_{\text{det}}\rangle$ & $\overline{\epsilon}_{X,D}$ & $\overline{\epsilon}_{X,B}$ &$\overline{\epsilon}_{X,W}$ & $N_{D,\text{eff}}$& $N_{B,\text{eff}}$ & $N_{W,\text{eff}}$ \\
    \hline
    MAT & $192\text{ m}$ & $38\text{ m}$ & $0.013$ & $0.018$ & 0.02& $5.7\cdot 10^{14}$ & $5.4\cdot 10^{13}$ & $1.2\cdot 10^{10}$ \\
    \hline
    SHiP & $50\text{ m}$ & $50\text{ m}$ & $0.09$ & $0.12$ &-- & $1.4\cdot 10^{17}$ & $1.3\cdot 10^{13}$ & -- \\
    \hline
  \end{tabular}
  \caption{Parameters of the SHiP and MATHUSLA experiments: the average length from the interaction point to the decay volume $\langle l_{\text{target-det}}\rangle$, the average length of the decay volume $\langle l_{\text{det}}\rangle$ (see Appendix~\ref{sec:geometry-experiments} for details), values of the overall efficiencies~\eqref{eq:overall-efficiency} averaged over the probed mass range of $X$ for the particles $X$ produced from $D$ and $B$ mesons, the effective number of the $D$ and $B$ mesons and $W$ bosons defined by $N_{M,\text{eff}} = N_{M}\times \bar{\epsilon}_{X,M}$. }
  \label{tab:ship-mathusla}
\end{table}

\subsection{Lower bound}
\label{sec:lower-bound-estimation}

Let us first compare the relevant parameters of the experiments summarized in Tables~\ref{tab:b-d-parameters},~\ref{tab:ship-mathusla}. One sees that the effective number of $D$ mesons is approximately two orders of magnitude larger at SHiP,\footnote{By the effective number of the mesons we call the production of the number of mesons multiplied by the overall efficiency, $N_{\text{meson}} \cdot \epsilon_{X,\text{meson}}$.} the effective numbers of $B$ mesons are comparable between the experiments, and the average momenta (and therefore the $\gamma$ factors) of the mesons produced in the direction of the decay volume are $\simeq 7-8$ times smaller at MATHUSLA. The latter is caused by (i) different beam configurations (colliding beams for MATHUSLA, fixed target for SHiP) (ii) their different geometric orientation relative to the proton beam direction (the decay volume of the SHiP experiment is located in the forward direction, while the one of MATHUSLA's is about $20^\circ$ off-axis.)

Using the numbers from the Tables~\ref{tab:b-d-parameters},~\ref{tab:ship-mathusla}, for the ratio of the mixing angles at the lower bound~\eqref{eq:events-ratio} we have
\begin{equation}
  \label{eq:events-ratio-explicit}
  \frac{U^{2}_{\text{lower},\ship}}{U^{2}_{\text{lower},\mat}}\bigg|_{M_{N}\lesssim m_{D}}\approx \frac{1}{5},\quad \frac{U^{2}_{\text{lower},\ship}}{U^{2}_{\text{lower},\mat}}\bigg|_{M_{N}\gtrsim m_{D}}\simeq \frac{\theta^{2}_{\text{lower},\ship}}{\theta^{2}_{\text{lower},\mat}}\bigg|_{M_{S}\gtrsim m_{K}} \approx 5
\end{equation}
Qualitatively, for particles produced in the decays of the $B$ mesons (HNLs with masses $M_N > m_D$ and scalars with masses $M_S > m_K$) MATHUSLA can probe mixing angles a factor $\simeq 5$ smaller than SHiP due to the smaller $\gamma$ factor of the $B$ mesons and larger effective number of $B$ mesons (\textit{i.e.}\ the total number of $B$ mesons times the overall efficiency~\eqref{eq:overall-efficiency}). 
For the HNLs in the mass range $m_K \lesssim M_N \lesssim m_{D}$ the smallness of $\gamma$ factor of the $D$ mesons at MATHUSLA and the suppression of the number of events at SHiP by the overall efficiency cannot compensate the difference of two orders of magnitude in the effective numbers of the $D$ mesons, and therefore the SHiP reaches a sensitivity which is about half an order of magnitude lower in $U^2$. We note again that the result~\eqref{eq:events-ratio-explicit} was obtained under the optimistic condition $\epsilon_{\det}=1$ for MATHUSLA; after using a realistic efficiency the lower bound of the sensitivity at MATHUSLA will be changed by a factor $1/\sqrt{\epsilon_{\det}}$, which will affect the ratio~\eqref{eq:events-ratio-explicit}.

\subsection{Upper bound}
\label{sec:upper-bound-results}

\begin{figure}
      \begin{minipage}{0.5\textwidth}
    \centering
    \includegraphics[width=\textwidth]{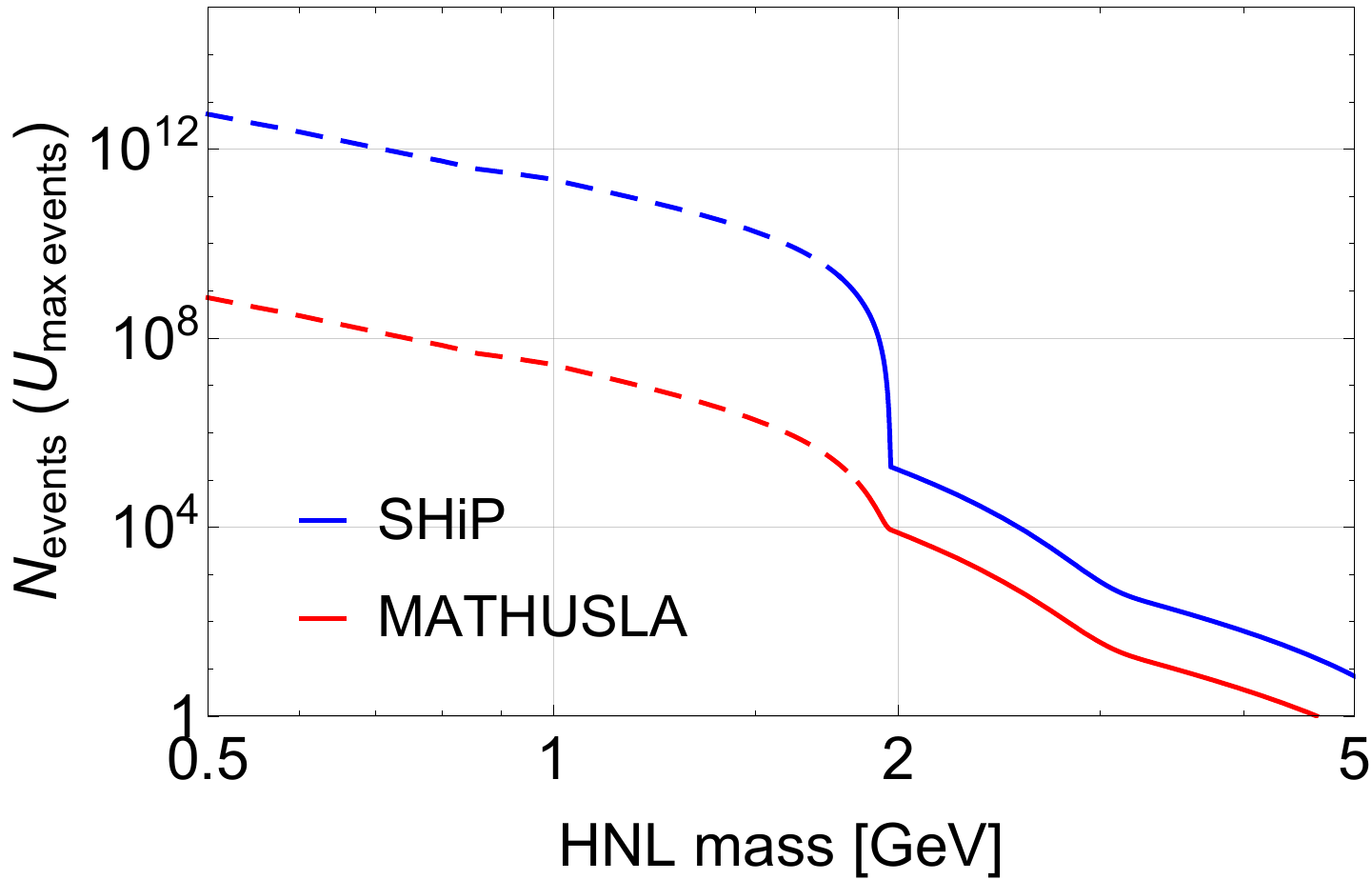}
  \end{minipage}\hfill
  \begin{minipage}{0.5\textwidth}
    \centering
    \includegraphics[width=\textwidth]{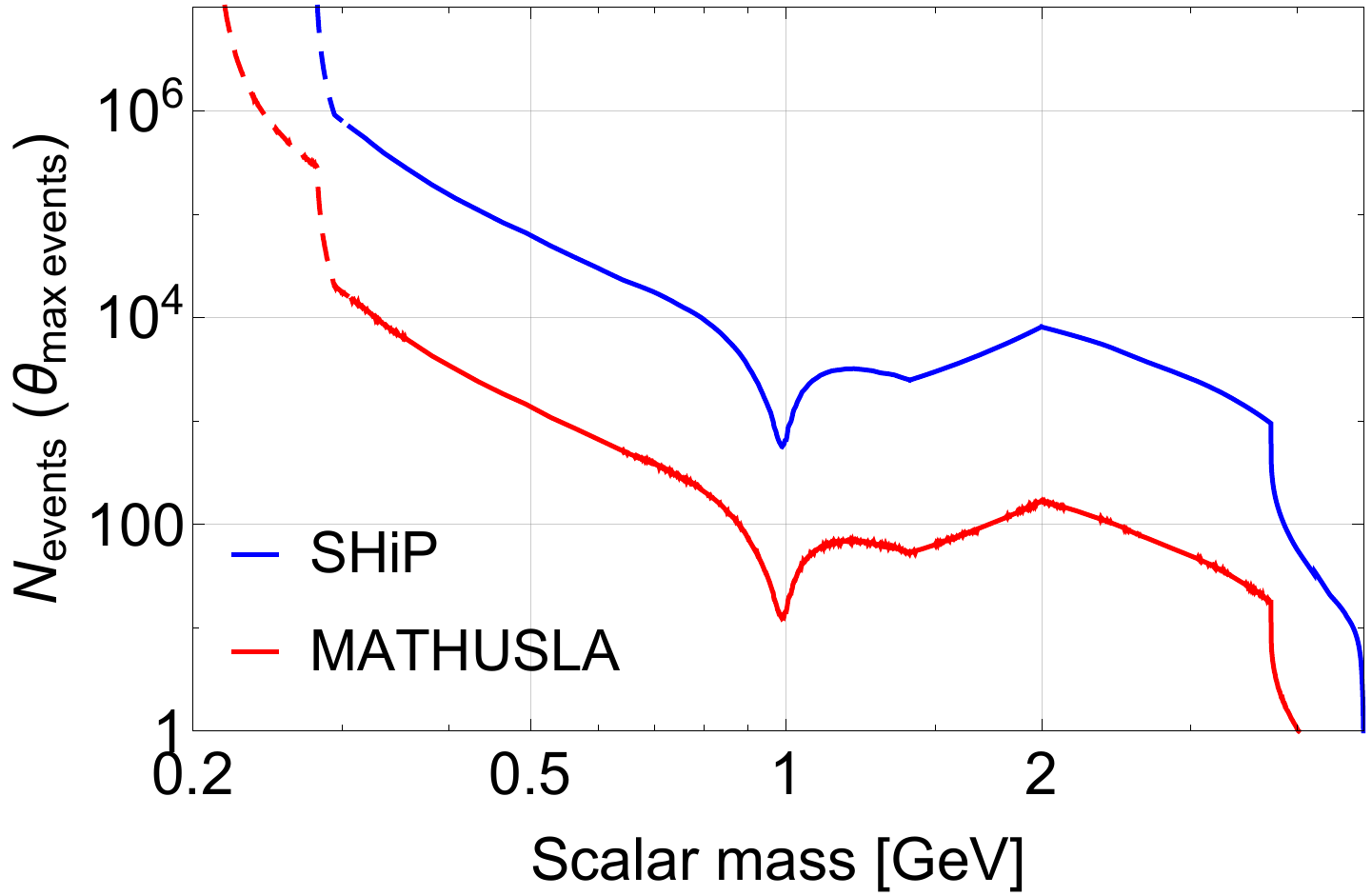}
  \end{minipage}
    \caption{The dependence of the number of events at SHiP and MATHUSLA evaluated at $U^{2} = \theta^{2}_{\text{max}}$ for the HNLs mixing with $\nu_{e}$ (left) and for scalars (right). Dashed lines denote the values for $U^{2}_{\text{max}}$ for which the sensitivity of SHiP and MATHUSLA intersects the domain that has been closed by previous experiments (see, e.g.,~\cite{Alekhin:2015byh}).}
    \label{fig:nevents-umax}
\end{figure}

We show the dependence of the number of events at $\theta^{2}_{X} = \theta^{2}_{\text{max}}$ as a function of the mass for the HNLs mixing with $\nu_{e}$ and the scalars in Fig.~\ref{fig:nevents-umax}. We see that by the maximal number of events the SHiP experiment is much better than the MATHUSLA experiment, which is explained by the shorter length to the decay volume and higher value of the average gamma factor.

With the energy distributions of the mesons and the $W$ bosons obtained in Sec.~\ref{sec:energy-distributions}, let us now estimate their effect on the upper bound of the sensitivity. To do this, we introduce the width of the upper bound defined by
\begin{equation}
    R = \theta^{2}_{\text{upper}}/\theta^{2}_{\text{max}}
    \label{eq:upper-bound-width}
\end{equation}

We take the HNLs as an example, commenting later on the difference with the scalar. We will be interested in the HNLs with $M_N \gtrsim 2$~GeV (for smaller masses $\theta^{2}_{\text{upper}}$ lies deep inside the region excluded by the previous searches, see e.g.~\cite{Bondarenko:2018ptm}). The HNLs in question are produced from the decays of $B$ mesons and $W$ bosons.

Our procedure of the estimation of the upper bound width is based on~\eqref{eq:decay-probability-upper-bound}. As we already mentioned at the beginning of this Section, in the case of the production from $B$ mesons we approximate the spectra of the HNLs by the spectra of the $B$ mesons (so that the HNLs fly in the same direction as the $B$ mesons) and take into the account the relation~\eqref{eq:5} between the $B$ meson and the energies of HNLs. In the case of the production from the $W$ bosons, we use the energy spectrum of the HNLs from Fig.~\ref{fig:hnl-energy-spectrum-w}. We approximate the shapes of the high-energy tails of these spectra by simple analytic functions. For the $B$ mesons at SHiP, the fit is an exponential function, for the $B$ mesons at MATHUSLA the fit is a power law function, while for the HNLs from the $W$ bosons the fit is a linear function, see Appendix~\ref{sec:spectra-fit}. Using the fits, we calculate the upper bound $\theta^{2}_{\text{upper}}$ using the steepest descent method for the evaluation of the integral~\eqref{eq:decay-probability-upper-bound}. The derivation of $\theta^{2}_{\text{upper}}$ is given in Appendix~\ref{sec:upper-bound-details}.

\begin{figure}[!t]
  \begin{minipage}{0.5\textwidth}
    \centering
    \includegraphics[width=\textwidth,draft=false]{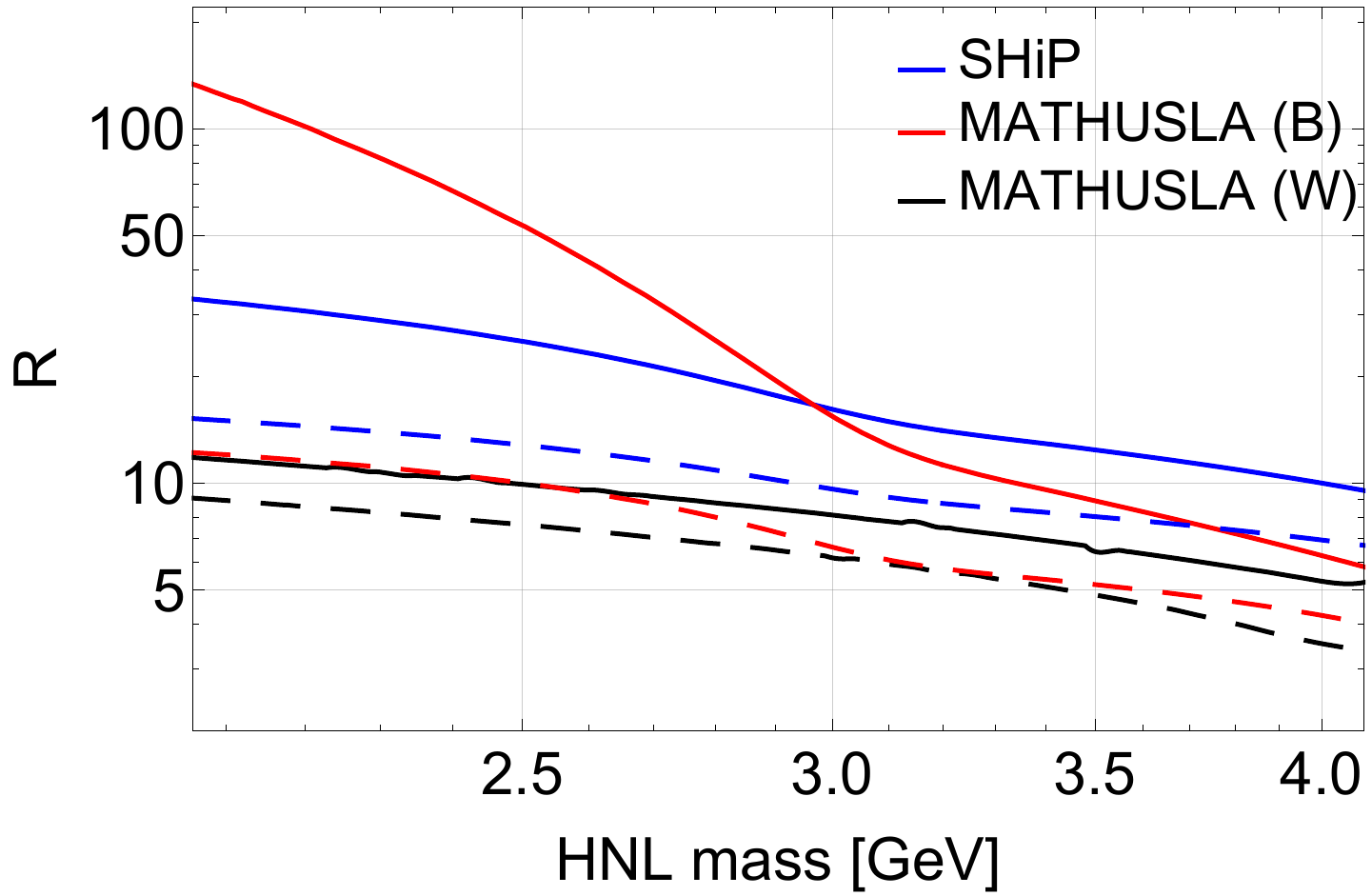}
  \end{minipage}\hfill
  \begin{minipage}{0.5\textwidth}
    \centering
    \includegraphics[width=\textwidth,draft=false]{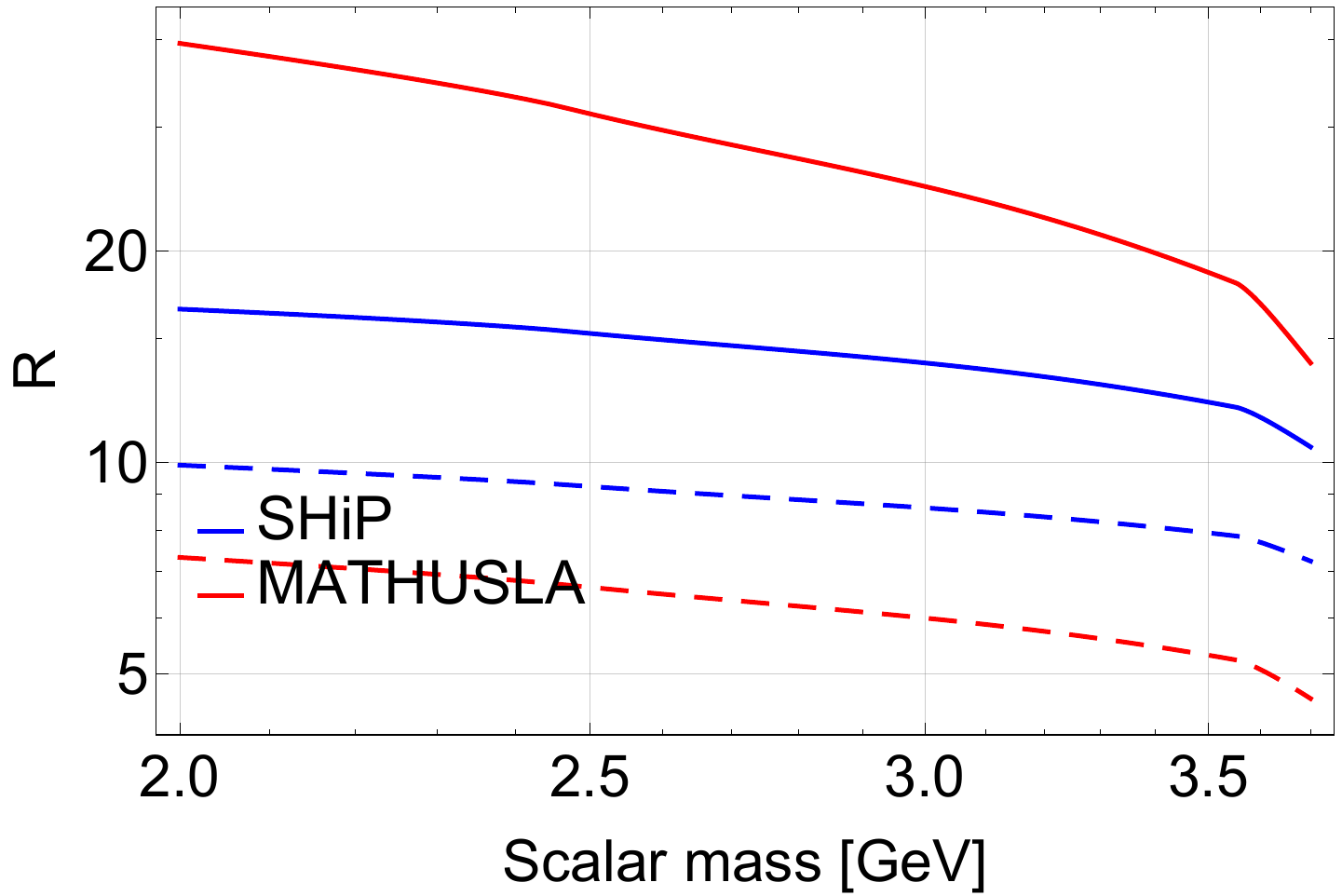}
  \end{minipage}
  \caption{Ratios $R = \theta^{2}_{\text{upper}} / \theta^{2}_{\text{max}}$ for the HNLs mixing with $\nu_{e}$ (left) and for scalars (right) at SHiP and MATHUSLA experiments. Solid lines are obtained by taking the account the energy distribution of the mother particles ($B$ mesons and $W$ bosons). Dashed lines are obtained under an assumption that all particles have the same average energy.}
  \label{fig:upper-bound-width}
\end{figure}

Using $\theta^{2}_{\text{upper}}$, we present the upper bound width~\eqref{eq:upper-bound-width} in Fig.~\ref{fig:upper-bound-width}. We also show there the prediction of the estimations of the upper bound width which assume that all of the produced particles have the same energy, see Eq.~\eqref{eq:upper-bound-estimation}.

We see that for the particles from $B$ mesons at SHiP and for the HNLs from the $W$ bosons at MATHUSLA the broadening of the width due to the distribution is small, while for the particles from $B$ mesons the distribution contributes significantly. This is a direct consequence of the behavior of the shape of the high-energy tails of the distributions. Namely, for the $B$ mesons at SHiP, the number of high-energy HNLs is exponentially suppressed. For the HNLs originating from the $W$ bosons the tail falls linearly, and naively the upper bound would be significantly improved. However, the distribution becomes zero not very far from $\langle p_{N}\rangle$, and the effect of the contribution is insignificant. Only for the $B$ mesons at MATHUSLA the tail causes significant improvement of the width of the upper bound.

Finally, let us comment on the difference between the shapes of the width between the HNL and scalar cases. The lifetime $\tau_{S}$ is changed with the mass slower than $\tau_{N}$, see the discussion in Sec.~\ref{sec:decay}. In addition, $\text{Br}_{B\to S}$ behaves with the mass monotonically, while for HNLs new production channels appear at different masses. Therefore the upper bound of the sensitivity region for the scalars changes less steeply and more smoothly with their mass, see Fig.~\ref{fig:upper-bound-width} (right panel).

The comparison of the upper bound of the sensitivity for the HNLs originating from $W$ bosons and $B$ mesons is shown in Fig.~\ref{fig:mathusla-comparison-b-w}. Our method of obtaining the sensitivity is summarized in Appendix~\ref{sec:simulations}. We see that the $W$s determine the upper bound. The reason for this is that the HNLs from $W$s have sufficiently larger average momentum, which compensates the production suppression (see Eq.~\eqref{eq:w-b-comparison}).

\begin{figure}[!t]
    \centering
    \includegraphics[width=0.7\textwidth,draft=false]{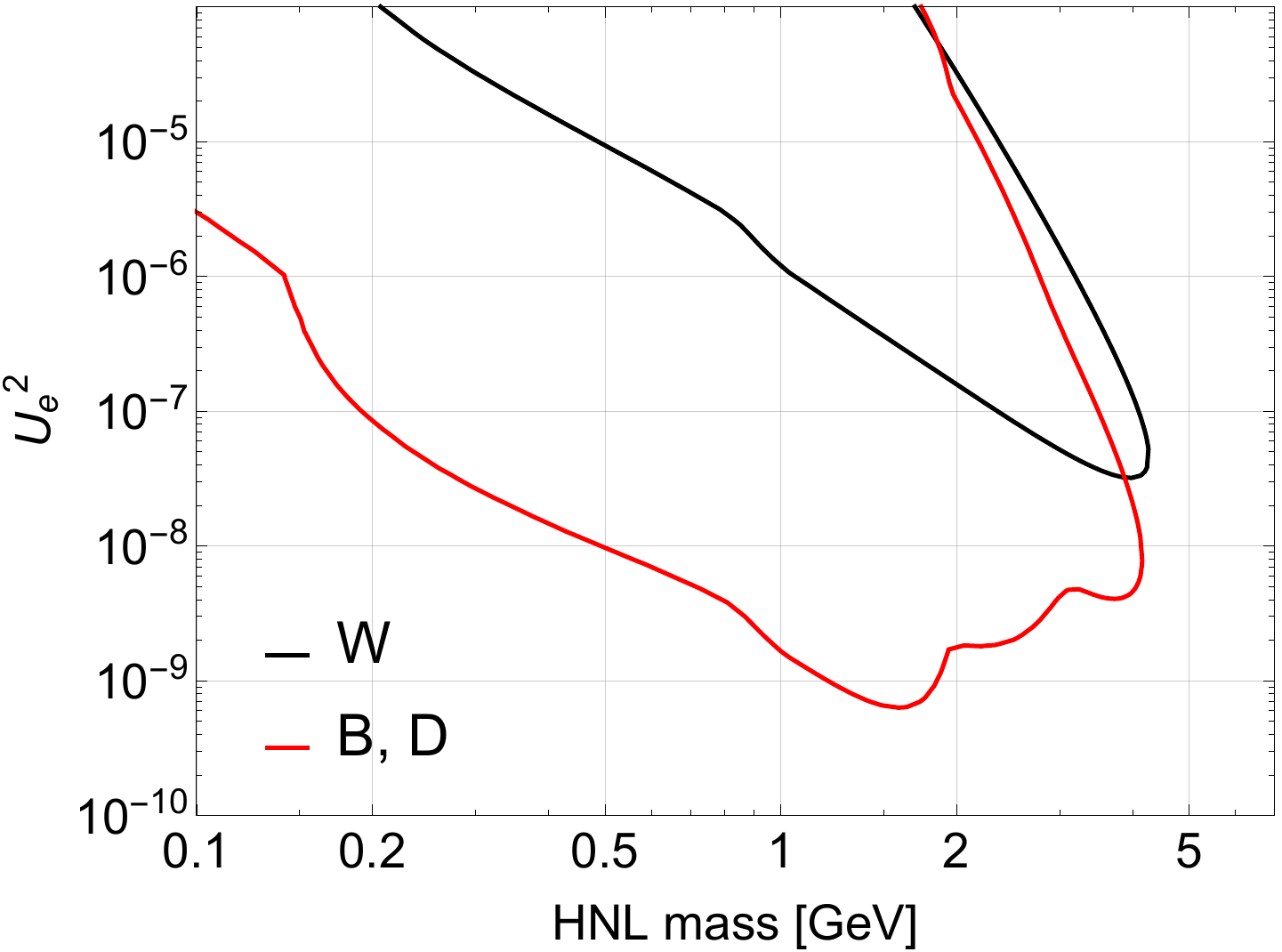}
  \caption{Comparison of the sensitivity of the MATHUSLA experiment to the HNLs that are produced in decays of $D$ and $B$ mesons (including $B_c$) and in decays of $W$ for the mixing with $\nu_{e}$.}
  \label{fig:mathusla-comparison-b-w}
\end{figure}

\subsection{Maximal mass probed}
The smaller $\gamma$ factor of the mesons at MATHUSLA adversely affects the upper bound of the sensitivity curve and thus the maximal mass probed. In particular, for the HNLs mixing with $\nu_{e/\mu}$, the maximal mass probed ratio~\eqref{eq:maximally-probed-mass-ratio} becomes
\begin{equation}
  \label{eq:max-mass-explicit}
  M_{\max}^{N,\ship}/M_{\max}^{N,\mat} \approx 1.3,
\end{equation}
which agrees well with the sensitivity plot from Fig.~\ref{fig:sensitivity-HNL}. For the other cases --- the HNLs mixing with $\nu_{\tau}$ and the scalars --- the estimation of the maximal mass for the SHiP experiment based on the definition above exceeds the kinematic threshold, and therefore the result~\eqref{eq:maximally-probed-mass-ratio} is not valid. However, for the scalars the maximal mass for the MATHUSLA experiment is smaller than the kinematic threshold, which is still a consequence of smaller $\gamma$ factor.

\section{Comparison with simulations}
\label{sec:comparison-simulations}
Next we compare our sensitivity estimates with the results of the SHiP and MATHUSLA collaborations. 
Different groups used different phenomenology for HNL and, especially, for scalars. Therefore, we will use different prescriptions for production and/or decay in different sections below, in order to facilitate the comparison of our approach with the Monte Carlo results of other groups.
Our current view of the HNL phenomenology is summarized in~\cite{Bondarenko:2018ptm} and for scalar in~\cite{Boiarska:2019jym}.
Our method of obtaining the sensitivity curves is summarized in Appendix~\ref{sec:simulations}.

\subsection{HNLs}
\label{sec:HNLs-sensitivity}
\begin{figure}[!t]
  \centering \includegraphics[width=0.5\textwidth]{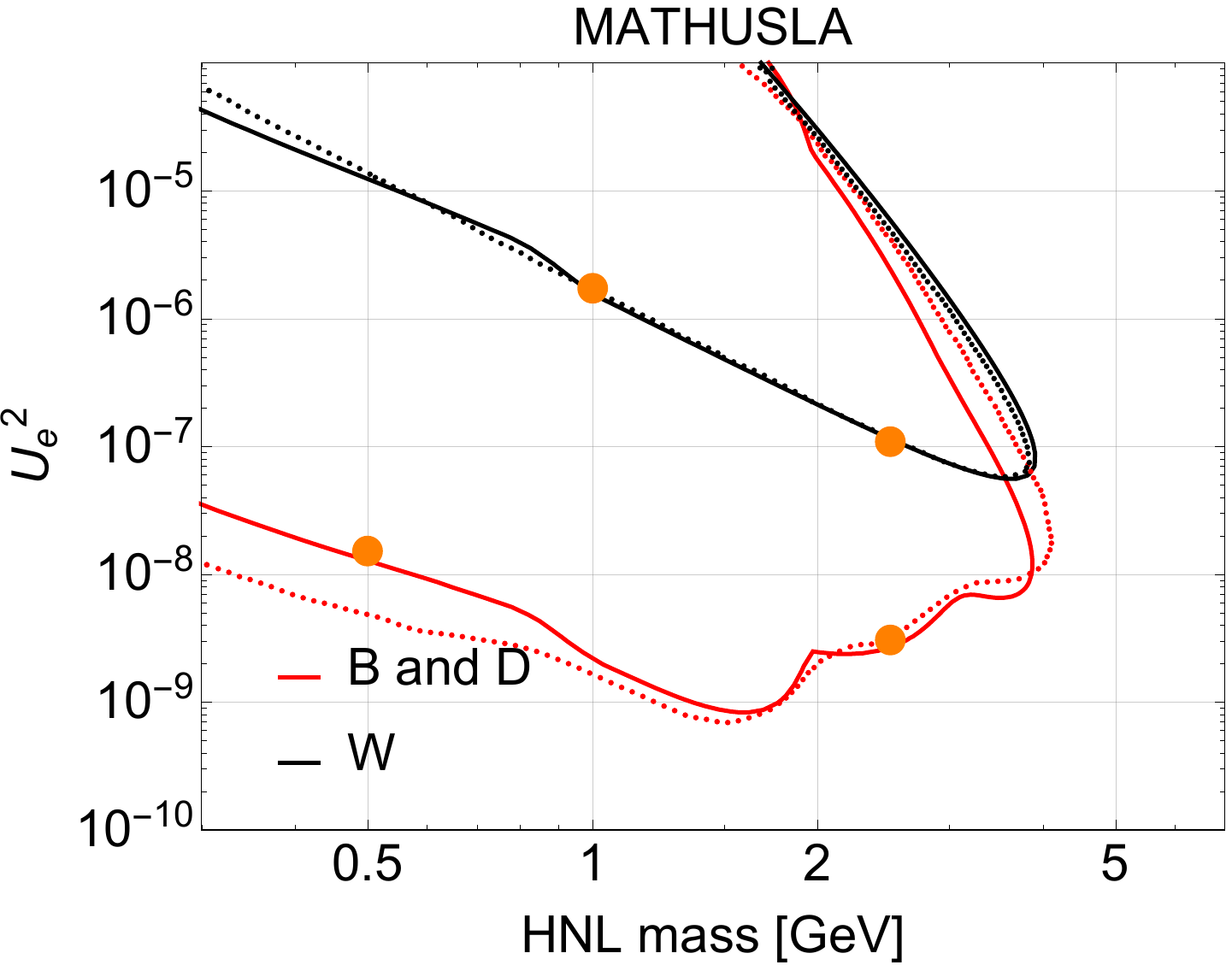}~
  \includegraphics[width=0.5\textwidth]{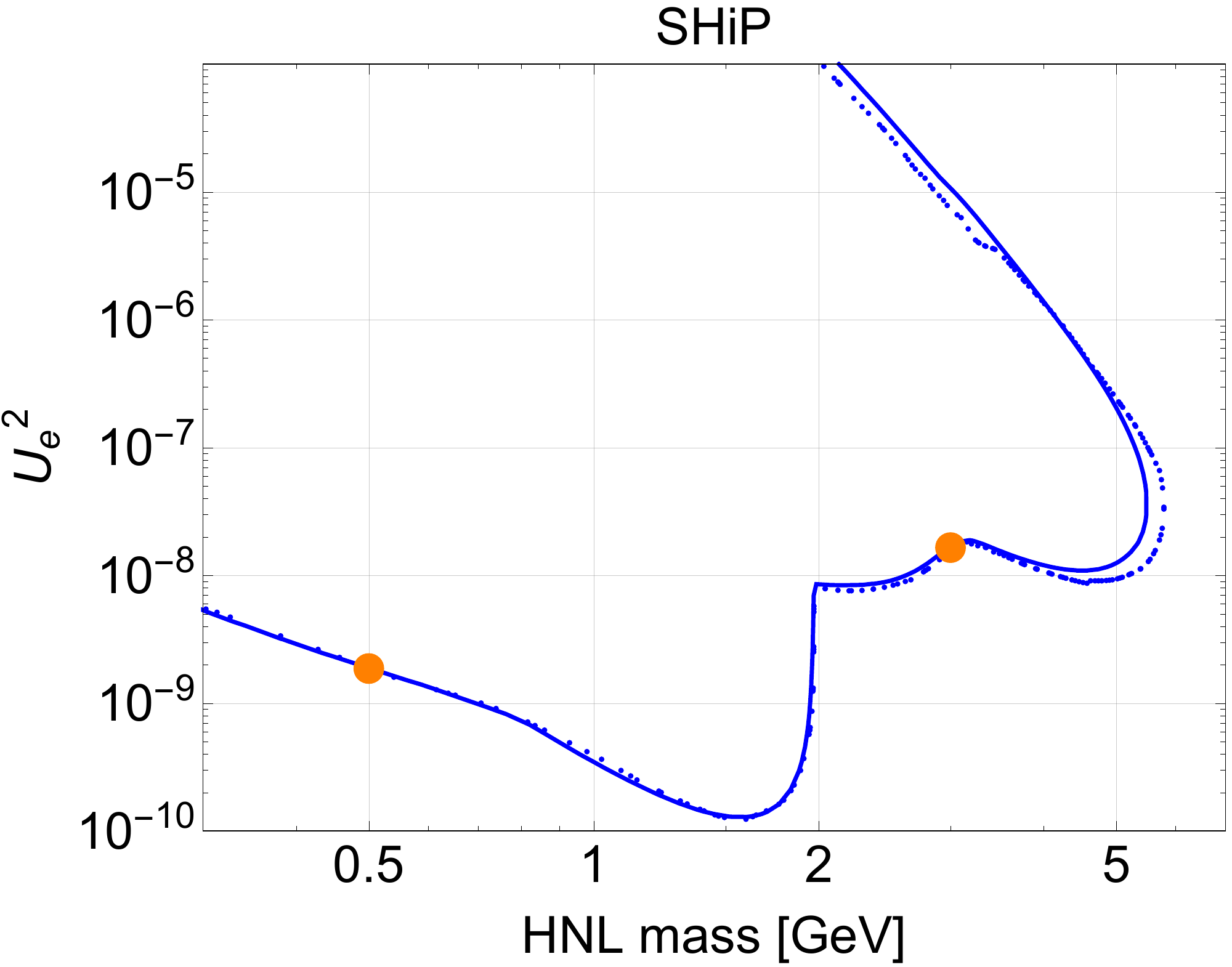}
  \caption[HNL mixing with electron flavour]{Comparison of the sensitivities to the HNLs mixing with the electron flavor obtained in this paper (solid lines) with the results of the SHiP~\cite{SHiP:2018xqw} and MATHUSLA~\cite{Curtin:2018mvb} collaborations y(dotted lines). For the MATHUSLA experiment, the contributions from both $B,D$ mesons and from $W$ bosons are shown separately. For the SHiP experiment, we consider the case of maximally possible contribution of $B_{c}$ mesons, given by the fragmentation fraction $f_{B_{c}} = 2.6\cdot 10^{-3}$ measured at LHC energies $\sqrt{s} = 13\text{ TeV}$~\protect\cite{Aaij:2017kea}. Orange points, based on analytic estimates of the lower boundary, allow for simple cross-check of our results, see Appendix~\ref{sec:nevents-lower-bound-prediction} for details. Possible origins of the discrepancy at low masses at the left panel are discussed in Appendix~\protect\ref{app:HNLsatMATHUSLA}.}
  \label{fig:sensitivity_comparison_HNLs}
\end{figure}

The results for the HNLs are shown in Fig.~\ref{fig:sensitivity_comparison_HNLs}. 
To facilitate the cross-check of our results, we also provide simple analytic estimates of the lower boundary for several HNL masses (see Appendix~\ref{sec:nevents-lower-bound-prediction}). 
Small discrepancies between the simple estimation of the lower bound and numeric result are caused by the difference in the values of $1/\langle p_{\text{meson}}\rangle$ and $\langle 1/p_{\text{meson}}\rangle$, which actually defines the lower bound.

For the sensitivity of the SHiP experiment, there is good agreement of the sensitivity curves, with a slight difference in the maximal mass probed. We think that this is due to the difference in the average $\gamma$ factors used in our estimation and those obtained in Monte Carlo simulations by the SHiP collaboration. Indeed, using the SHiP simulations results available in~\cite{SHiP:2018xqw,zenodo}, we have found that for the masses $M_{N}\simeq m_{B_{c}}$ the ratio of average $\gamma$ factors is $\langle \gamma_{N}^{\text{analytic}}\rangle/\langle\gamma_{N}^{\text{simulations}}\rangle \simeq 0.8$, which seems to explain the difference.

For the sensitivity of MATHUSLA~\cite{Curtin:2018mvb} to the HNLs produced in  $W$ decays there is good agreement for the entire mass range probed. For the sensitivity to the HNLs from $B$ and $D$ mesons, the situation is somewhat different. In the mass range $M_{N}\gtrsim m_{D_{s}}$, where the main production channel is the decay of the $B$ mesons, there is reasonable agreement with our estimate. The discrepancy can be caused by higher average energy of the HNLs in the simulations, which simultaneously lifts up the lower and upper bounds of the sensitivity. The reason for the difference at masses $M_N<1$~GeV is not known, see a discussion in Appendix~\ref{app:HNLsatMATHUSLA}.

\subsection{Scalars}
\label{sec:scalar-comparison}
The comparison of our sensitivity estimates with the results of the SHiP and MATHUSLA experiments is presented in Fig.~\ref{sec:scalar-phenomenologies-comparison}. We also show the results of a simple analytic estimate of the lower bound for particular masses from Appendix~\ref{sec:nevents-lower-bound-prediction}. For the comparison with the sensitivity provided by the MATHUSLA collaboration we used the model of scalar production and decay given in~\cite{Curtin:2018mvb}, while comparing with the results of the SHiP collaboration -- from~\cite{Beacham:2019nyx}. A description of the models is given in Appendix~\ref{sec:scalar-phenomenologies-comparison}.

\begin{figure}[!t]
  \centering
  \includegraphics[width=0.49\textwidth]{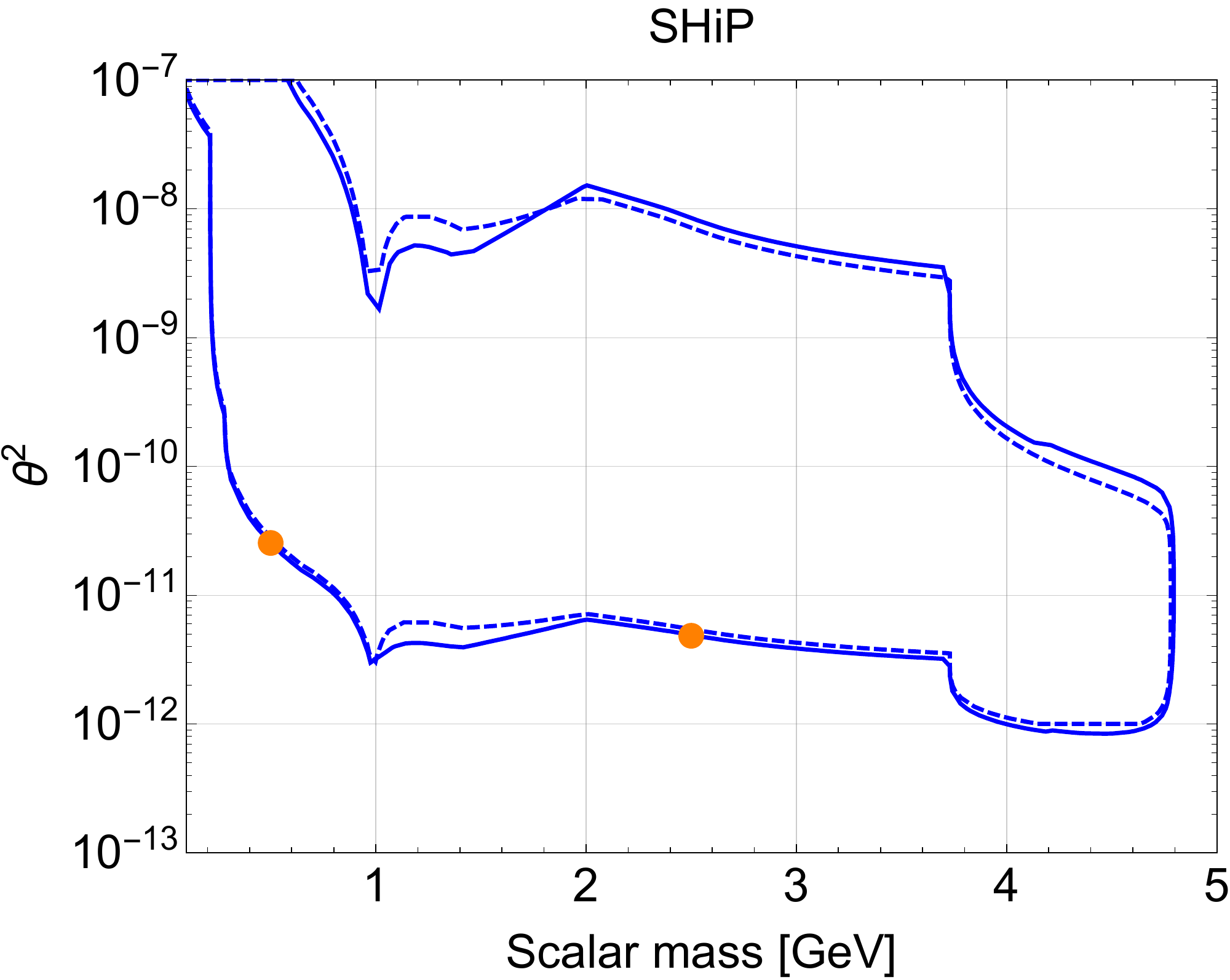}~
  \includegraphics[width=0.49\textwidth]{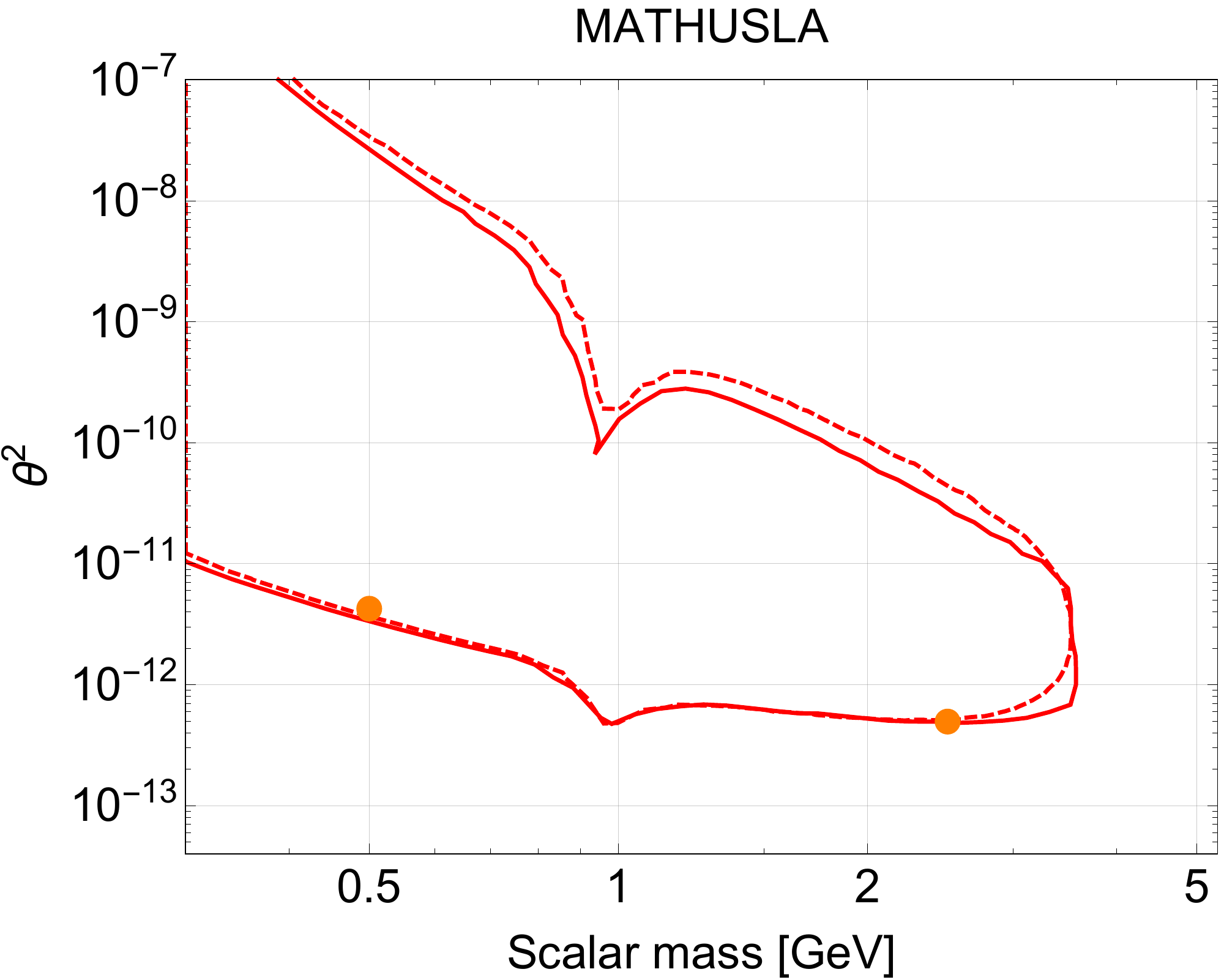}
  \caption[Scalar sensitivity]{Sensitivity to the scalar portal particles for the SHiP (\textit{left panel}) and MATHUSLA (\textit{right panel}) experiments. 
  Solid lines -- results obtained in this work. Dashed lines -- simulations of SHiP~\cite{Beacham:2019nyx} (left panel) and MATHUSLA~\cite{Curtin:2018mvb} (right panel). In order to facilitate the comparison with collaboration results \textit{we have used different scalar production and decay models} in left and right panels: for comparison with MATHUSLA results we took the model from~\cite{Curtin:2018mvb}, while for comparison with SHiP we used the model from~\cite{Krnjaic:2015mbs}, see Section~\protect\ref{sec:scalar-comparison}. Orange points, based on analytic estimates of the lower boundary, allow for simple cross-check of our results, see Appendix~\protect\ref{sec:nevents-lower-bound-prediction}.
  }
  \label{fig:sensitivity_comparison_scalars}
\end{figure}
The sensitivity curves are in good agreement. Small differences in the position of the maximal mass probed can be explained by different energy distributions of the scalars used in our estimate and in~\cite{Curtin:2018mvb} and  in~\cite{Beacham:2019nyx}.

\section{Conclusions}
\label{sec:conclusion}

In this work, we investigated the sensitivity of Intensity Frontier experiments to two models of new super-weakly interacting physics: \textit{heavy neutral leptons} and  \textit{dark scalars}. 
We explored analytically the characteristic features of the experiment's sensitivity regions: upper and lower boundaries and the maximal mass of new particles that can be probed.
Our analytic analysis allows identifying the parameters responsible for the positions of  the main ``features'' of these curves and to cross-check/validate the results of the Monte Carlo simulations.
We analyse a number of experimental factors that contribute to the sensitivity estimates:
\textit{(i)} the number of heavy flavour mesons traveling in the direction of the detector;
\textit{(ii)} their average momentum and the high-energy tail of the momentum distribution;
\textit{(iii)} geometry of the experiment;
\textit{(iv)} efficiency.
We use SHiP and MATHUSLA as examples of the fixed target and LHC-accompanying Intensity Frontier experiments, respectively. 
Our analytic estimates agree well with the Monte Carlo-based sensitivities provided by the SHiP~\cite{SHiP:2018xqw} and MATHUSLA~\cite{Curtin:2018mvb} collaborations under similar assumptions about the overall efficiencies of the experiments.

\begin{figure}[!t]
  \centering \includegraphics[width=0.5\textwidth]{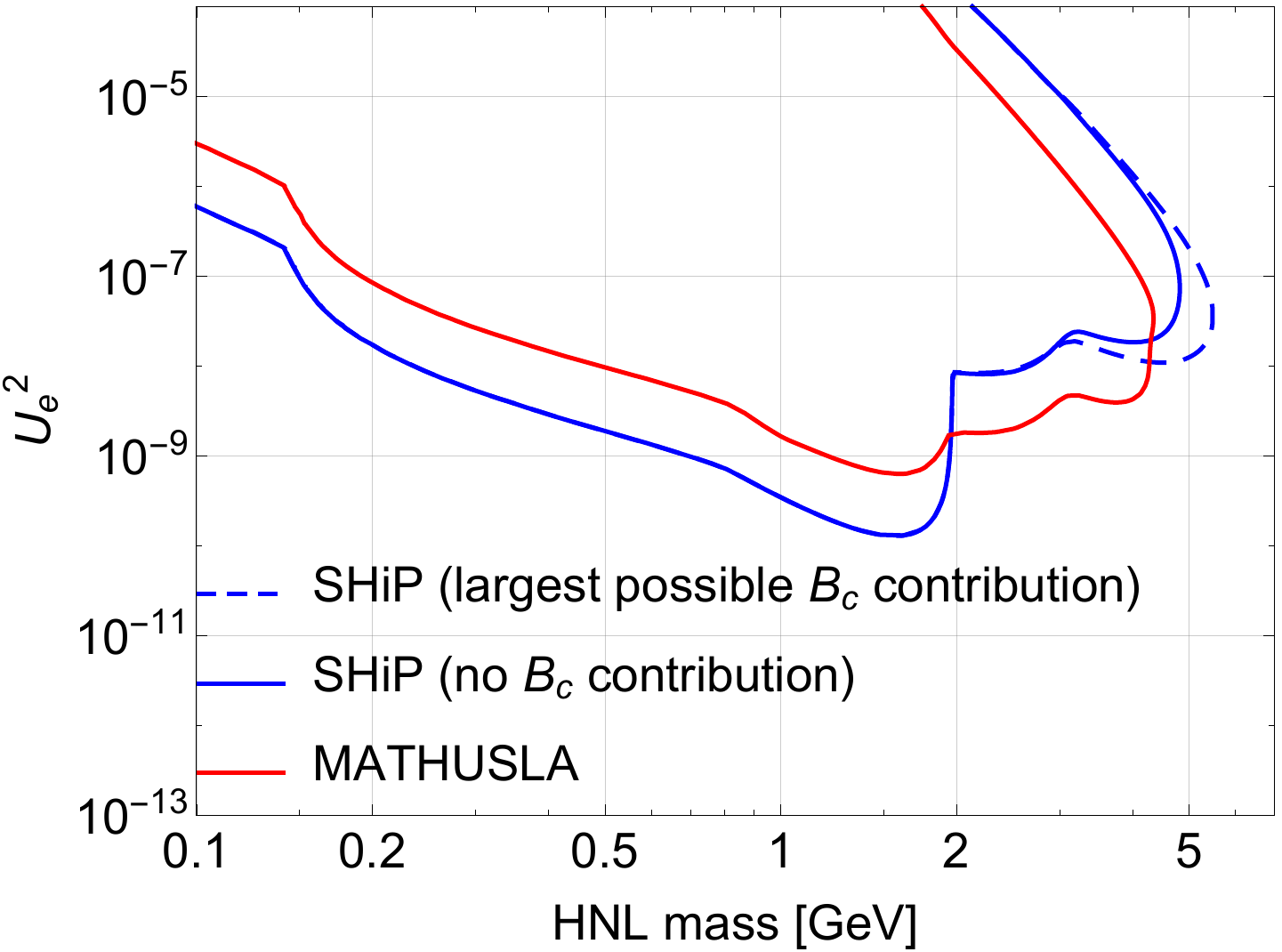}~
  \includegraphics[width=0.5\textwidth]{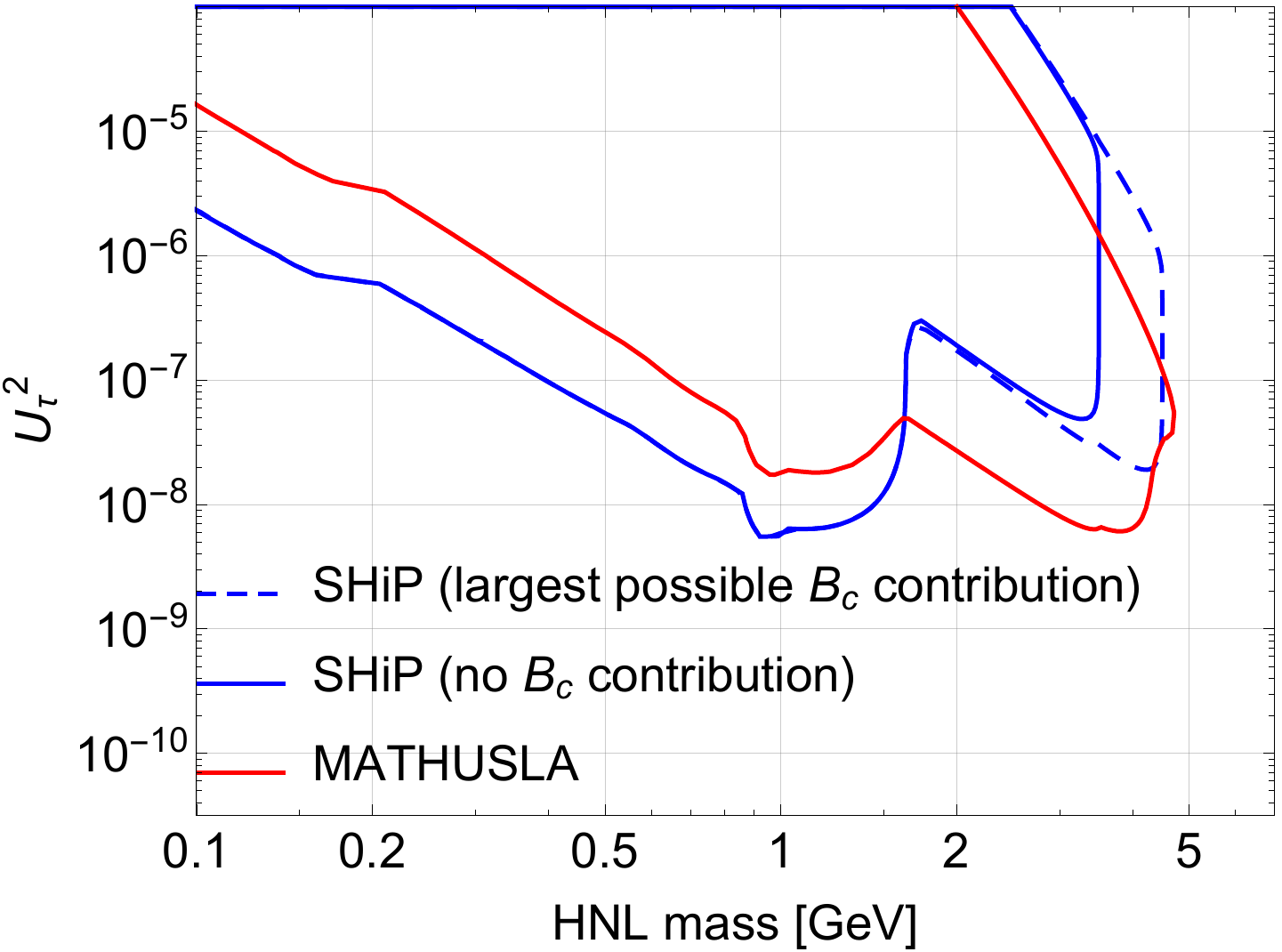}
  \caption{Comparison of the sensitivity of SHiP and MATHUSLA for the HNL. The production fraction of $B_c$ mesons at SHiP energies $\sqrt{s} \approx 28\text{ GeV}$ is not known, and the largest possible contribution is based on the production fraction measured at the LHC, $f(b\to B_c)= 2.6\times 10^{-3}$. In the case of the SHiP experiment we used the overall efficiency calibrated against the Monte Carlo simulations~\cite{SHiP:2018xqw} and also selected only those channels where at least two charged tracks from the HNL decay appear. In the case of the MATHUSLA experiment we optimistically used $\epsilon_{\det} = 1$ for the detection efficiency.}
  \label{fig:sensitivity-HNL}
\end{figure}

\begin{figure}[!t]
  \centering \includegraphics[width=0.5\textwidth]{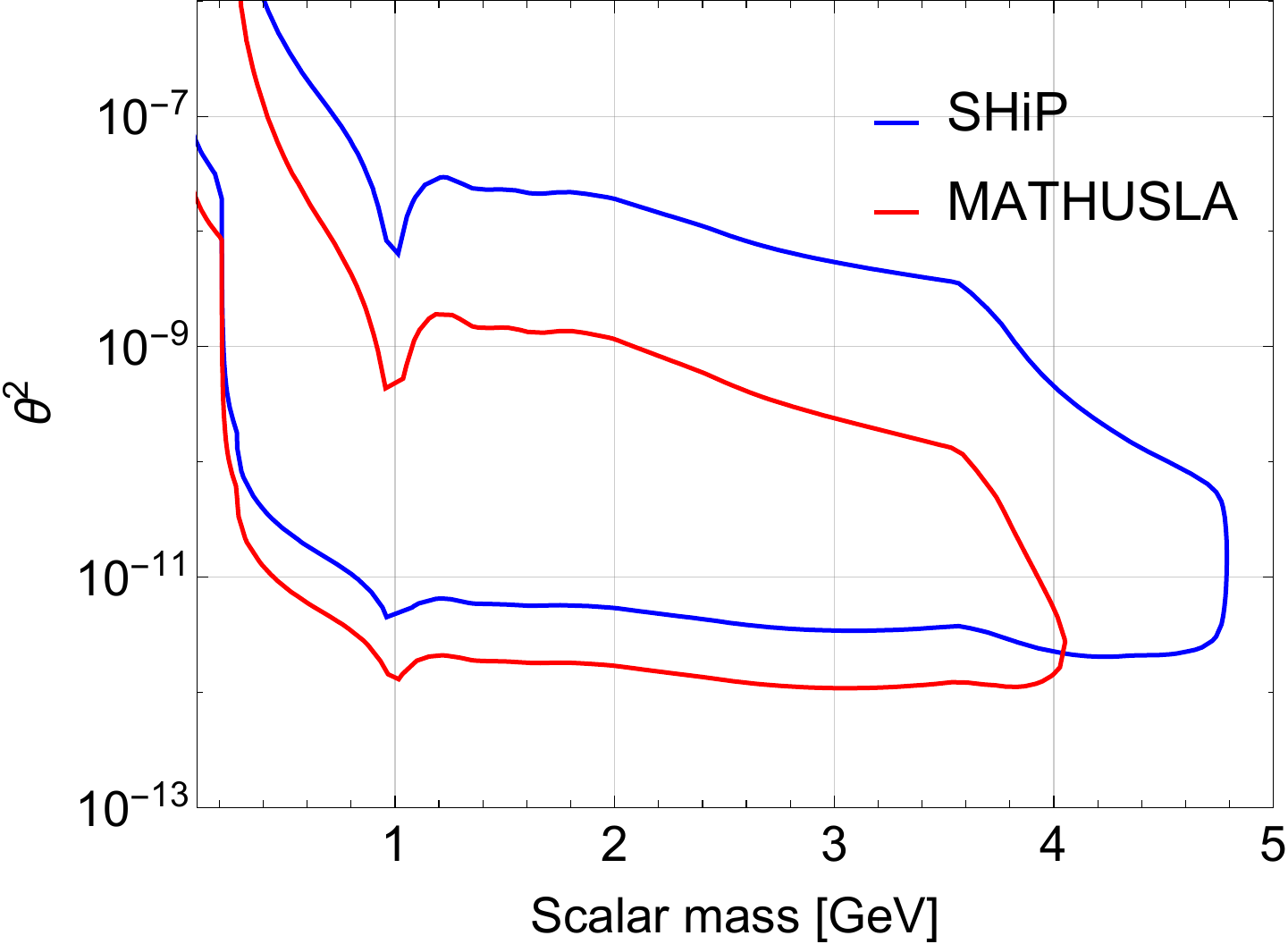}
  \caption{Comparison of sensitivities of the SHiP and MATHUSLA experiments for the scalar portal model. In the case of the SHiP experiment we used the overall efficiency $\overline{\epsilon} = 0.2$, see the text for details. In the case of the MATHUSLA experiment we optimistically used $\epsilon_{\det} = 1$ for the detection efficiency. We used the scalar phenomenology described in~\cite{Boiarska:2019jym}.}
  \label{fig:sensitivity-scalar}
\end{figure}

\paragraph{Our main results are as follows.} 
Our estimates of the sensitivities of the SHiP and MATHUSLA experiments to the HNLs are shown in Figs.~\ref{fig:sensitivity-HNL} and to the dark scalars in Fig.~\ref{fig:sensitivity-scalar}.

Qualitatively both experiments can probe similar ranges of parameters. 
The SHiP has higher average $\gamma$ factors of the mesons ($\langle\gamma_{\text{meson}}^{\ship}\rangle/\langle\gamma_{\text{meson}}^{\mat}\rangle \simeq \mathcal{O}(10)$) and, as a result, significantly higher upper boundary of the sensitivity region than MATHUSLA (as the upper boundary is exponentially sensitive to the $\gamma$ factor). As the consequence, the SHiP can probe higher masses for both HNLs and scalars than MATHUSLA (except of HNLs with dominant mixing with tau flavor). However, the $W$ boson decays at the LHC would produce some highly boosted HNLs traveling to the MATHUSLA decay volume, partly mitigating this difference. 

The SHiP experiment is able to probe lower mixing angles for HNLs with $M_N \lesssim m_{D_s}$ owing to the larger number of $D$ mesons. 
MATHUSLA can probe lower mixing angles for the HNLs with $M_{N}\gtrsim m_{D_{s}}$ and the scalars for all masses, owing to the larger number of the $B^{+/0}$ mesons at the LHC (as charmed mesons contribute negligibly to the scalar production).

\paragraph{Uncertainties.} According to the theoretical predictions the $d\sigma/dp_{T}$ distribution of $B$ mesons at the LHC has a maximum at  $p_{T}\sim \unit{GeV}$, see Fig.~\ref{fig:b-d-mesons-simulations}. The region of low $p_T$ is complicated for the theoretical predictions because of limitations of the applicability of the perturbative QCD. At the same time, these cross-sections have not been measured by neither the ATLAS, nor the CMS collaborations in the required kinematic range. 
The increase in the overall amount of low-momentum mesons  shifts leftwards the position of the peak of the  $d\sigma/dp_{T}$ distribution, thus decreasing their average momentum. 
Both factors lead to the increase of the number of events at the lower boundary. 
Therefore  the uncertainty in the position of the lower boundary of the sensitivity region depends on both of these numbers such that the uncertainty in the position of the peak enters into the sensitivity estimate \textit{squared}.

Another uncertainty comes from the background estimates. For the SHiP experiment, comprehensive background studies have proven that the yield of background events passing the online and offline event selections is negligible~\cite{Anelli:2015pba,SHIP:2018yqc}.
For MATHUSLA such an analysis is not available at the time of writing. 
The Standard Model background at MATHUSLA is non-zero (due to neutrinos from LHC and atmosphere, cosmic rays, muons, etc), however, it is claimed to be rejected with high efficiency based on the topology of the events~\cite{Curtin:2017izq,Curtin:2018mvb}. 
It is not known how much this rejection affects the detection efficiency, $\epsilon_{\det}$. 
In this work, we conservatively assumed $\epsilon_{\det}=1$ for MATHUSLA, while for SHiP it was taken from the actual Monte Carlo simulations~\cite{SHiP:2018xqw}.
More detailed analysis of the MATHUSLA background should be performed, which could influence the sensitivities.

In case of the SHiP experiment, the main uncertainty for HNLs is the unknown production fraction of the $B_c$ mesons at $\sqrt s \approx 28$~GeV. It changes the position of the lower bound and consequently the maximal mass probed in a significant way, see Fig.~\ref{fig:sensitivity-HNL}. 

\paragraph{Comparison with other works.}
We have compared our sensitivity estimates with the results of the Monte Carlo simulations presented by the SHiP collaboration~\cite{SHiP:2018xqw,Beacham:2019nyx,SHiP:2019} and with the estimates of the MATHUSLA physics case paper~\cite{Curtin:2018mvb} (Figs.~\ref{fig:sensitivity_comparison_HNLs}--\ref{fig:sensitivity_comparison_scalars}). For the HNLs, the estimates are in good agreement with the results of the SHiP collaboration. In the case of MATHUSLA, there is a difference for HNLs with mass smaller than $1$~GeV. It can be attributed to different branching for the HNL production used in our estimates and in the Monte Carlo simulations of~\cite{Curtin:2018mvb}, see discussion in Appendix~\ref{app:HNLsatMATHUSLA}. For the scalars, our estimates are in good agreement with the results from the SHiP and MATHUSLA collaboration. Small discrepancies between the sensitivities at the upper bound can be explained mainly by the difference in the meson energy spectrum used in our estimation and obtained in the Monte Carlo simulations.

\subsubsection*{Acknowledgements.}

We thank D.~Curtin, J.~Evans, R.~Jacobsson and W.~Valkenburg for fruitful discussions and comments on the manuscript. This project has received funding from the European Research Council (ERC) under the European Union's Horizon 2020 research and innovation programme (GA 694896) and from the Netherlands Science Foundation (NWO/OCW).

\appendix

\section{Portals}
\label{sec:coupling}

New particles with masses much lighter than the electroweak scale can couple to the Standard Model fields via renormalizable interactions with small dimensionless coupling constants (sometimes called ``\emph{portals}'' as they can mediate interactions between the Standard Model and ``hidden sectors''). In this work, we considered two {\em renormalizable} portals: scalar (or Higgs) portal and neutrino portal.

The scalar portal couples a gauge-singlet scalar $S$ to the gauge invariant combination $H^\dagger H$ made of the Higgs doublet:
\begin{equation}
  \label{eq:6}
  \mathcal{L}_{\text{scalar}} = \CL_{\rm SM} + \frac12 (\partial_\mu S)^2 - \frac{M_S^2}2 S^2 + g S H^\dagger H + \CL_{\rm int}
\end{equation}
where $g$ is the coupling constant and $\CL_{\rm int}$ are interaction terms that play no role in our analysis. After the spontaneous symmetry breaking the cubic term in~\eqref{eq:6} gives rise to the Higgs-like interaction of the scalar $S$ with all massive particles with their mass times a small
\emph{mixing parameter}
\begin{equation}
  \label{eq:scalar-mixing}
  \CL_{S, {\rm int}} = \theta S\biggl[\sum_f m_f \bar f f + M_W W^+_\mu W^-_\mu + \dots\biggr] \quad \theta \equiv \frac{g v}{m_H} \ll 1
\end{equation}
where $g$ is the coupling in~\eqref{eq:6}; $v$ is the Higgs VEV; $m_H$ is the Higgs mass; sum in~\eqref{eq:scalar-mixing} goes over all massive fermions (leptons and quarks); $W^\pm_\mu$ is the $W$ boson and $\cdots$ denote other interaction terms, not relevant for this work. The details of the phenomenology of the scalar portal are provided in~\cite{Boiarska:2019jym}
(see
also~\cite{Patt:2006fw,Batell:2009di,Clarke:2013aya,Schmidt-Hoberg:2013hba}).
The computation of hadronic decay width of $S$ is subject to large
uncertainties at masses $M_S\sim{}$few GeV, where neither chiral perturbation theory not perturbative QCD can provide reliable results (see a discussion in \cite{Monin:2018lee}).

We have also considered the neutrino portal where one adds to the Standard Model new gauge-singlet fermion -- heavy neutral lepton $N$ -- that couples to the $\epsilon_{ab} \bar L_a H_b$ where $L_a$ is the SU(2) lepton doublet and $\epsilon_{ab}$ is absolutely antisymmetric tensor in 2 dimensions. Phenomenologically, HNL is massive Majorana particle that possesses ``neutrino-like'' interactions with $W$ and $Z$ bosons (the interaction with
the Higgs boson does not play a role in our analysis and will be ignored). The interaction strength is suppressed as compared to that of ordinary neutrinos by a flavour-dependent factors (known as \emph{mixing angles}) $U_\alpha \ll 1$ ($\alpha=\{e,\mu,\tau\}$).
\section{Production and detection of portal particles}

\subsection{Production in proton-proton collisions}
\label{sec:production}

The number of mesons is determined by the number of produced $q\bar{q}$ pairs and fragmentation fractions $f_{\text{meson}}$, that can be extracted from the experimental data~\cite{Aaij:2011jp,Aaij:2017kea,Kling:2018wct}.
We summarize the fragmentation fractions that we use for MATHUSLA in the Table~\ref{tab:br-fractions}.
For the SHiP experiment, all fragmentation fractions except for $B_{c}$ meson are known to be close to the MATHUSLA's ones~\cite{Graverini:2133817}.
The $B_{c}$ meson fragmentation fraction at the energy of the SHiP experiment is unknown.
In our estimations, we take it the same as for the MATHUSLA experiment.
\begin{table}[!t]
  \centering
  \begin{tabular}{l|c|c|c|c|c|c|c}
    \toprule
   \textbf{Meson M} & $B^{+}$ & $B^{0}$ & $B_{s}$ & $B_{c}^{+}$ & $D^{+}$& $D^{0}$& $D^{+}_{s}$ \\
    \toprule
    \textbf{MATHUSLA} & 0.324 & 0.324& 0.088 & $2.6\cdot 10^{-3}$ &0.225 & 0.553& 0.105 \\
    \midrule
    \textbf{SHiP} & 0.417 & 0.418 & 0.09 & ? &0.207 & 0.632& 0.088 \\
    \bottomrule
  \end{tabular}
  \caption{The fragmentation fractions for heavy mesons at the LHC
    energies~\cite{Aaij:2011jp,Aaij:2017kea,Kling:2018wct} and of the SHiP
    experiment~\protect\cite{Graverini:2133817,Bondarenko:2018ptm}.
    For SHiP the contribution of flavoured baryons or quarkonia states can be neglected, see~\cite{Bondarenko:2018ptm}.
  For the LHC energies, the remaining 20-25\% come of all heavy flavour quarks hadronize into baryons, mostly $\Lambda_{b}$ states~\cite{Aaij:2011jp}.}
  \label{tab:br-fractions}
\end{table}

\subsubsection{HNL production}
\label{sec:hnl-production}
\begin{figure}[!htb]
  \begin{minipage}{0.51\textwidth}
    \centering
    \includegraphics[width=\textwidth,draft=false]{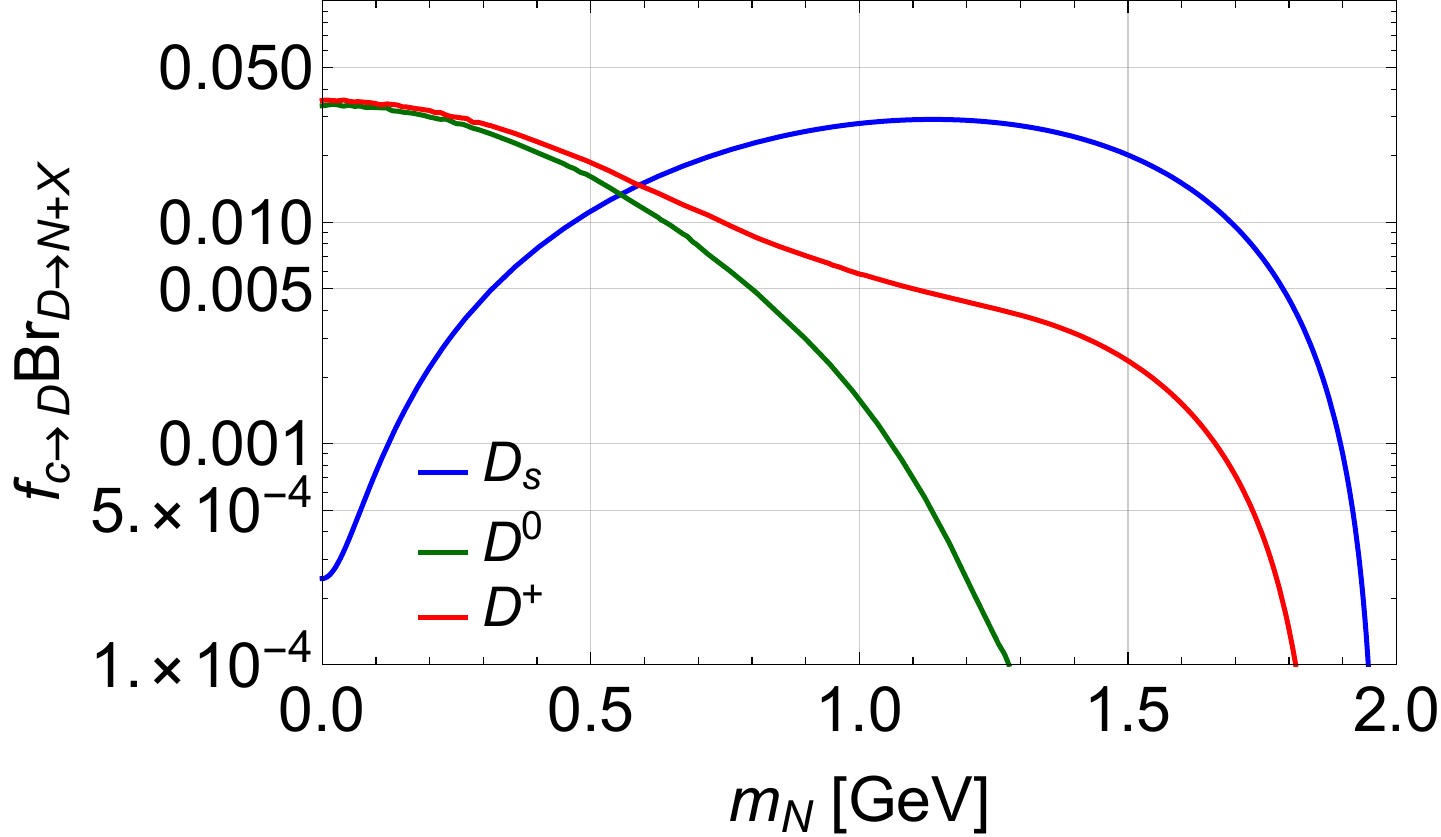}
  \end{minipage}\hfill
  \begin {minipage}{0.49\textwidth}
    \centering
    \includegraphics[width=\textwidth,draft=false]{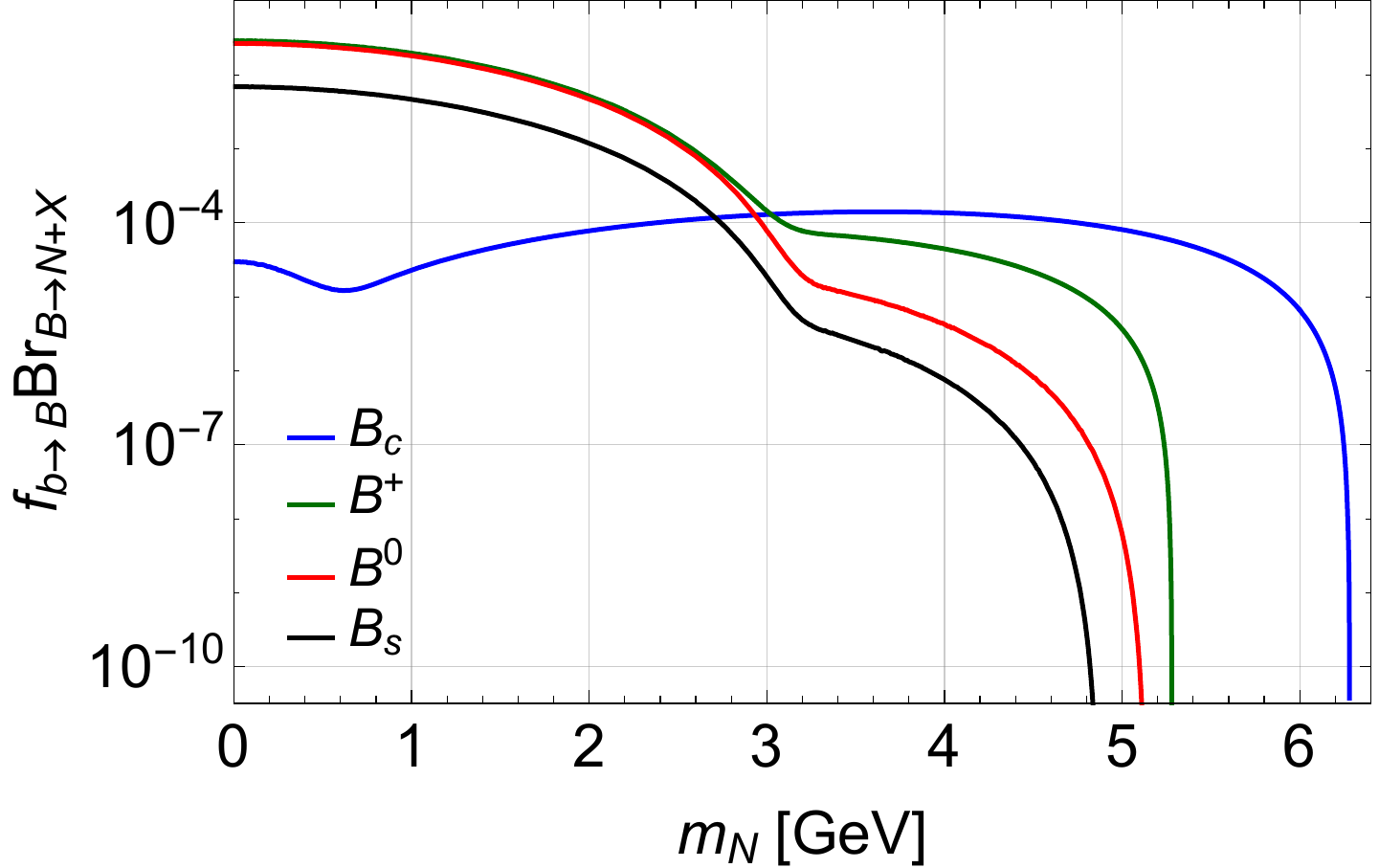}
  \end{minipage}
  \begin{minipage}{0.51\textwidth}
    \centering
    \includegraphics[width=\textwidth,draft=false]{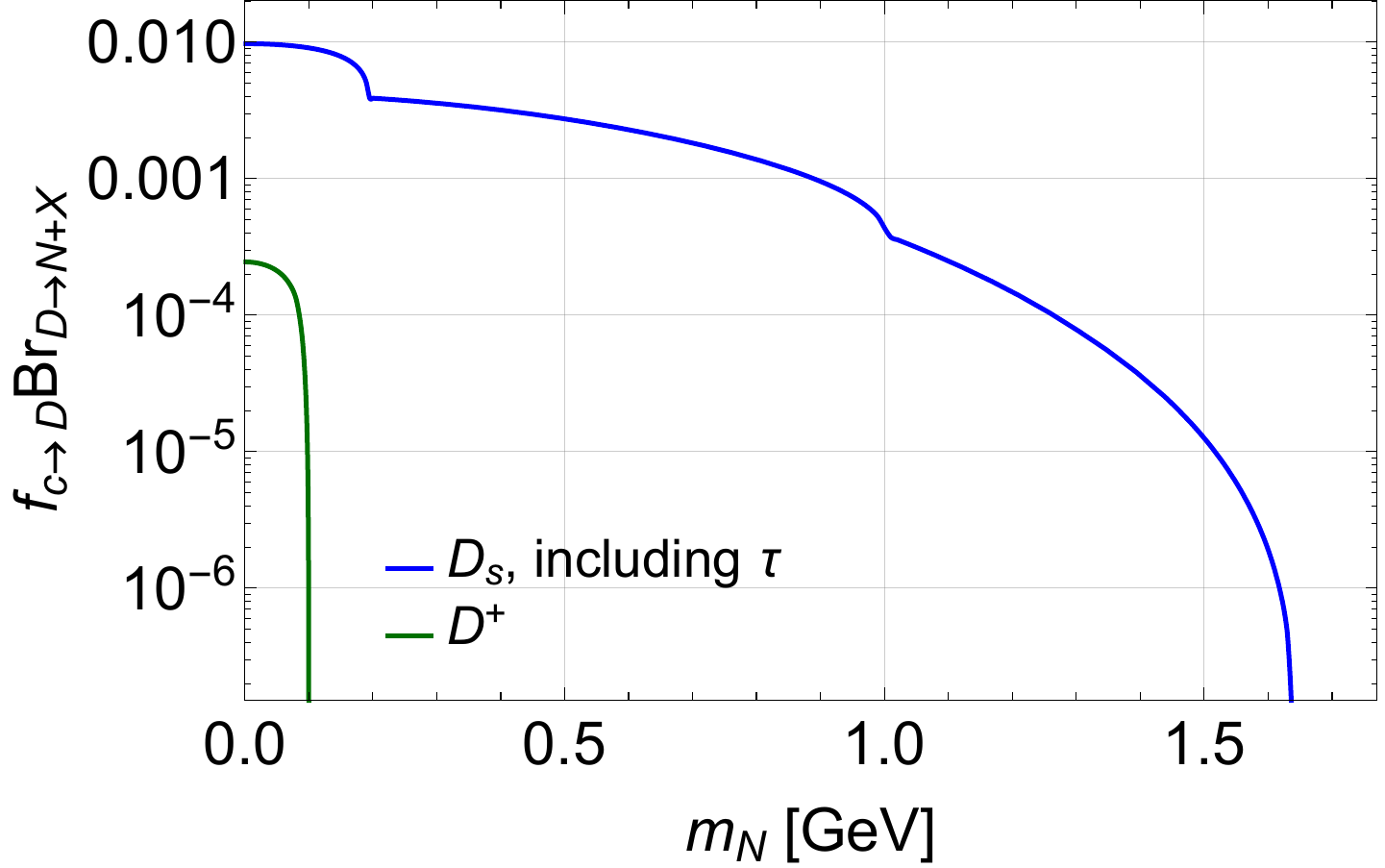}
  \end{minipage}\hfill
  \begin{minipage}{0.49\textwidth}
    \centering
    \includegraphics[width=\textwidth,draft=false]{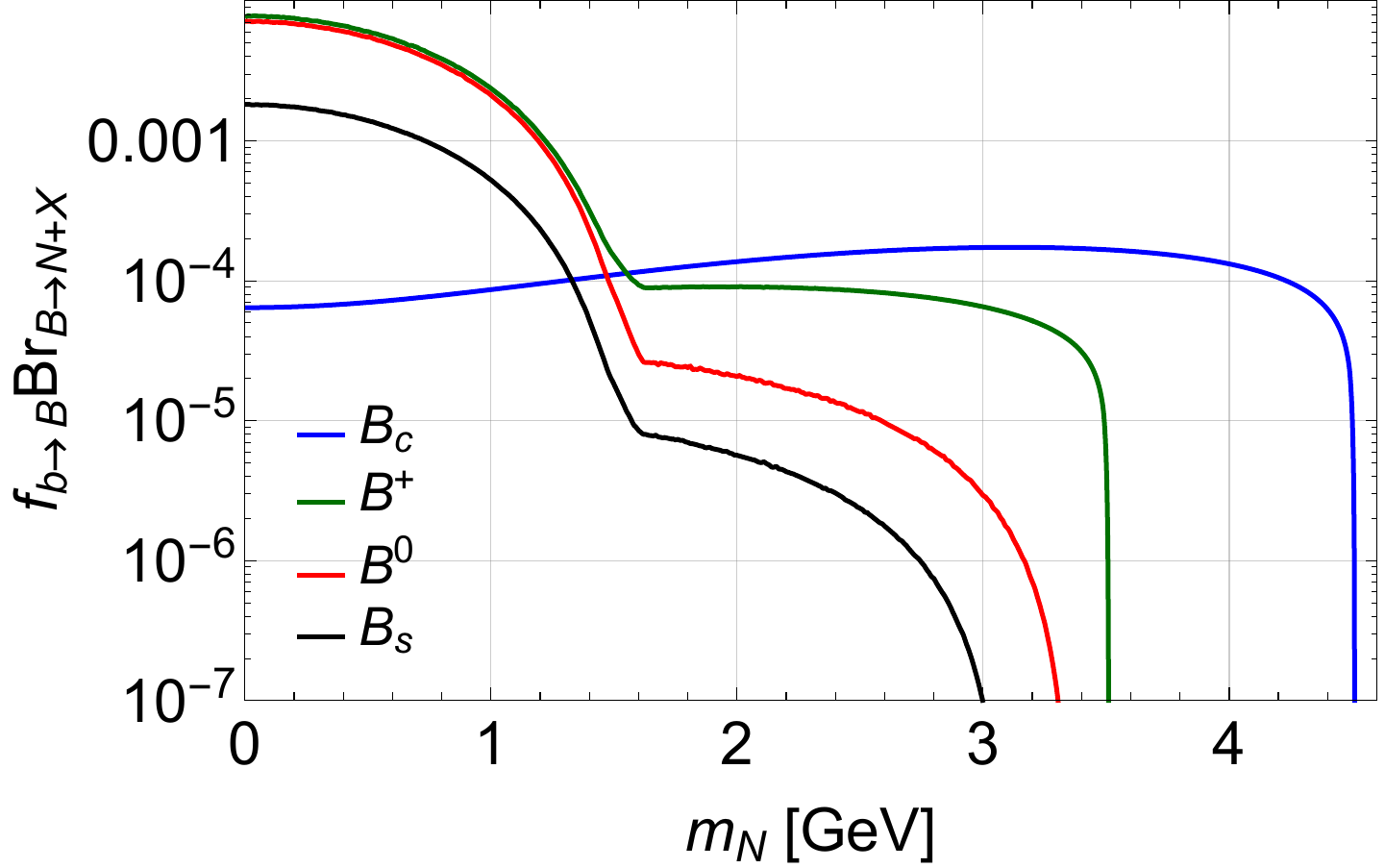}
  \end{minipage}
  \caption{Branching ratios multiplied by fragmentation fractions of $D$
    and $B$ mesons decaying into HNL through e-type mixing (upper panel) and into the HNL through $\tau$-type mixing (lower panel) for $U^{2}=1$. The values of fragmentation fractions are taken at LHC energies $\sqrt{s} = 13\text{ TeV}$, see Table~\ref{tab:br-fractions}.}
  \label{fig:mesons-br-app}
\end{figure}
The production of the HNL in the decay of charmed and beauty mesons have been
considered in~\cite{Gorbunov:2007ak,Atre:2009rg},
see~\cite{Bondarenko:2018ptm} for the recent review and summary of the results. The branching ratios multiplied by fragmentation fractions of $D$ and $B$ for the most relevant channels and the values of the fragmentation fractions from the Table~\ref{tab:br-fractions} are presented at the Fig.~\ref{fig:mesons-br-app}. We see that for the HNL mass range $m_N\gtrsim 3.5$~GeV the main production channel is $B_c$ meson decay $B_{c} \to N+ l$. This is a quite surprising fact, taking into account that $B_c$ fragmentation fraction is of order $10^{-3}$. To understand this result let us compare HNL production from $B_c$ with production from the two-body $B^+$ decay. The decay widths for both cases are given by
\begin{equation}
  \text{BR}(h\to \ell_{\alpha} N) \approx \frac{G_F^2 f_h^2 m_h m_N^2}{8\pi \Gamma_{h}} |V_h^{\text{CKM}}|^2 |U_\alpha|^2 K(m_N/m_h),
\end{equation}
where we take $m_N\gg m_{\ell}$, and $K$ is a kinematic
suppression. Neglecting them, for the ratio for the numbers of HNLs produced by $B_{c}$ and $B^{+}$ we obtain
\begin{equation}
  \frac{N_{\text{HNL}}(B_c\to \ell N)}{N_{\text{HNL}}(B^+\to \ell N)} \approx \underbrace{\frac{f_{B_c}}{f_{B^+}}}_{\approx 0.008}
  \times
  \underbrace{\frac{\Gamma_{B^+}}{\Gamma_{B_c}}}_{\approx 0.3}
  \times
  \underbrace{\left(\frac{f_{B_c}}{f_{B^+}}\right)^2}_{\approx 5}
  \times
  \underbrace{\frac{m_{B_c}}{m_{B^+}}}_{\approx 1.2}
  \times
  \underbrace{\left(\frac{V^{\text{CKM}}_{cb}}{V^{\text{CKM}}_{ub}}\right)^2}_{\approx 100} \approx 1.44.
\end{equation}
We see that the small fragmentation fraction of $B_c$ meson is compensated by the ratio of the CKM matrix elements and 
meson decay constants.

HNLs can also be produced in the decays of the $W$ bosons, $W \to N + l$. The corresponding branching ratio is
\begin{equation}
\label{eq:BrW-Nl}
    \BR(W\to N+\ell_\alpha) \approx \frac{1}{\Gamma_{W}}\frac{G_{F}m_{W}^{3}}{6\sqrt{2}\pi} \approx 0.1 U_{\alpha}^{2},
\end{equation}
where we have neglected the HNL and the lepton masses.

\subsubsection{Quarkonia and heavy flavour baryons}

Quarkonia states (especially $\Upsilon$ meson) can produce HNLs reaching $10$~GeV in mass above the beauty meson threshold.
The contribution of quarkonia decays to the production were found negligible at SHiP energies, see~\cite{Bondarenko:2018ptm}.

The LHC experiments  have measured $\Upsilon$ production at both ATLAS and CMS~\cite{Aad:2012dlq,Chatrchyan:2013yna}.
In the rapidity range $|y|<2$, relevant for MATHUSLA, the cross-section is given by~\cite{Hu:2017pat}
\begin{equation}
  \label{eq:Upsilon}
  \sigma(pp\to \Upsilon(nS))\times \BR(\Upsilon\to\mu^+\mu^-) \sim \unit[10]{nb}
\end{equation}
(as this is an order of magnitude estimate, we combine production of $1S$, $2S$ and $3S$ bottomonium states and neglected both statistical and systematic uncertainties of the cross-section measurement).
Using $\BR(\Upsilon\to\mu^+\mu^-) \simeq 2.4\times 10^{-2}$~\cite{Tanabashi:2018oca} we find that during the high luminosity phase  one can expect $N_\Upsilon \sim 10^{12}$. Large fraction of this mesons are traveling into the direction of the fiducial decay volume of MATHUSLA, as their distribution is sufficiently flat in the $|y|<2   $ rapidity range.
This number should be multiplied by the branching ration $\BR(\Upsilon \to
N\nu)$, estimated in~\cite{Bondarenko:2018ptm} to be at the level
$\BR(\Upsilon \to N\nu) \sim 10^{-5} U^2$, so that overall one expects in
MATHUSLA detector about
$10^7 U^2$ HNLs from $\Upsilon$ decays. 

This number should be compared with those, produced from $W$-bosons (as we are
above the $B$-meson threshold): $N_{W}\times \BR(W\to N+l)\times
\epsilon_{N}$, where $N_W$ is given by~\eqref{eq:nw-lhc}, $\epsilon_{N}
\approx 0.02$ is the geometric acceptance for the HNLs produced from $W$ and
flying into the MATHUSLA fiducial volume and the branching fraction  is given
by~\eqref{eq:BrW-Nl}.    The resulting number is $\sim 6\times 10^8 U^2$ --
exceeding the number of HNLs from $\Upsilon$-mesons by about 2 orders of
magnitude.

As Table~\ref{tab:br-fractions} demonstrates, about $25\%$ of $b$-quarks at the LHC hadronize into the $\Lambda_b^0$ baryons.
These baryons produce HNLs in the 3-body semi-leptonic decay $\Lambda_b^0 \to B + \ell + N$ where $B$ is a baryon.
The mass of the $\Lambda_b^0$ is $m_{\Lambda_b^0} \simeq \unit[5.62]{GeV}$.
The decays $\Lambda_b^0 \to p + \ell^- + N$ are suppressed by the CKM matrix element $V_{bu}$, while the decays $\Lambda_b^0 \to \Lambda_c^+ + \ell^- + N$ can only produce HNLs with $M_N < m_{\Lambda_b^0} - m_{\Lambda_c^+} \simeq \unit[3.35]{GeV}$.
HNLs of this mass are produced from more copious $B$-mesons and therefore $\Lambda$ baryons can be neglected. 

The contribution of heavy flavour baryon decays to the production were found
negligible at SHiP energies, see~\cite{Bondarenko:2018ptm}.

\subsubsection{Scalar production}
\label{sec:scalar-production}

The main difference in the phenomenology of the Higgs-like scalar $S$ in comparison to HNLs is that the interaction of $S$ with fermions is proportional to their mass, see Sec.~\ref{sec:coupling}. Therefore, its production at the mass range $M_S>M_K$ is dominated by the decay of the $B^{+}, B^{0}$, while the contribution from $D$ mesons is negligible~\cite{Bezrukov:2009yw,Boiarska:2019jym}. The main production process is the 2-body decay
\begin{equation}
    B \to X_{s/d}+S,
    \label{eq:scalar-production-process}
\end{equation}
where $X_{q}$ is a hadron that contains the quark $q$. The branching ratios for these states are discussed in~\cite{Boiarska:2019jym}. Here we only state the main points. For $B\to X_{s}+S$ we choose two lightest resonances $X_{s}$ for each given spin and parity. Exceptions are pseudo-scalar and tensor mesons (there is only one known meson that has these properties, see~\cite{Tanabashi:2018oca}). We have found that each heavier meson from the ``family'' gives a smaller contribution to the branching ratio than the lighter one. For $B\to X_{d}+S$ we take only one meson $X_{d} = \pi$ since this channel has the largest kinematic threshold $m_{B} - m_{\pi}$. We summarize the list of the final states below:
\begin{compactitem}[--]
     \item Spin 0, odd parity: $X_{q} = \pi, K$;
    \item Spin 0, even parity: $X_{q} = K_{0}^{*}(700), K_{0}^{*}(1430)$;
    \item Spin 1, odd parity: $K_{1}(1270), K_{1}(1400)$;
    \item Spin 1, even parity: $K^{*}(892), K^{*}(1410)$;
    \item Spin 2, even parity: $K_{2}^{*}(1430)$.
\end{compactitem}
The main source of the uncertainty is unknown quark squad of the $K_{0}^{*}(700)$ meson: it can be either a di-quark or a tetra-quark (see e.g.~\cite{Cheng:2013fba}). In the second case, the $K_{0}^{*}(700)$ contribution to the scalar production is unknown, which causes an uncertainty up to 30\%. We consider it as the di-quark state.

\begin{figure}[h!]
    \centering
    \includegraphics[width=0.8\textwidth]{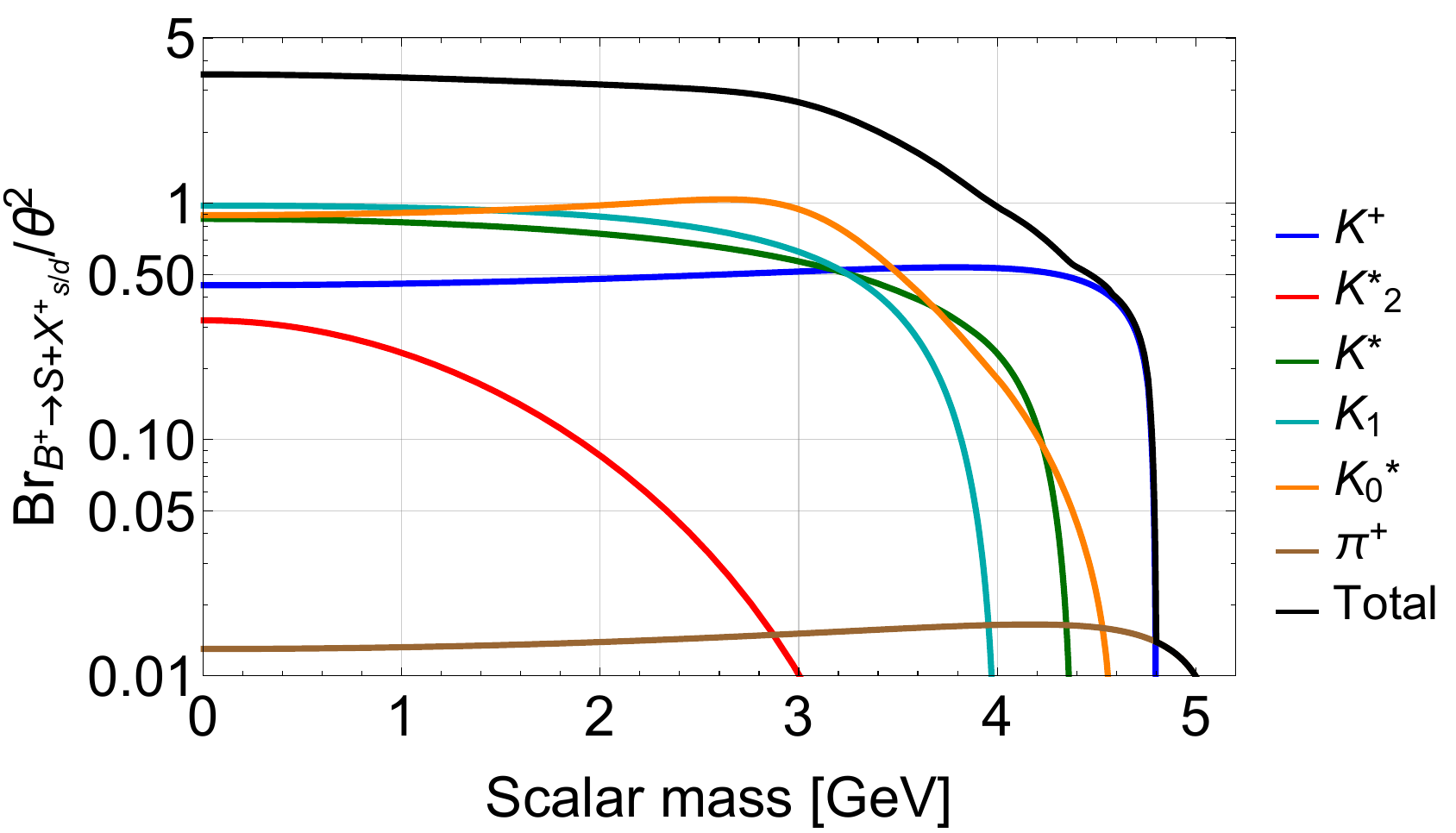}
    \caption{Branching ratios of the scalar production in the process $B \to S + X$, where $X$ denotes one of the mesons from the caption~\eqref{eq:scalar-production-process}.}
    \label{fig:branching-ratio-scalar}
\end{figure}

The dependence of the branching ratios of the process~\eqref{eq:scalar-production-process} on the scalar mass is shown in Fig.~\ref{fig:branching-ratio-scalar}.

We estimate the production of the scalars from the $W$ bosons by the decay $W\to S+f+\bar{f'}$, where the summation over all the SM fermions species $f = l,q$ is taken. We obtained $\BR_{W\to S}/\theta^{2} \simeq 4\cdot 10^{-3}$.

We mention in passing that the production of scalars from $\Upsilon$ (due to $b \to s$ transition) is not playing essential role, as the mass difference $m_\Upsilon - m_{B} < m_{B}$ and therefore one should compare the number of scalars produced from the bottomonium decays with the number of scalars from $B$-meson decays.
  The latter of $B$-mesons is several orders of magnitude higher (see Table~\ref{tab:b-d-parameters}).
  In addition to that the branching ratio of $\Upsilon \to B +S$ is much
  smaller than $B \to K+S$ because the width of $\Upsilon$ is dominated by
  electromagnetic decays.

  The $b\to s$ transitions also generate decays $\Lambda_b^0  \to \Lambda^0 +
  S$. However, the mass of thus produced scalar, $M_S < m_{\Lambda_b^0}  -
  m_{\Lambda^0} \simeq \unit[4.5]{GeV}$ and thus is subdominant to the
  production from $B$ mesons.
  
\subsection{Main decay channels}
\label{sec:decay}

\subsubsection{HNL}
\label{sec:hnl-decay}
The HNL has 3-body leptonic decays and different semileptonic
modes. Following the paper~\cite{Bondarenko:2018ptm}, we estimate the decay width of HNL into hadronic states as a sum of decay widths of specific channels for the HNL with a mass lower as 1~GeV and use the decay into the quarks with QCD corrections for larger masses. In the latter mass region, the decay width of HNLs mixing with the flavor $\alpha$ can be approximated by the formula
\begin{equation}
  \Gamma_N \approx g_{\text{eff}} |U_{\alpha}|^2 \frac{G_F^2 m_N^5}{192\pi^3},
  \label{eq:HNLdecaywidth}
\end{equation}
where $g_{\text{eff}}$ is a dimensionless factor that depends on the mass of HNL and changes from 1 to $\sim 10$, see e.g.~\cite{Bondarenko:2018ptm} for details.

The dependence of the proper lifetime $c\tau_{N}$ on the HNL mass at $U^{2}= 1$ is given at the left panel of Fig.~\ref{fig:ctau}.

\subsubsection{Scalar}
\label{sec:scalar-decay}
The decay width of the scalar particle has large uncertainty in the scalar mass region $0.5\text{ GeV}<M_S<2\text{ GeV}$ because of resonant nature of $S\to 2\pi$ decay, see~\cite{Monin:2018lee} for the recent overview. 
At higher masses the decay width is determined by perturbative QCD calculations~\cite{Spira:1997dg}. 
We omit the problem of pion resonance in this work using continuous interpolation between
the sum of the decay channel at low masses and perturbative QCD at high ones.

For scalar mass region above 2~GeV one can naively estimate $S$ decay width as $\Gamma_S\propto \sum_f \theta^2 y_f^2 M_S$. This estimation does not take into account three effects:
\begin{enumerate}
\item For the decay into quarks parameter $y_q$ depends on scalar mass as
  $y_q \equiv M_q(M_S^2)/v$, where $M_q(M_S^2)$ is quark running mass, which
  gives logarithmic correction;
\item The decay into gluons has different $M_S$ dependence,
  $\Gamma_S\propto \theta^2 M_S^3/v^2$, and dominates in the region
  $2\text{ GeV}<M_S<3.5\text{ GeV}$~\cite{Bezrukov:2009yw};
\item In the region $M_S$ near $3.5\text{ GeV}$ new decay channels appear
  (into $\tau$ and $c$ quark), and the kinematical factor is important.
\end{enumerate}
Taking them into account, for the mass domain $3.5\text{ GeV}<M_S<5\text{ GeV}$, near the threshold of production from $B$ mesons, we made a fit to the total $\Gamma_{S}$ and found that its behavior is $\Gamma_S \propto M_S^2$.

\begin{figure}[t!]
\begin{minipage}{0.5\textwidth}
    \centering \includegraphics[width=\textwidth]{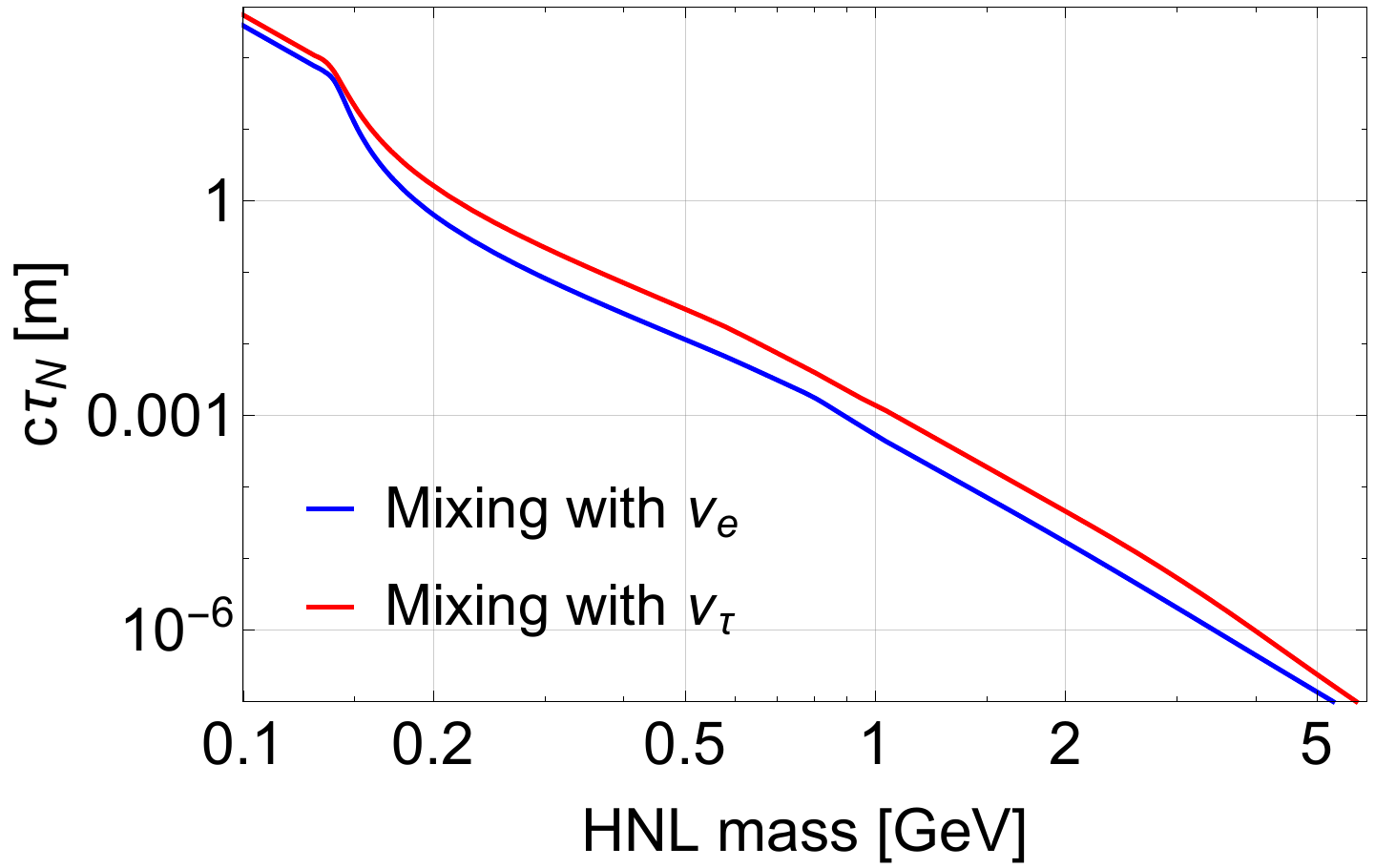}~
\end{minipage}
  \begin{minipage}{0.5\textwidth}
    \centering \includegraphics[width=\textwidth]{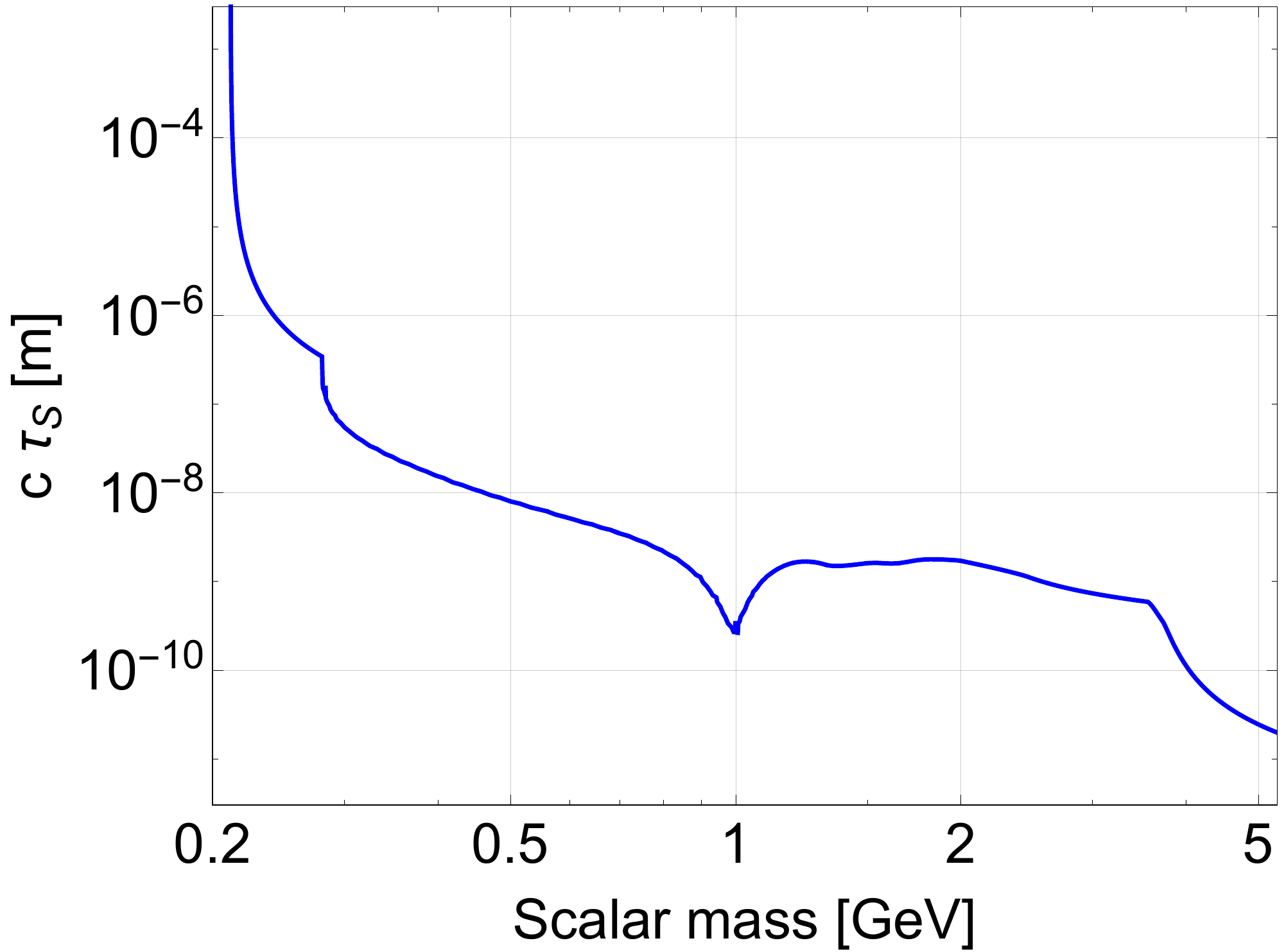}~
\end{minipage}
  \caption{The dependence of the proper lifetime $c\tau$ on the mass for the HNL (left panel, based on~\cite{Bondarenko:2018ptm}) and the scalar (right panel, based on~\cite{Boiarska:2019jym}).}
  \label{fig:ctau}
\end{figure}

The dependence of the proper lifetime $c\tau_{S}$ on the scalar mass for $\theta^2 = 1$ is shown in Fig.~\ref{fig:ctau} (right panel).

\subsubsection{Visible branching ratio}
\label{sec:visible-br}
We define the ``visible'' decay channels as those that contain at least two charged particles $\alpha$ in the final state. The corresponding decays are 
\begin{equation}
X \to \alpha \alpha' Y, \quad X \to F\tilde{Y}
\end{equation}
where $Y$ is arbitrary state, $F$ is uncharged state that decays to $n$ charged particles and $\tilde{Y}$ is a state with at least $2-n$ charged particles (assuming $n<2$). Using this definition, the decay $N \to 3\nu$ is identified as invisible decay, the decay $N \to \bar{\mu}\nu_{\mu} e$ --- as visible decay, while the decay $N \to \eta \nu$ --- as the visible decay if $\eta$ meson decays into two charged particles, and as invisible decay otherwise. To take into account only visible decays of $F$, we include the factor $\BR_{F\to \text{vis}}$ to the partial decay width $\Gamma_{X\to F\tilde{Y}}$. We take the values of $\BR_{F\to\text{vis}}$ from~\cite{Tanabashi:2018oca}.
For HNL/scalar masses $M > 2$~GeV we describe hadronic decays as having quarks and gluons in the final states. In this case we assume that any such decay will contain at least 2 charged tracks and therefore the whole hardonic width is visible.

\subsection{Comparison with scalar models used by SHiP and MATHUSLA collaborations}
\label{sec:scalar-phenomenologies-comparison}
Our model of the scalar production and decay described above differs from those used in~\cite{Curtin:2018mvb,Beacham:2019nyx} for estimating the sensitivity.
Namely, for scalar production in Ref.~\cite{Boiarska:2019jym} has summed over main exclusive channels $B \to X_{s/d}+S$.\footnote{There is $30\%$ level uncertainty in the total production rate because of the $B\to K^*_0(700) + S$ channel. The meson $K_{0}^{*}(700)$ is not observed experimentally and it could be either a di-quark or a tetra-quark state (see, e.g.~\cite{Cheng:2013fba} and references therein). We did an estimation assuming that $K^*_0(700)$ is the di-quark state, in the other case this production channel is absent.} In Ref.~\cite{Curtin:2018mvb} the production from $B$ mesons is estimated using the free quark model, while~\cite{Beacham:2019nyx} considers only the channel $B\to KS$.
This causes differences in the magnitude of the branching ratio and kinematic production thresholds.
In particular, we note that the free quark model breaks down for large scalar masses $M_{S}\simeq 3\text{ GeV}$ since the QCD enters the non-perturbative regime, and therefore it gives meaningless predictions for the production rate of heavy scalars. 

As for the scalar decay width, because of theoretical uncertainty for the mass range $2m_{\pi}\lesssim M_{S} \lesssim 2 m_{D}$ there is no agreement in the literature how to describe the scalar decays in this domain, see~\cite{Monin:2018lee}. Our decay width differs significantly from the decay width used in~\cite{Curtin:2018mvb,Beacham:2019nyx}.

\section{Relation between momentum of HNL and meson momentum}
\label{sec:hnl-momentum-meson-momentum}
The energy of the particle $X$ at lab
frame, $E_{X}$, is related to energy of $X$ at meson's rest frame, $E_{X}^{\text{rest}}$, and meson energy $E_{\text{meson}}$ at lab frame as
\begin{equation}
  \label{eq:lab-energy}
  E_{X}(\theta, E_{X}^{\text{rest}}) = \frac{E_{\text{meson}}}{m_{\text{meson}}}\left(E_{X}^{\text{rest}} + |\bm{p}_{X}^{\text{rest}}|\frac{|\bm{p}_{\text{meson}}|}{E_{\text{meson}}}\cos(\theta)\right),
\end{equation}
where $\theta$ is the angle between the direction of motion of meson in lab frame and the direction of  motion of the particle $X$ in the meson's rest frame. 
At meson frame the angle distribution is isotropic, so for the average energy we obtain
\begin{equation}
  \label{eq:gamma_X}
  \langle E_X\rangle = \gamma_{\text{meson}}\langle E_X^{\text{rest}}\rangle,
\end{equation}
where $\gamma_{\text{meson}} \equiv E_{\text{meson}}/m_{\text{meson}}$.

\section{Geometry of the experiments}
\label{sec:geometry-experiments}

\subsection{SHiP}

The SHiP experiment~\cite{SHIP:2018yqc} is a fixed-target experiment using the proton beam of the Super Proton Synchrotron (SPS) at CERN. SPS can deliver
$N_{\text{p.o.t.}}  = 2\times 10^{20}$ protons with the energy $400$~GeV over a $5$~year term. The SHiP will be searching for new physics in the largely unexplored domain of very weakly interacting particles with masses below $\mathcal{O}(10)$~GeV and $c\tau$ exceeding tens of meters. The overview of the experiment is as follows.

The proton beam hits a
target~\cite{Alekhin:2015byh,Anelli:2015pba}. The target will be followed by a $5~\text{m}$ hadron stopper, intended to stop all $\pi^{\pm}$ and $K$ mesons before they decay, and by a system of shielding magnets called active muon shield, constructed to sweep muons away from the fiducial decay volume. The whole active muon shield system is $34~\text{m}$ long.

The decay volume is a long pyramidal frustum vacuum chamber with the length
\begin{equation}
  l_{\text{det}} = 50\text{ m}
\end{equation}
and the cross-section $5\text{ m}\times 10\text{ m}$. It begins at
\begin{equation}
  l_{\text{target-det}} = 50 \text{ m}
\end{equation}
downstream of the primary target respectively. The SHIP spectrometer downstream of the decay volume consists of a four-station tracker, timing detector, and an electromagnetic calorimeter and muon detector for particle identification. The detectors are seen from the interaction point at an angle
$\theta \approx 25\text{ mrad}$.
\subsection{MATHUSLA}

\begin{figure}[h!]
  \begin{center}
    \begin{overpic}[width=0.6\textwidth,unit=1mm]{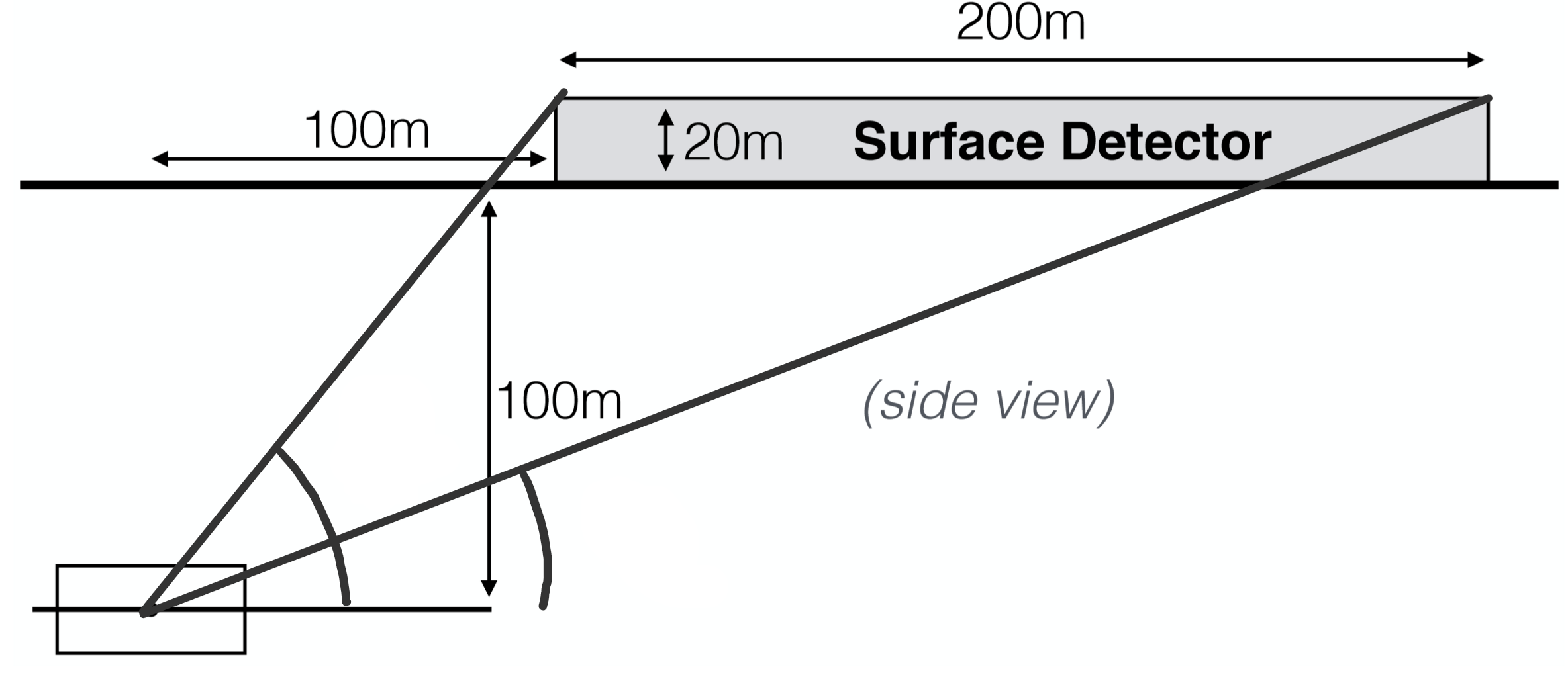}
    \put(19,11){$\theta_1$}
    \put(33,7){$\theta_2$}
    \end{overpic}
  \end{center}
  \caption{MATHUSLA experiment geometry. Adapted from~\cite{Chou:2016lxi}}
  \vspace{-0.4cm}
  \label{fig:mathusla-geometry}
\end{figure}
MATHUSLA (MAssive Timing Hodoscope for Ultra Stable neutraL pArticles) is a proposed experiment~\cite{Evans:2017lvd,Curtin:2017izq} that consists of a
$20\text{ m}\times 200\text{ m}\times 200\text{ m} $ surface detector, installed above ATLAS or CMS detectors (see Fig.~\ref{fig:mathusla-geometry}). The long-lived particles, created at the LHC collisions, travel 100+ meters of rock and decay within a large decay volume ($8\times 10^{5}\text{ m}^{3}$).  Multi-layer tracker at the roof of the detector would catch charged tracks, originating from the particle decays.  The ground between the ATLAS/CMS and MATHUSLA detector would serve as a passive shield, significantly reducing the Standard Model background (with the exception of neutrinos, muons and $K_L^0$ created near the surface). Assuming isotropic angular distribution of a given particle traveling to MATHUSLA, the average distance that it should travel to reach the MATHUSLA decay volume is equal to
\begin{equation}
  \label{eq:ltardet}
  \bar{l}_{\text{target-det}} \equiv \left\langle\frac{L_{\text{ground}}}{\sin\theta}\right\rangle = 192.5\text{ m}
\end{equation}
where $L_{\text{ground}} = 100\text{ m}$. The average distance a particle travels inside the decay volume, $\bar L_{\text{det}}$, is given by
\begin{equation}
  \label{eq:ldet}
  \bar{l}_{\text{det}} \equiv \left\langle\frac{20\text{ m}}{\sin\theta}\right\rangle = 38.5\text{ m}
\end{equation}
Geometrical parameters of MATHUSLA experiment are summarized in
Table~\ref{tab:mathusla-geometry}.
\begin{table}[!t]
  \centering
  \begin{tabular}{|c|c|c|c|c|c|c|c|}
    \hline
    Parameter & $\theta_{1}$& $\theta_{2}$& $\eta_{1}$ & $\eta_{2}$ & $\bar{l}_{\text{target-det}}$, m & $\bar{l}_{\text{det}}$, m & $\Delta \phi$ \\
    \hline
    Value & $44.3^{\circ}$ & $22.9^{\circ}$ & 0.9 & 1.6 & 192.5 & 38.5 & $\pi/2$ \\
    \hline
  \end{tabular}
  \caption{Parameters of MATHUSLA experiment~\cite{Curtin:2018mvb}. For the
    definition of angles $\theta_{1,2}$ see Fig.~\ref{fig:mathusla-geometry},
    and $\Delta \phi$ is the azimuthal size of MATHUSLA}
  \label{tab:mathusla-geometry}
\end{table}

\section{Analytic estimation of the upper bound: details}
\label{sec:widths-appendix}
In this Section we estimate the ratio between $\theta_{\text{max}}$ and
$\theta_{\text{upper}}$ -- the quantities defined in Section~\ref{sec:upper-bound}.

\subsection{Fits of the spectra}
\label{sec:spectra-fit}
The high-energy tail of the $B$ mesons distribution function at SHiP is well described by the exponential distribution, see the left Fig.~\ref{fig:b-spectra-fits}:
\begin{figure}[h!]
  \begin{minipage}{0.5\textwidth}
    \centering
    \includegraphics[width=\textwidth,draft=false]{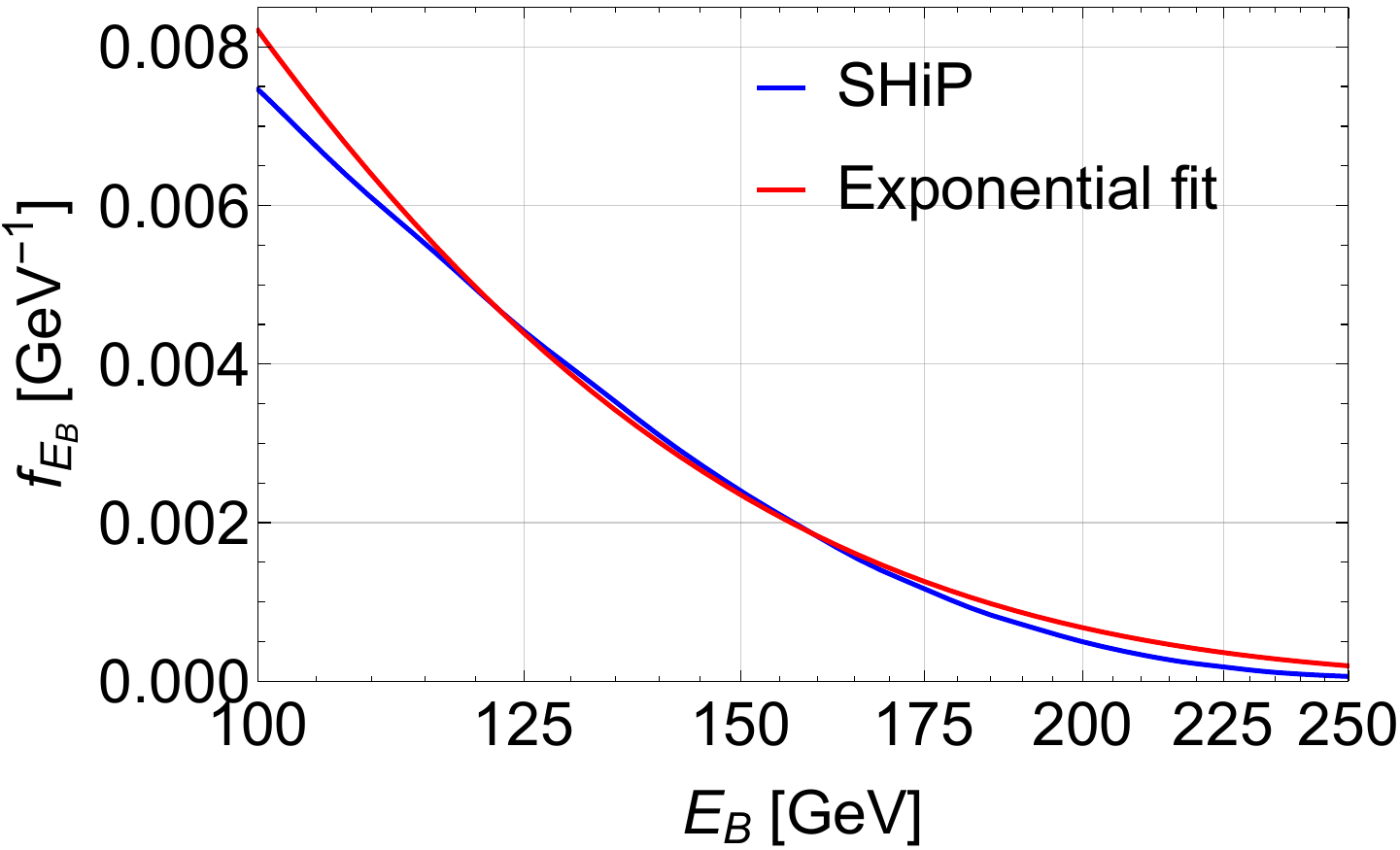}
  \end{minipage}\hfill
  \begin{minipage}{0.5\textwidth}
    \centering
    \includegraphics[width=\textwidth,draft=false]{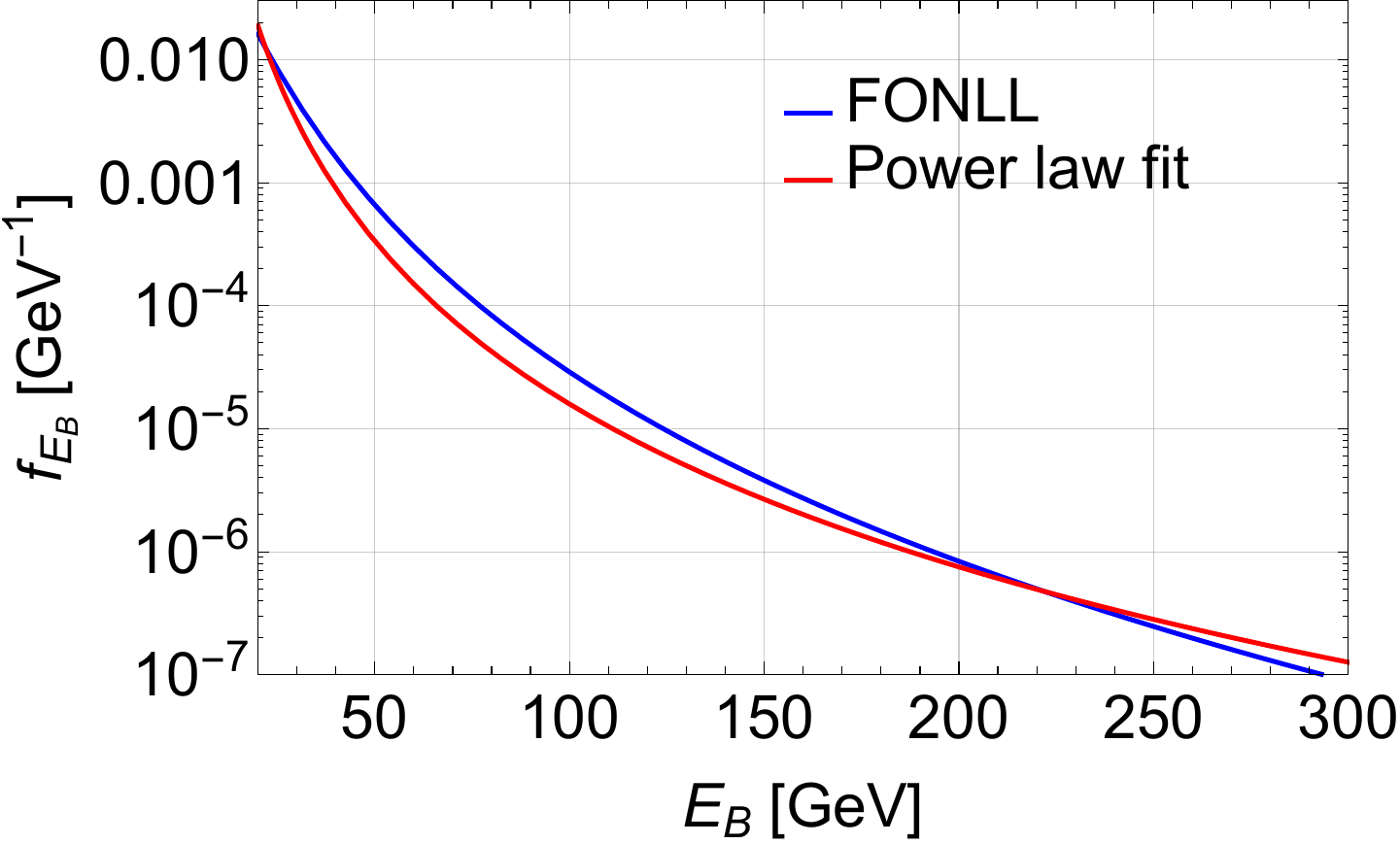}
  \end{minipage}
  \caption{Fits of the high energy tail of the distributions of the $B$ mesons. \emph{Left panel:} SHiP data is taken from~\cite{CERN-SHiP-NOTE-2015-009}. \emph{Right panel:} the FONLL simulations are performed for $\sqrt s = 13$~TeV and $|\eta| \sim 1$.}
  \label{fig:b-spectra-fits}
\end{figure}
\begin{equation}
  \label{eq:ship-B}
  \frac{dN_B^\ship}{dE} = f_{0} e^{-E\delta}, \quad \delta \approx 3\cdot 10^{-2}\text{ GeV}^{-1}\quad\text{and}\quad f_{0}\approx 0.3\text{ GeV}^{-1}
\end{equation}
The distribution of the high energy $B$ mesons at the LHC for energies $ E_{B}\lesssim 300\text{ GeV}$ can be approximated by the power law function, see the right panel in Fig.~\ref{fig:b-spectra-fits}:
\begin{equation}
  \label{eq:mathusla-B}
  \frac{dN_B^\mat}{dE}\approx \tilde{f}_{0}E^{-\alpha}, \quad \tilde{f}_{0}\approx \unit[1.6\times 10^{4}]{GeV}^{\alpha-1}\quad  \text{and}\quad\alpha \simeq 4.6
\end{equation}
Finally, the distribution of the HNLs originating from the $W$ bosons can be approximated by the expression
\begin{equation}
      \label{eq:mathusla-W}
  \frac{dN_{N,W}}{dE_{N}}\approx F_{0}\left(\frac{E_{1} - E_{N}}{E_{2}}\right), \quad F_{0} \approx 0.3\text{ GeV}^{-1}, \quad E_{1} = 60\text{ GeV}, \quad E_{2}\approx 107\text{ GeV}
\end{equation}

\subsection{Upper bound estimation}
\label{sec:upper-bound-details}
We start from the number of the events
\begin{equation}
    N_{\event}(M_{X},\theta_{X}^{2}) = \tilde{N}_{\text{prod}}(M_{X},\theta_{X}^{2}) \times \tilde{P}_{\decay}
\end{equation}
Here for simplicity we defined quantities $\tilde{P}_{\decay}$ and $\tilde{N}_{\decay}$ defined by
\begin{equation}
    P_{\decay} = \epsilon_{\det}\times \BR_{\text{vis}}\tilde{P}_{\decay}, \quad \tilde{N}_{\pr} = N_{\pr}\times \epsilon_{\det}\times \BR_{\text{vis}}
\end{equation}
The decay probability~\eqref{eq:decay-probability-upper-bound} can be rewritten in the form
\begin{equation}
  \tilde{P}_{\decay} =
  \int dE e^{-g(E)},
\end{equation}
where $g(E) = l_{\rm target-det}\Gamma_X M_X/E-\log\left(\frac{dN_X}{dE}\right)$
(for clarity we assumed $\frac{dN_X}{dE}$ to be dimensionless).
The integral~\eqref{eq:decay-probability-upper-bound} can be evaluated as
\begin{equation}
  \tilde{P}_{\decay}\approx \sqrt{\frac{2\pi}{-g''(E_{\mathrm{peak}})}}e^{-g(E_{\mathrm{peak}})}
\end{equation}
where $E_\mathrm{peak}$ is the minimum of $g(E) = 0$, here we used the steepest descent approximation.

For the exponential spectrum~\eqref{eq:ship-B} the peak energy and the probability are, correspondingly,
\begin{equation}
  E_{\mathrm{peak}} = \sqrt{\frac{l_{\text{target-det}}\Gamma_XM_X}{\delta}},\quad P_{\decay}\approx \sqrt{\pi}f_{0}e^{-2E_{\mathrm{peak}}\delta}\sqrt{\frac{E_{\mathrm{peak}}}{\delta}},
\end{equation}
while for the power law spectrum $f_{E} = f_{0}E^{-\alpha}$ they are
\begin{equation}
  E_{\mathrm{peak}} = \frac{l_{\text{target-det}}\Gamma_XM_X}{\alpha},\quad P_{\decay}\approx \sqrt{\frac{2\pi}{\alpha}}f_{0}e^{-\alpha}(E_{\mathrm{peak}})^{-\alpha}
\end{equation}
Expressing then $\Gamma_X \propto U^{2}$ and
$\theta^{2}_{X} \equiv \theta^{2}_{\text{max}}(M_X)\times R$, for the upper bound given by the particles produced from $B$ mesons one immediately arrives to
\begin{align}
  \label{eq:r-distribution}
  \theta^{\text{2,SHiP}}_{X,\text{upper}}(M_{X}) &\approx \theta^{\text{2,SHiP}}_{X,\text{max}}(M_{X})\frac{\log^{2}\left(f_{0}\sqrt{\pi}(\delta^{-3}\langle p_{B}\rangle)^{\frac{1}{4}}\tilde{N}_{\text{prod}}(M_{X},\theta^{\text{SHiP}}_{X,\text{max}}(M_{X})\right)}{4\langle p_{B}\rangle \delta},
  \\
  \theta^{\text{2,MAT}}_{X,\text{upper}}(M_{X}) &\approx 
  \theta^{\text{2,MAT}}_{X,\text{max}}(M_{X})
\frac{\alpha}{\langle p_{B}\rangle}\left(\sqrt{\frac{2\pi}{\alpha^{3}}}\tilde{N}_{\text{prod}}(M_{X},\theta^{\text{MAT}}_{X,\text{max}}(M_{X}))\right)^{\frac{1}{\alpha-2}}.
\end{align}
Similarly we can estimate the upper bound for the $W$ bosons.

We estimate $\theta_{\text{max}}$ for the production from $B$ mesons using the HNLs momenta given by Eq.~\eqref{eq:5} with $\langle p_{B}\rangle = 12\text{ GeV}$ for MATHUSLA and $\langle p_{B}\rangle = 88\text{ GeV}$ for SHiP (see Table~\ref{tab:ship-mathusla}), while for the production for $W$ we use $\langle p_{N}\rangle \approx 62\text{ GeV}$, see Sec.~\ref{sec:w-boson-distributions}.

\section{Details of the sensitivity curve drawing}
\label{sec:simulations}
We draw the sensitivity curve for HNLs and scalars requiring
\begin{equation}
  \label{eq:decay-events-accurate}
  N_{\event}(\theta_{X},M_X) =\sum_{\text{meson}} N_{\event,\text{meson}}+N_{\event,W} \geqslant 2.3,
\end{equation}
where numbers of decay events of particles produced from $B,D$ mesons and $W$ bosons are estimated as
\begin{equation}
   N_{\event,\text{meson}} = N_{q\bar{q}}\times f_{q\to \text{meson}}\text{Br}_{\text{meson} \to X} \times \epsilon_{\text{meson}}\int dp_{\text{meson}}f_{p_{\text{meson}}} \times \tilde{P}_{\decay}(p_{\text{meson}}),
\end{equation}
\begin{equation}
   N_{\event,W} = N_{W,\text{LHC}}\times \text{Br}_{W \to X} \times \epsilon_{W}\times \int dp_{X} \ f_{p_{X},W}\times \tilde{P}_{\decay}(p_{X})
\end{equation}
Here, $N_{q\bar{q}}$ is the total number of the $q\bar{q}$ pairs that are produced in $pp$ collisions at the high luminosity LHC or in $p-\text{target}$ collisions at SHiP, $f_{p_{\text{meson}}}$ is the momentum distribution of the mesons that fly to the decay volume of the experiment (see Sec.~\ref{sec:meson-distributions}), and $\epsilon$ is the overall efficiency (see Sec.~\ref{sec:efficiency}). In the expression for the decay probability $\tilde{P}_{\decay}$ the $\gamma$ factor of the $X$ particle is related to the meson momentum by the relation~\eqref{eq:gamma_X}. $N_{W,\text{LHC}}$ is the number of the $W$ bosons produced at the high luminosity LHC, and $f_{p_{X},W}$ is the momentum distribution function of the $X$ particles (see Sec.~\ref{sec:w-boson-distributions}).
\section{Analytic estimation of the lower bound for particular masses}
\label{sec:nevents-lower-bound-prediction}
Here we make an analytic estimation of the lower bound for particular masses using the formula~\eqref{eq:nevents-lower-bound-explicit}. The relevant parameters and the value $\theta^{2}_{\text{lower}}$ for the SHiP and MATHUSLA experiments are given in Tables~\ref{tab:analytic-estimation-ship},~\ref{tab:analytic-estimation-mathusla}.

\begin{table}[h!]
    \centering
    \begin{tabular}{|c|c|c|c|c|c|c|c|}
    \hline
        $X/M$ & $M_{X},\text{ GeV}$ &  $f_{M}\times \BR_{M\to X}$ & $\langle \gamma_{X}\rangle$ & $c\tau_{X}, \text{ m}$ & $\epsilon$ & $U^{2}_{e,\text{lower}}$ 
         \\
            \hline
         $N,\nu_{e}/D$ & $0.5$ &  $4.5\cdot 10^{-2}$ & $30.1$ & $1.1\cdot 10^{-2}$ & $6.5\cdot 10^{-2}$ & $1.9\cdot 10^{-9}$
         \\
         \hline
         $N,\nu_{e}/B$ & $3$ & $4\cdot 10^{-4}$ & $20.2$ & $2\cdot 10^{-6}$ & $0.13$ & $1.7\cdot 10^{-8}$
         \\
         \hline
         $S/B$ &$0.5$ &  $88.2$ & $30.1$ & $4.3\cdot 10^{-9}$ & $0.14$ & $2.5\cdot 10^{-11}$
         \\
         \hline
         $S/B$ & $2.5$ & $21.4$ & $20.2$ & $1.\cdot 10^{-9}$ & $0.2$ & $4.8\cdot 10^{-12}$
         \\
         \hline
    \end{tabular}
    \caption{Table of parameters used in simple analytic estimation of the lower bound~\eqref{eq:nevents-lower-bound-explicit} for the SHiP experiment for particular masses of the HNLs with the pure electron mixing and the scalars. The columns are as follows: the type of the particle $X$ and the mother particle, the branching ratio of the production of $X$ at $\theta_{X}^{2} = 1$, average $\gamma$ factor, proper decay length $c\tau_{X}$ at $\theta_{X}^{2} = 1$, overall efficiency~\eqref{eq:overall-efficiency}, the mixing angle at the lower bound estimated as $N_{\event,\text{lower}} = 2.3$, where $N_{\event,\text{lower}}$ is given by~\eqref{eq:nevents-lower-bound-explicit}.}
    \label{tab:analytic-estimation-ship}
\end{table}

\begin{table}[h!]
    \centering
    \begin{tabular}{|c|c|c|c|c|c|c|c|}
    \hline
         X/M & $M_{X},\text{ GeV}$ & $f_{M}\times \BR_{M\to X}$ & $\langle \gamma_{X}\rangle$ & $c\tau_{X}, \text{ m}$ & $\epsilon$ & $\theta^{2}_{\text{lower}}$ 
         \\
            \hline
         $N, \nu_{e}/D$ & $0.5$ & $4.5\cdot 10^{-2}$ & $2.4$ & $1.1\cdot 10^{-2}$ & $6.5\cdot 10^{-2}$ & $1.2\cdot 10^{-8}$
         \\
         \hline
         $N, \nu_{e}/B$ & $2.5$ & $2.6\cdot 10^{-3}$ & $2.2$ & $5.4\cdot 10^{-6}$ & $1.8\cdot 10^{-2}$ & $2.4\cdot 10^{-9}$
         \\
         \hline
         $N, \nu_{e}/W$& $1$ & $0.11$ & $62$ &  $6.5\cdot 10^{-2}$ & $1.9\cdot 10^{-2}$ & $1.7\cdot 10^{-6}$
         \\
         \hline
         $N, \nu_{e}/W$ & $2.5$ & $0.11$ & $24.8$ & $5.4\cdot 10^{-6}$ & $1.9\cdot 10^{-2}$ & $1.1\cdot 10^{-7}$
         \\
                  \hline
         $S/B$ & $0.5$ & $5.3$ & $9$ & $4.3\cdot 10^{-9}$ & $1.8\cdot 10^{-2}$ & $4.2\cdot 10^{-12}$
         \\
                  \hline
         $S/B$ & $2.5$ & $2.7$ & $2$ & $1.\cdot 10^{-9}$ & $0.1$ & $4.9\cdot 10^{-13}$
         \\
                  \hline
    \end{tabular}
\caption{Table of parameters used in simple analytic estimation~\eqref{eq:nevents-lower-bound-explicit} for the MATHUSLA experiment for particular masses of the HNLs and scalars. We use the description of the scalar phenomenology from~\cite{Curtin:2018mvb}. The columns are as follows: the type of the particle $X$ and the mother particle, the branching ratio of the production of $X$ at $\theta_{X}^{2} = 1$, average $\gamma$ factor, proper decay length $c\tau_{X}$ at $\theta_{X}^{2} = 1$, overall efficiency~\eqref{eq:overall-efficiency}, the mixing angle at the lower bound estimated as $N_{\event,\text{lower}} = 4$, where $N_{\event,\text{lower}}$ is given by~\eqref{eq:nevents-lower-bound-explicit}.}
    \label{tab:analytic-estimation-mathusla}
\end{table}

\section{HNLs at MATHUSLA for small mass}
\label{app:HNLsatMATHUSLA}

For HNLs with $M_{N}\lesssim m_{D_{s}}$, where the sensitivity curve is
determined by the production  from $D$ mesons our sensitivity curve reproduces that of the MATHUSLA collaboration~\cite{Curtin:2018mvb} in the range up to 1~GeV (Fig.~\ref{fig:sensitivity_comparison_HNLs}).
For smaller masses our estimates of the lower boundary differ by a factor $\sim 3$ (which would corresponds to the order-of-magnitude difference between the number of decay events).
Moreover, the shapes of the sensitivity curves also differ.

Below we list several possible reasons for this discrepancy:
\begin{compactenum}[\bf a)]
\item different estimate of the number of the parent  $D$ mesons produced
\item HNLs that are produced from the mesons
  that \emph{do not fly} into the fiducial decay volume of MATHUSLA (they were not taken into
  account in our estimate)
\item different HNL phenomenology (production and decay) used in comparison.
\item Production from $K$-mesons was not taken into account in our estimates
\end{compactenum}

\medskip

Since the positions of the lower bounds in the mass range $\unit[1]{GeV}\lesssim M_{N} \lesssim m_{D_{s}}$ are in good agreement, we conclude that case a) with different amounts of $D$ mesons is less probable, as it should shift the lower bound for the whole mass range $M_{N}\lesssim m_{D}$.

In order to estimate the amount of  light HNLs produced from  $D$ mesons, we performed a MadGraph 5 simulation of a process $pp \to \bar{c} e^{+}\nu_{e}s$, whose kinematics corresponds to the main process for a production of light HNLs $M_{N}\lesssim \unit[0.5]{GeV}$ --- $D \to N e^{+}K$~\cite{Bondarenko:2018ptm}. We computed the ratio
\begin{equation}
\chi = \left(\frac{\sigma_{\bar{c}e^{+}\nu_{e}s, \ \mat}}{\sigma_{\bar{c}e^{+}\nu_{e}s,\text{ tot}}}\right) / \left(\frac{\sigma_{c\bar{c},\ \mat}}{\sigma_{c\bar{c},\text{ tot}}}\right),
\end{equation}
where the first fraction is the amount of the HNLs that fly in the decay volume of MATHUSLA, while the second one is the amount of $c\bar{c}$ pairs that fly in the same direction. We found $\chi \approx 1.7$, which is not enough to explain the discrepancy. 

For c), we compared the decay widths of the HNLs used in our analysis with those, used in~\cite{Curtin:2018mvb} (based on~\cite{Helo:2010cw}). 
We  found them to be different by $20-40\%$ (for $U_e : U_\mu : U_\tau = 1$)
with the decay width from~\cite{Helo:2010cw} being smaller than that
from~\cite{Bondarenko:2018ptm}. The difference can reach up to 80\% at small
masses (below $\mathcal{O}(500)$~MeV).

Finally, production from $K$ mesons would not explain why the discrepancy starts close to 1~GeV, much higher than production threshold from kaons.

We did not find the information about the HNL production ratios used in~\cite{Curtin:2018mvb}. As we see, the cases b) -- d) are not enough to explain a factor $10$ in the number of events, and we assume that the main reason for the discrepancy is different production branching ratios adopted in~~\cite{Curtin:2018mvb}.

\bibliographystyle{JHEP} %
\bibliography{ship}

\end{document}